\documentclass[10pt,journal,compsoc]{IEEEtran}
\ifCLASSINFOpdf
\else
\fi
\usepackage{amsmath,amsfonts}
\usepackage{algorithmic}
\usepackage{algorithm}
\usepackage{array}

\usepackage[caption=false,font=normalsize,labelfont=sf,textfont=sf]{subfig}
\usepackage{subcaption}
\usepackage{textcomp}
\usepackage{stfloats}
\usepackage{url}
\usepackage{epigraph} 
\usepackage{pifont}
\usepackage{array}

\newcommand{\cmark}{\ding{51}}  
\newcommand{\xmark}{\ding{55}}  

\usepackage[table,xcdraw]{xcolor}
\usepackage[normalem]{ulem}

\usepackage{tikz}
\usepackage{forest}
\usetikzlibrary{trees,positioning,shapes,shadows,arrows.meta}

\usepackage{verbatim}
\usepackage{graphicx}
\usepackage{cite}
\usepackage[T1]{fontenc}
\usepackage[utf8]{inputenc}
\usepackage{makecell}
\usepackage{multirow}
\usepackage{xcolor}
\usepackage{enumitem}

\usepackage[]{mdframed}
\usepackage{xurl}
\usepackage[bookmarks=true, hidelinks, breaklinks,colorlinks,citecolor=blue]{hyperref}

\usepackage[skins]{tcolorbox}

\def\ie{{\em i.e.}}
\def\eg{{\em e.g.}}

\def\x{\mathbf{x}}

\begin{document}

\title{Detecting Multimedia Generated by 
Large AI Models: A Survey}


\author{Li Lin, Neeraj Gupta, Yue Zhang, Hainan Ren, Chun-Hao Liu, Feng Ding, Xin Wang,~\IEEEmembership{Senior Member,~IEEE,}
Xin Li,~\IEEEmembership{Fellow,~IEEE,} Luisa Verdoliva,~\IEEEmembership{Fellow,~IEEE,}
Shu Hu$^{*}$,~\IEEEmembership{Member,~IEEE}
\IEEEcompsocitemizethanks{
\IEEEcompsocthanksitem Li Lin, Neeraj Gupta, and Shu Hu are with the Department of Computer and Information Technology, Purdue University, West Lafayette, IN, 47907, USA. e-mail:(\{lin1785, gupt1031, hu968\}@purdue.edu)
\IEEEcompsocthanksitem Yue Zhang and Feng Ding are with the School of Software, Nanchang University, Nanchang, Jiangxi, 330031, China. e-mail:(\{yuezhang, fengding\}@ncu.edu.cn)
\IEEEcompsocthanksitem Hainan Ren. e-mail:(hnren666@gmail.com)
\IEEEcompsocthanksitem Chun-Hao Liu is with Amazon Prime Video, Sunnyvale, CA, 94089, USA. e-mail:(chunhaol@amazon.com)
\IEEEcompsocthanksitem Xin Wang is with the Department of Epidemiology and Biostatistics, School of Public Health, and Xin Li is with the Department of Computer Science, College of Nanotechnology, Science, and Engineering, both at University at Albany, SUNY, Albany, NY 12222, USA. e-mail:(\{xwang56, xli48\}@albany.edu)
\IEEEcompsocthanksitem Luisa Verdoliva is with the Department of Electrical Engineering and Information Technology, University Federico II of Naples,  Naples 80125, Italy. e-mail:(verdoliv@unina.it)
}
\thanks{$^*$ Shu Hu is the corresponding author.}
}



\IEEEtitleabstractindextext{%
\begin{abstract}
The rapid advancement of Large AI Models (LAIMs), particularly diffusion models and large language models, has marked a new era where AI-generated multimedia is increasingly integrated into various aspects of daily life. 
Although beneficial in numerous fields, this content presents significant risks, including potential misuse, societal disruptions, and ethical concerns. Consequently, detecting multimedia generated by LAIMs has become crucial, with a marked rise in related research. Despite this, there remains a notable gap in systematic surveys that focus specifically on detecting LAIM-generated multimedia. Addressing this, we provide the first survey to comprehensively cover existing research on detecting multimedia (such as text, images, videos, audio, and multimodal content) created by LAIMs. Specifically, we introduce a novel taxonomy for detection methods, categorized by media modality, and aligned with two perspectives: \textit{pure detection} (aiming to enhance detection performance) and \textit{beyond detection} (adding attributes like generalizability, robustness, and interpretability to detectors). Additionally, we have presented a brief overview of generation mechanisms, public datasets, online detection tools, and evaluation metrics to provide a valuable resource for researchers and practitioners in this field. Most importantly, we offer a focused analysis from a social media perspective to highlight their broader societal impact. Furthermore, we identify current challenges in detection and propose directions for future research that address unexplored, ongoing, and emerging issues in detecting multimedia generated by LAIMs. Our aim for this survey is to fill an academic gap and contribute to global AI security efforts, helping to ensure the integrity of information in the digital realm.
The project link is \url{https://github.com/Purdue-M2/Detect-LAIM-generated-Multimedia-Survey}.

\end{abstract}

}

\maketitle
\IEEEdisplaynontitleabstractindextext

%
\IEEEpeerreviewmaketitle

\section{Introduction}


\IEEEPARstart{L}{arge} AI Models (LAIMs) are characterized by their exceptionally high parameter counts, often reaching billions, as highlighted in the work of \cite{zhao2023survey, lin2022large, qiu2023large}. These models typically include \textit{Diffusion Models} (DMs) and \textit{Large Language Models} (LLMs), both of which are trained on extensive datasets and require significant computational resources. 
A key distinction of LAIMs, especially DMs, from traditional AI models like Generative Adversarial Networks (GANs) and Variational Autoencoders (VAEs), lies in their scalability and generation quality. 
More specifically, LAIMs excel in tackling intricate tasks such as language comprehension, pattern recognition, and the generation of highly realistic multimedia content encompassing text, images, videos, and audio. Since 2020, as documented by \cite{lin2022large}, these models have attracted substantial interest due to their consistent improvements in performance and versatility across various domains, including natural language processing \cite{zhao2023survey}, computer vision \cite{li2023multimodal}, and health informatics \cite{qiu2023large}.

Recent advances in LAIMs offer numerous benefits for humans in various aspects of life and work. Specifically, the benefits of LAIM-generated multimedia span a wide range of fields, from enhancing education~\cite{malinka2023educational} and healthcare~\cite{zhao2023chatcad+} to boosting creativity, and optimizing business processes~\cite{chui2022generative}, and improving accessibility~\cite{AI-disabilities-2023}. These advances improve task efficiency and unlock new avenues for innovation and problem-solving.


Meanwhile, multimedia generated by LAIMs also carries significant risks of misuse and societal upheaval, as discussed in \cite{saetra2023generative, napitupulu2023implication}. Specifically, these models can be employed to craft convincing fake news, deepfakes (highly realistic fake images, audios, and videos created using AI algorithms \cite{masood2023deepfakes}), and other forms of misinformation, thereby challenging
the information accuracy and public trust \cite{bohacek2024making, tan2020detecting, shao2023detecting}. For instance, LAIM-generated multimedia can be weaponized for political propaganda or manipulative advertising, exploiting its ability to create persuasive and tailored content. There are also ethical dilemmas regarding the use of LAIMs to produce art or content that imitates human creativity, sparking debates over originality, intellectual property rights, and the intrinsic value of human artistic expression \cite{jiang2023ai}. Furthermore, LAIMs' ability to generate multimedia could potentially impact employment in creative fields, fueling concerns about the displacement of human workers in journalism, the arts, and entertainment. A notable example of this tension was the months-long strike in the summer of 2023, where writers and performers fought against major Hollywood studios \cite{bohacek2024making}.

Such challenges have highlighted the urgent need for effective detection methods for multimedia produced by LAIMs. In recent years, there has been a significant increase in research focused on this area. However, to the best of our knowledge, there is a notable lack of systematic surveys specifically addressing the detection of LAIM-generated multimedia, in contrast to the numerous surveys focusing on multimedia generation using LAIMs. To bridge this gap, we present this first comprehensive survey, which not only fills a critical academic void but also aligns with AI security initiatives of various governments worldwide, such as the AI Safety Summit 2023 \cite{AI-safety-summit-2023} and the United States government’s ``AI Executive Order” \cite{biden2023executive}. 

In this survey, we provide a detailed and thorough review of the existing research on identifying multimedia generated by LAIMs, particularly emphasizing Diffusion Models and Large Language Models. Our goal is to guide researchers toward understanding the current challenges and exploring potential future directions in this field, aiming to reinstate trust in digital content among users. Furthermore, we endeavor to show that, despite the high degree of realism in LAIM-generated multimedia, it can still be identified, which is crucial for its ethical use and for maintaining the integrity of information in the digital world. The dynamic and ongoing interplay between generating and detecting multimedia using LAIMs is shown in Fig. \ref{fig:interplay_intro}.

\newcommand{\coloredTextBox}[4]{
  \colorbox{#3}{\parbox[c][#2][c]{#1}{\centering #4}}
}

\definecolor{textcolor}{HTML}{97D077}
\definecolor{imagecolor}{HTML}{FF9999}
\definecolor{videocolor}{HTML}{FF8000}
\definecolor{audiocolor}{HTML}{CDA2BE}
\definecolor{mmcolor}{HTML}{FFCE9F}
\begin{figure*}[t]
    \centering
    \includegraphics[width=1\textwidth]{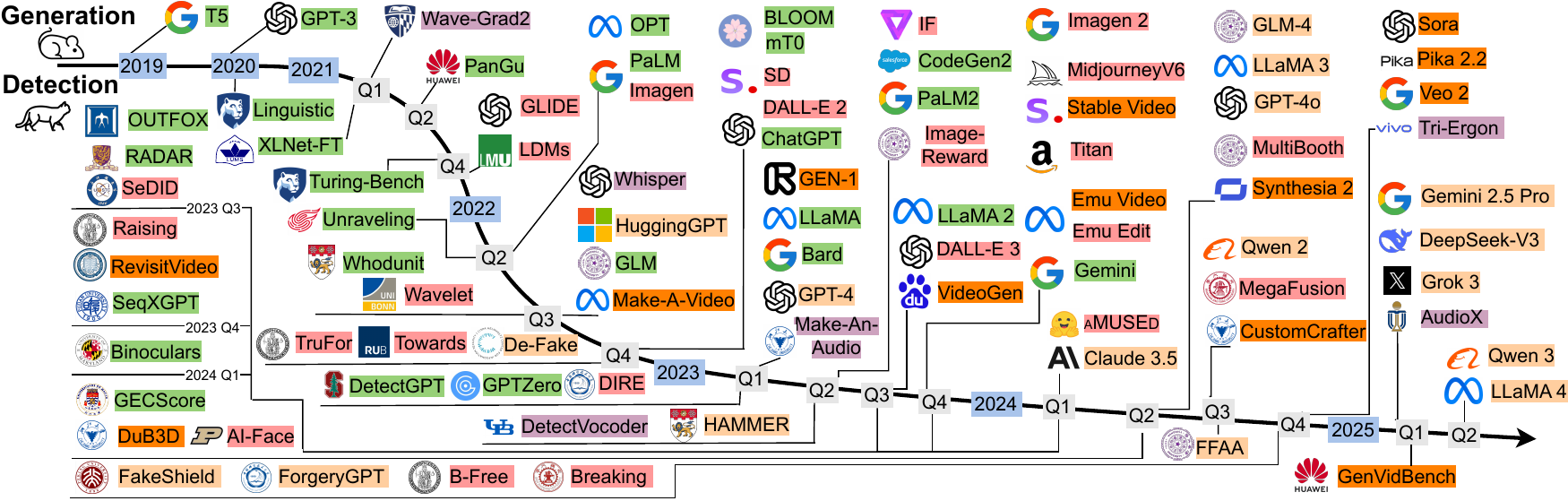}
    \vspace{-6mm}
    \caption{\small A cat-and-mouse game between generating and detecting multimedia (\coloredTextBox{0.5cm}{0.18cm}{textcolor}{text}, \coloredTextBox{0.81cm}{0.18cm}{imagecolor}{image}, \coloredTextBox{0.75cm}{0.18cm}{videocolor}{video}, \coloredTextBox{0.75cm}{0.18cm}{audiocolor}{audio}, and \coloredTextBox{1.55cm}{0.18cm}{mmcolor}{multimodal}) using LAIMs, showcasing only representative works. Q1 represents from Jan to Mar, Q2: Apr-Jun, Q3: Jul-Sep, Q4: Oct-Dec.}
    \label{fig:interplay_intro}
    \vspace{-4mm}
\end{figure*}

\subsection{Related Works}

While there are several surveys  \cite{yang2023survey,wu2023survey,ghosal2023towards, crothers2023machine, kwon2025comprehensive, xu2025recent, cardenuto2023age, zou2025survey, nguyen2024passive} addressing the detection of multimedia generated by LAIMs, like Diffusion Models (DMs) and Large Language Models (LLMs), their scopes are very limited. Specifically, surveys like \cite{yang2023survey,wu2023survey,ghosal2023towards,crothers2023machine, kwon2025comprehensive} mainly concentrate on the detection of LAIM-generated text, overlooking other multimedia forms such as images, videos, and audio. Additionally, while these surveys provide insights into detection techniques,
they tend to focus solely on detection mechanisms without delving into broader aspects or applications of these technologies beyond detection or without discussing real-world challenges like social media adaptation, which are discussed in our survey.

Xu et al.~\cite{xu2025recent} present a focused review on generalizable diffusion-generated image detection. They introduce a dual taxonomy—data-driven and feature-driven—and classify methods into six subcategories based on their design principles. While their survey offers deep insights into generalization and detection robustness for diffusion models, it only addresses a single modality (images) and does not cover other forms of LAIM-generated media such as text, video, or audio.

A few works~\cite{cardenuto2023age, zou2025survey, nguyen2024passive} extend the scope to multimodal detection. Cardenuto et al. \cite{cardenuto2023age} offer a high-level overview of generative and detection pipelines but fall short of deep technical analysis. Their scope is not limited to large AI models and omits many recent LAIM-specific detection approaches. Moreover, they do not cover detection datasets in sufficient detail or address multimodal detection settings. Zou et al.~\cite{zou2025survey} and Nguyen-Le et al.~\cite{nguyen2024passive} both examine multimodal detection, with the former focusing on the transition from Non-MLLM to MLLM-based frameworks and the latter on passive deepfake detection across image, video, audio, and multimodal content. Zou et al. introduce a taxonomy centered on authenticity verification, explainability, and localization, while Nguyen-Le et al. emphasize robustness, generalization, attribution, and real-world deployment. However, both surveys fall short in providing detailed modality-specific analysis, such as concrete challenges, benchmarks, or future directions tailored to each modality.



\textit{In summary, current surveys are either not comprehensive in modality coverage or do not address social media–processed multimedia detection in detail, such as datasets, detection methods, challenges, and future directions in a modality-specific manner. Our survey fills this gap by providing the first comprehensive and systematic review of LAIM-generated multimedia detection across five modalities—text, image, video, audio, and multimodal, with an additional dedicated focus on real-world social media settings.}

\subsection{Contributions}
\begin{enumerate}[leftmargin=*]
    \item This is the first comprehensive survey on detecting multimedia generated by LAIMs, covering text, images, videos, audio, and multimodal content. We review the most up-to-date detection methodologies and propose a novel taxonomy that categorizes methods within each modality into two main groups: \textit{pure detection} and \textit{beyond detection}. This taxonomy offers a unique perspective that has not been previously explored. Within these two categories, we further classify the methods into more specific subcategories based on their common and distinct characteristics.
    \item We provide a brief overview of LAIMs, their generation mechanisms, and the content they generate, shedding light on the current status of the cat (generation)-and-mouse (detection) game in this field. Additionally, we present public datasets and evaluation metrics tailored for detection tasks.
    \item We are the first to systematically examine the detection of LAIM-generated multimedia in the context of social media, a major source of AI-generated misinformation. Throughout the survey, we review social media-specific datasets across all modalities, analyze detection methods evaluated on in-the-wild content, and identify unique challenges and future directions tailored to social media environments. Given that social media is a primary channel through which AI-generated content spreads and influences public perception, understanding detection in this context is essential for maintaining societal trust.
    \item We have conducted a comparative summary of online detection tools. This provides a valuable resource for researchers, developers, and stakeholders in the field. Additionally, we pinpoint current challenges faced by recent detection methods and provide directions for future work.
\end{enumerate}

\smallskip
\noindent
The remainder of this survey is organized as follows:
In Section \ref{sec:LAIMs_generated_Multimedia}, we briefly introduce LAIMs, including generation mechanisms and datasets. In Section \ref{sec:detection}, we classify detection by their functionality and organize them into our newly defined taxonomy. We summarize online detection tools and evaluation metrics in Section \ref{sec:ToolsCompetitionsMetrics}. Critical challenges faced by detectors and potential future directions are discussed in Section \ref{sec:Discussion}. Finally, we conclude the survey in Section \ref{sec:Conclusion}.

\begin{figure*}[t]
    \centering
    \includegraphics[width=1\textwidth]{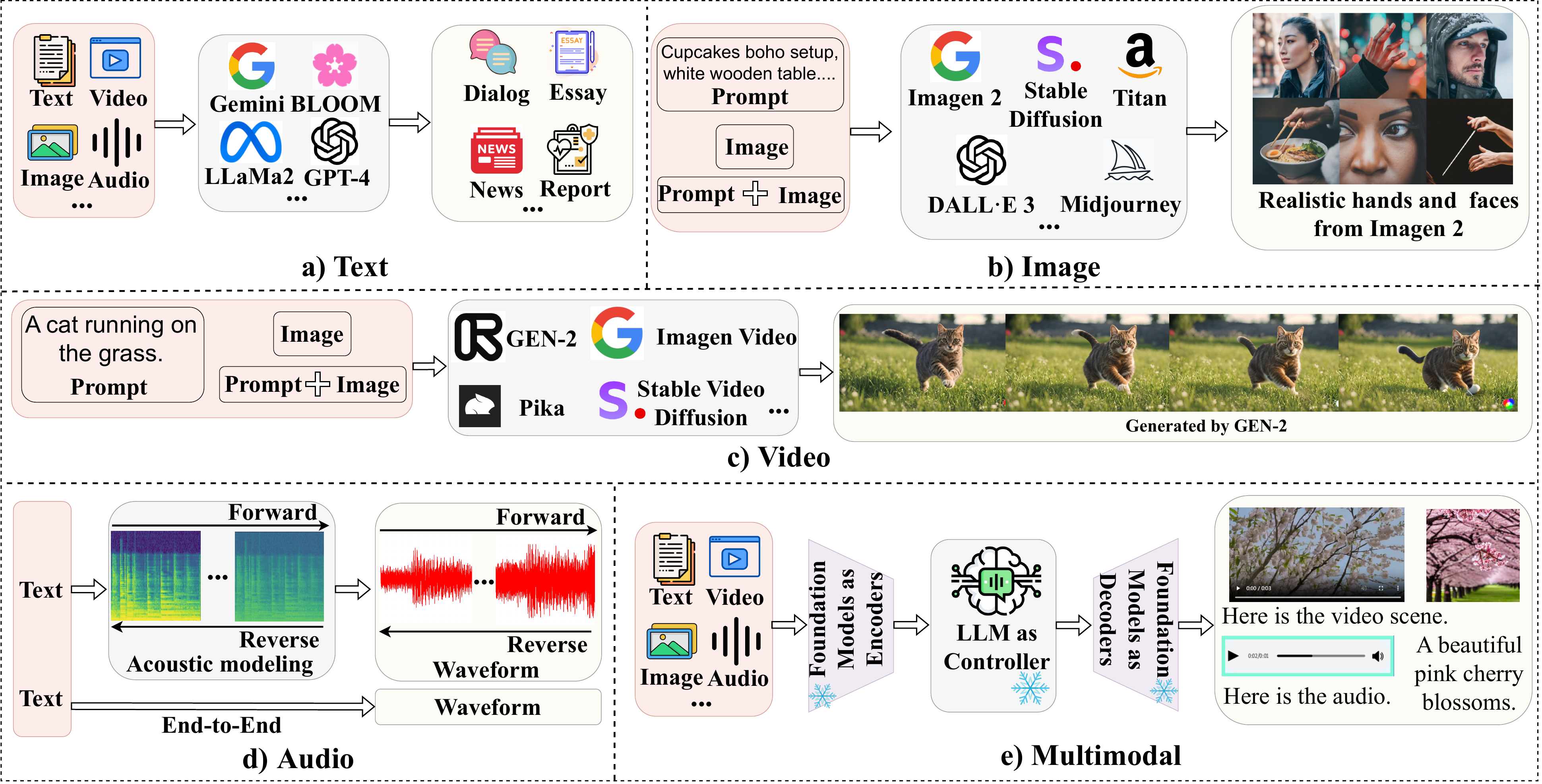}
    \vspace{-6mm}
    \caption{\small Illustrations of different types of multimedia generation process based on LAIMs.}
    \label{fig:generration_overview}
    \vspace{-4mm}
\end{figure*}

\begin{table*}[htb]
\centering
\caption{ \small{
Summary of \textbf{public} datasets that are generated by LAIMs. I2O: Input-to-Output, T2T: Text-to-Text, V2T: Video-to-Text, T/I2I: Text-to-Image,Image-to-Image, T2A: Text-to-Audio, I.A2V: (Image conditioned with Audio)-to-Video. BM: Benchmark. Note that the "Year" corresponds to the initial public release (\eg, on arXiv), not necessarily the final publication year. Note that not all works in ''Source of Real Media'' and ``Generative Method'' are listed due to space limitations}.}
\vspace{-2mm}
\scalebox{0.65}{
\begin{tabular}{|c|c|c|c|c|c|c|c|c|c|c|}
\hline
\renewcommand\arraystretch{1.2}
\textbf{Modality}                    & \textbf{Dataset}    & \textbf{Year}  & \textbf{BM}    & \textbf{Content} & \textbf{Link} &  \textbf{I2O} &\textbf{\#Real}                                                  & \textbf{\#Generated}      & \textbf{Source of Real Media}                                                                                                                                                                           & \textbf{Generative Method}                                                                                                                                                                                                  \\ \hline
\multirow{28}{*}{{\cellcolor[HTML]{FFFFFF}}\textbf{Text}}  &  \cellcolor[HTML]{C8E4B2}TuringBench\cite{uchendu2021turingbench}  & \cellcolor[HTML]{C8E4B2}2021      &\cellcolor[HTML]{C8E4B2}\checkmark                   & \cellcolor[HTML]{C8E4B2}News                &       \cellcolor[HTML]{C8E4B2}\href{https://turingbench.ist.psu.edu/}{\color[HTML]{0000EE} {Link}}&  \cellcolor[HTML]{C8E4B2}T2T                      & \cellcolor[HTML]{C8E4B2}8,854                                                            & \cellcolor[HTML]{C8E4B2}159,758               &         
                                     \cellcolor[HTML]{C8E4B2}News Media & \cellcolor[HTML]{C8E4B2}GPT-1\&2\&3, CTRL~\cite{keskar2019ctrl}, GROVER~\cite{zellers2019defending}             \\ \cline{2-11}    \rowcolor[HTML]{C8E4B2}  
                                     \cellcolor[HTML]{FFFFFF}&Paraphrase\cite{wahle2022large}  & 2022  &\checkmark    & Essays                &        \href{https://huggingface.co/datasets/jpwahle/autoregressive-paraphrase-dataset}{\color[HTML]{0000EE} {Link}}&  T2T                                                                 & 
                                     98,280 & 
                                     163,710& Arxiv, Wikipedia, Theses                                                                                                             & GPT-3, T5\cite{raffel2020exploring}                   \\ \cline{2-11}  & \cellcolor[HTML]{C8E4B2}SynSciPass \cite{rosati-2022-synscipass}     &\cellcolor[HTML]{C8E4B2}2022        &\cellcolor[HTML]{C8E4B2}      & \cellcolor[HTML]{C8E4B2}Passages              & \cellcolor[HTML]{C8E4B2}\href{https://github.com/domenicrosati/synscipass/tree/main/data}{\color[HTML]{0000EE}{Link}}& \cellcolor[HTML]{C8E4B2}T2T      & \cellcolor[HTML]{C8E4B2}99,989                                                           &\cellcolor[HTML]{C8E4B2} 10,485               &\cellcolor[HTML]{C8E4B2} Scientific papers                                           &\cellcolor[HTML]{C8E4B2} GPT-2, BLOOM\cite{workshop2022bloom}                     \\ \cline{2-11} 
                                     & \cellcolor[HTML]{C8E4B2} MAGE\cite{li2023deepfake}    & \cellcolor[HTML]{C8E4B2}2023  &\cellcolor[HTML]{C8E4B2}\checkmark   &\cellcolor[HTML]{C8E4B2} General               & \cellcolor[HTML]{C8E4B2}\href{https://github.com/yafuly/DeepfakeTextDetect}{\color[HTML]{0000EE}{Link}}&\cellcolor[HTML]{C8E4B2}T2T                           & \cellcolor[HTML]{C8E4B2}154,078                                                          &\cellcolor[HTML]{C8E4B2} 294,381              &\cellcolor[HTML]{C8E4B2} Reddit, ELI5\cite{fan2019eli5},Yelp, XSum~\cite{narayan2018don}                                                 & \cellcolor[HTML]{C8E4B2}27 LLMs             \\ \cline{2-11} &\cellcolor[HTML]{C8E4B2}  Stu.Essays\cite{verma2023ghostbuster}        &\cellcolor[HTML]{C8E4B2}2023 &\cellcolor[HTML]{C8E4B2}  &\cellcolor[HTML]{C8E4B2} Essays                & \cellcolor[HTML]{C8E4B2}\href{https://github.com/vivek3141/ghostbuster}{\color[HTML]{0000EE}{Link}}& \cellcolor[HTML]{C8E4B2}T2T                                                                                       &\cellcolor[HTML]{C8E4B2} 1,000                                                            &\cellcolor[HTML]{C8E4B2} 6,000                &\cellcolor[HTML]{C8E4B2} IvyPanda~\cite{ivypanda}                                                                                                                                                                                        &\cellcolor[HTML]{C8E4B2} ChatGPT                                                                                                                                                                                                                              \\ \cline{2-11}   &\cellcolor[HTML]{C8E4B2} Writing\cite{verma2023ghostbuster}    &\cellcolor[HTML]{C8E4B2}2023  &  \cellcolor[HTML]{C8E4B2}  &\cellcolor[HTML]{C8E4B2} Stories                & \cellcolor[HTML]{C8E4B2}\href{https://github.com/vivek3141/ghostbuster}{\color[HTML]{0000EE}{Link}}&\cellcolor[HTML]{C8E4B2}T2T                                                                                     & \cellcolor[HTML]{C8E4B2}1,000                                                            &\cellcolor[HTML]{C8E4B2} 6,000                &\cellcolor[HTML]{C8E4B2} Reddit WritingPrompts~\cite{reddit_writing_prompts}                                                                                  & \cellcolor[HTML]{C8E4B2}ChatGPT                                                                              \\ \cline{2-11}\rowcolor[HTML]{C8E4B2} 
                                     \cellcolor[HTML]{FFFFFF}& News\cite{verma2023ghostbuster}   & 2023    &      & News                & \href{https://github.com/vivek3141/ghostbuster}{\color[HTML]{0000EE}{Link}}&  T2T                                                                                     & 1,000                                                            & 6,000                & Reuters 50-50~\cite{houvardas2006n}              & ChatGPT                       \\ \cline{2-11}\rowcolor[HTML]{C8E4B2} 
                                     \cellcolor[HTML]{FFFFFF}& OUTFOX~\cite{koike2023outfox} & 2023 &  & Essays                &           \href{https://github.com/ryuryukke/OUTFOX}{\color[HTML]{0000EE} { Link}}&  T2T                                                                           & 15,400                                                           & 15,400               & Feedback Prize~\cite{feedbackprize2022}                                             & ChatGPT, GPT-3.5, T5~\cite{raffel2020exploring}                                              \\ \cline{2-11}\rowcolor[HTML]{C8E4B2} 
                                     \cellcolor[HTML]{FFFFFF}& MULTITuDE\cite{macko2023multitude}  & 2023       &\checkmark           & News                  &   \href{https://zenodo.org/records/10013755}{\color[HTML]{0000EE} { Link}}&  T2T                                                                   & 7,992                                                            & 66,089               & MassiveSumm~\cite{varab2021massivesumm}                               & GPT-3\&4, ChatGPT                             \\ \cline{2-11}\rowcolor[HTML]{C8E4B2} 
                                     \cellcolor[HTML]{FFFFFF}& MGTDetect-CoCo\cite{liu2023coco}    & 2023    &     & News                  &    \href{https://huggingface.co/datasets/ZachW/MGTDetect_CoCo}{\color[HTML]{0000EE} { Link}}&  T2T                                                                                         & 10,486                                                            & 10.484               & 
                                    News Outlets & GPT-3.5                                          \\ \cline{2-11}\rowcolor[HTML]{C8E4B2}\cellcolor[HTML]{FFFFFF} 
                                     & HPPT\cite{yang2023chatgpt}    & 2023        &          & Abstracts             &       \href{https://github.com/FreedomIntelligence/ChatGPT-Detection-PR-HPPT}{\color[HTML]{0000EE} { Link}}&  T2T                                                                                        & 6,050                                                            & 6,050                & ACL Anthology~\cite{aclanthology}        & ChatGPT           \\ \cline{2-11}
                                     & \cellcolor[HTML]{C8E4B2}HC-Var\cite{xu2023generalization} & \cellcolor[HTML]{C8E4B2}2023       &\cellcolor[HTML]{C8E4B2}\checkmark             & \cellcolor[HTML]{C8E4B2}General               & \cellcolor[HTML]{C8E4B2}\href{https://huggingface.co/datasets/hannxu/hc_var}{\color[HTML]{0000EE} { Link}}&\cellcolor[HTML]{C8E4B2}T2T        &\cellcolor[HTML]{C8E4B2} 90,096                                                           &\cellcolor[HTML]{C8E4B2} 45,000               & \cellcolor[HTML]{C8E4B2}XSum~\cite{narayan2018don}, IMDb, Yelp, FiQA~\cite{thakur2021beir}      & \cellcolor[HTML]{C8E4B2}ChatGPT                                         \\ \cline{2-11} 
                                     & \cellcolor[HTML]{C8E4B2} HC3~\cite{guo2023close}  & \cellcolor[HTML]{C8E4B2}2023      &\cellcolor[HTML]{C8E4B2}\checkmark               &\cellcolor[HTML]{C8E4B2} General               &\cellcolor[HTML]{C8E4B2}\href{https://huggingface.co/datasets/Hello-SimpleAI/HC3}{\color[HTML]{0000EE} { Link}}&\cellcolor[HTML]{C8E4B2}T2T                                                                                            &\cellcolor[HTML]{C8E4B2}  26,903                                                         & \cellcolor[HTML]{C8E4B2}58,546               &  \cellcolor[HTML]{C8E4B2}   FiQA\cite{thakur2021beir}, ELI5\cite{fan2019eli5}, Meddialog\cite{zeng2020meddialog}                                                                         & \cellcolor[HTML]{C8E4B2}ChatGPT                                                                                                                                                \\ \cline{2-11}
                                     &\cellcolor[HTML]{C8E4B2} M4\cite{wang2023m4}     &\cellcolor[HTML]{C8E4B2}2023     &\cellcolor[HTML]{C8E4B2}\checkmark               & \cellcolor[HTML]{C8E4B2}General               &   \cellcolor[HTML]{C8E4B2}\href{https://github.com/mbzuai-nlp/M4}{\color[HTML]{0000EE} { Link}}& \cellcolor[HTML]{C8E4B2}T2T                                                                                             &\cellcolor[HTML]{C8E4B2} 32,798                                                           &\cellcolor[HTML]{C8E4B2} 89,683               & \cellcolor[HTML]{C8E4B2} WikiHow~\cite{koupaee2018wikihow}, Arxiv, Reddit                                                & \cellcolor[HTML]{C8E4B2}ChatGPT, LLaMA,T5~\cite{raffel2020exploring},    BLOOM~\cite{workshop2022bloom}                                                                                                                                                 \\ \cline{2-11}    &\cellcolor[HTML]{C8E4B2} F3~\cite{lucas-etal-2023-fighting}     &\cellcolor[HTML]{C8E4B2}2023     &\cellcolor[HTML]{C8E4B2}\checkmark               & \cellcolor[HTML]{C8E4B2}Social Media               &   \cellcolor[HTML]{C8E4B2}\href{https://github.com/mickeymst/F3/tree/main}{\color[HTML]{0000EE} { Link}}& \cellcolor[HTML]{C8E4B2}T2T                                                                                             &\cellcolor[HTML]{C8E4B2}  12,723                                                         &\cellcolor[HTML]{C8E4B2} 27,667            & \cellcolor[HTML]{C8E4B2}Politifact~\cite{politifact}, Snopes~\cite{snopes}                                               & \cellcolor[HTML]{C8E4B2}GPT3.5                                                                                                                                                 \\ \cline{2-11}
                                       &\cellcolor[HTML]{C8E4B2} MixSet\cite{gao2024llm}      &\cellcolor[HTML]{C8E4B2}2024        &\cellcolor[HTML]{C8E4B2}\checkmark             & \cellcolor[HTML]{C8E4B2}General               &   \cellcolor[HTML]{C8E4B2}\href{https://github.com/Dongping-Chen/MixSet}{\color[HTML]{0000EE} { Link}}& \cellcolor[HTML]{C8E4B2}T2T                                                                                             &\cellcolor[HTML]{C8E4B2} 300                                                           &\cellcolor[HTML]{C8E4B2} 3,600               &\cellcolor[HTML]{C8E4B2}Email\cite{Enron_email},  BBC News\cite{greene2006practical}, ArXiv                                                & \cellcolor[HTML]{C8E4B2}\begin{tabular}[c]{@{}c@{}} GTP-4, LLaMA2\end{tabular}   \\ \cline{2-11} \rowcolor[HTML]{C8E4B2}\cellcolor[HTML]{FFFFFF} 
                                     & GPABench\cite{liu2024detectability}      & 2024     &\checkmark      & Writing               &       \href{https://github.com/liuzey/CheckGPT-v2}{\color[HTML]{0000EE} { Link}}&  T2T           & 150,000                                                          & 450,000              & Arxiv                  & GPT-3.5                 \\ \cline{2-11}\rowcolor[HTML]{C8E4B2}\cellcolor[HTML]{FFFFFF} 
                                     & M4GT-Bench~\cite{wang2024m4gt}    & 2024        &\checkmark          &    General          &       \href{https://github.com/mbzuai-nlp/M4GT-Bench}{\color[HTML]{0000EE} { Link}}&   T2T                                                                                       & 93,177                                                           &    110,388              &  Wikipedia, WikiHow, Reddit, ArXiv, News     &10 LLMs  \\  \cline{2-11}       \rowcolor[HTML]{C8E4B2}\cellcolor[HTML]{FFFFFF} 
                                     & RAID~\cite{dugan-etal-2024-raid}    & 2024        &\checkmark          &    General          &       \href{https://github.com/liamdugan/raid}{\color[HTML]{0000EE} { Link}}&   T2T                                                                                       & 14,971                                                            &   6,287,820                &  public datasets from 8 domains       &11 LLMs  \\  \cline{2-11}           \rowcolor[HTML]{C8E4B2}\cellcolor[HTML]{FFFFFF} 
                                     & DetectRL~\cite{wu2024detectrl}    & 2024        &\checkmark          &   General           &       \href{https://github.com/NLP2CT/DetectRL}{\color[HTML]{0000EE} { Link}}&   T2T                                                                                       & 100,800                                                            &   134,400               &  \begin{tabular}[c]{@{}c@{}}ArXiv, XSum~\cite{narayan2018don}, \\Writing Prompts~\cite{fan2018hierarchical}, Yelp\end{tabular}       & GPT-3.5, PaLM2~\cite{anil2023palm}, Claude, LLaMA2 \\ \cline{2-11}           \rowcolor[HTML]{C8E4B2}\cellcolor[HTML]{FFFFFF} 
                                     & MultiSocial~\cite{macko2024multisocial}    & 2024        &\checkmark          &    Social Media           & - &   T2T                                                                                       & ~58,000                                                           &   ~414,000               &  \begin{tabular}[c]{@{}c@{}}Telegram, Twitter,\\ Gab, Discord, WhatsApp\end{tabular}       & 7 LLMs     \\ \cline{2-11}           \rowcolor[HTML]{C8E4B2}\cellcolor[HTML]{FFFFFF} 
                                     & SM-D~\cite{sun2024we}    & 2024        &  \checkmark        &    Social Media           &-  &   T2T                                                                                       & -                                                           &  -               & Medium,
Quora, Reddit      & Sourced from social media                                                              
                                     
                                                          \\ \hline
\rowcolor[HTML]{ffcfd2}
\multirow{20}{*}{\cellcolor[HTML]{FFFFFF} \textbf{Image}}     & DFF\cite{song2023robustness}     & 2023      &     & Face                  &     \href{https://huggingface.co/datasets/OpenRL/DeepFakeFace}{\color[HTML]{0000EE} { Link}}&  T/I2I                                                                                 & 30,000                                                           & 90,000               & IMDB-WIKI\cite{rothe2015dex}                                                                                                                                                                                                  & SDMs, InsightFace~\cite{Insightface}                                                                                                                                                                                                                    \\ \cline{2-11}  \rowcolor[HTML]{ffcfd2}\cellcolor[HTML]{FFFFFF} 
                                     &RealFaces~\cite{papa2023use}   & 2023     &     & Face                  &     \href{https://github.com/LucaCorvitto/RealFaces_w_StableDiffusion}{\color[HTML]{0000EE} { Link}}&  T2I                                                                                         & 258                                                             & 25,800               & Prompts                                                                                                                                                                                                    & SDMs               \\ \cline{2-11} \rowcolor[HTML]{ffcfd2}\cellcolor[HTML]{FFFFFF}
                                     & 
                                    OHImg~\cite{may2023comprehensive}          & 2023    &         & Overhead        &    \href{https://stresearch.github.io/synthetic-overhead-dataset/}{\color[HTML]{0000EE} { Link}}& T/I2I                                                                                          & 6,475                                                            & 6,675                & MapBox~\cite {mapbox}, Google Maps                                                                                                  & GLIDE~\cite{nichol2021glide}, DDPM~\cite{ho2020denoising}                                   \\ \cline{2-11} &\cellcolor[HTML]{ffcfd2} Western Blot\cite{mandelli2022forensic}     &\cellcolor[HTML]{ffcfd2} 2022   &\cellcolor[HTML]{ffcfd2}     &\cellcolor[HTML]{ffcfd2} Biology               &  \cellcolor[HTML]{ffcfd2}\href{https://www.dropbox.com/sh/nl3txxfovy97b1k/AABqb-gkGBEfjS6pjke3a-d7a?dl=0}{\color[HTML]{0000EE} { Link}}&\cellcolor[HTML]{ffcfd2}  I2I                                                                                       &\cellcolor[HTML]{ffcfd2} $\sim$14,000                                                     &\cellcolor[HTML]{ffcfd2} $\sim$24,000         &\cellcolor[HTML]{ffcfd2} Western Blot                                                                                            &\cellcolor[HTML]{ffcfd2} DDPM, Pix2pix~\cite{isola2017image}, CycleGAN~\cite{Zhu_2017_ICCV}                                                                                                                                                                                                            \\ \cline{2-11} \rowcolor[HTML]{ffcfd2}\cellcolor[HTML]{FFFFFF} & Synthbuster\cite{bammey2023synthbuster}      & 2023  &            & General                  &     \href{https://zenodo.org/records/10066460}{\color[HTML]{0000EE} { Link}}&  T2I                                                                                  &   -                                                         & 9,000               &  Raise-1k\cite{dang2015raise}                                                                                                                                                                                                  & DALL·E 2\&3, Midjourney, SDMs, GLIDE\cite{nichol2021glide}                                                                                                                                                                                                               \\ \cline{2-11}  \rowcolor[HTML]{ffcfd2}\cellcolor[HTML]{FFFFFF}
                                     & GenImage\cite{zhu2023genimage}       & 2023    &\checkmark        & General               &      \href{https://github.com/GenImage-Dataset/GenImage}{\color[HTML]{0000EE} { Link}}&  T/I2I                                                                                  & 1,331,167                                                        & 1,350,000            & ImageNet                                                                    & SDMs, Midjourney, BigGAN~\cite{brock2018large}, 
                                                                       \\ \cline{2-11}\rowcolor[HTML]{ffcfd2}\cellcolor[HTML]{FFFFFF} 
                                     & CIFAKE\cite{bird2023cifake} & 2023      &                 & General               &     \href{https://www.kaggle.com/datasets/birdy654/cifake-real-and-ai-generated-synthetic-images}{\color[HTML]{0000EE} { Link}}&  T2I                                                                                        & 60,000                                                           & 60,000               & CIFAR-10                                                      & SD-V1.4                                                                   \\ \cline{2-11} \rowcolor[HTML]{ffcfd2}\cellcolor[HTML]{FFFFFF}
                                     & AutoSplice\cite{jia2023autosplice}    & 2023  &\checkmark         & General               &       \href{https://github.com/shanface33/AutoSplice_Dataset}{\color[HTML]{0000EE} { Link}}&  T2I                                                                                          & 2,273                                                            & 3,621                & Visual News~\cite{liu2020visual}                                                          & DALL·E 2                                                                       \\ \cline{2-11} \rowcolor[HTML]{ffcfd2}\cellcolor[HTML]{FFFFFF}
                                     & DiffusionDB\cite{wang2022diffusiondb}       & 2023   &         & General               &     \href{https://huggingface.co/datasets/poloclub/diffusiondb}{\color[HTML]{0000EE} { Link}}&  T2I                                                                                       & 3,300,000                                                        & 16,000,000           & DiscordChatExporter~\cite{discord}                                                                                              & SD                                                                                 \\ \cline{2-11} 
                                     & \cellcolor[HTML]{ffcfd2}ArtiFact\cite{awsafur2023artifact} & \cellcolor[HTML]{ffcfd2}2023          &\cellcolor[HTML]{ffcfd2}         & \cellcolor[HTML]{ffcfd2}General               &   \cellcolor[HTML]{ffcfd2}\href{https://www.kaggle.com/datasets/awsaf49/artifact-dataset/}{\color[HTML]{0000EE} { Link}}&  \cellcolor[HTML]{ffcfd2}T/I2I                                                                                            &\cellcolor[HTML]{ffcfd2} 964,989                                                          & \cellcolor[HTML]{ffcfd2}1,531,749            &\cellcolor[HTML]{ffcfd2}  COCO, FFHQ~\cite{karras2019style},  LSUN                                                          & \cellcolor[HTML]{ffcfd2}SDMs,  DDPM~\cite{ho2020denoising},LDM~\cite{rombach2021highresolution},  CIPS~\cite{anokhin2021image}           \\ \cline{2-11}
                                     & \cellcolor[HTML]{ffcfd2}HiFi-IFDL\cite{guo2023hierarchical}  & \cellcolor[HTML]{ffcfd2}2023        &\cellcolor[HTML]{ffcfd2}          & \cellcolor[HTML]{ffcfd2}General               & \cellcolor[HTML]{ffcfd2}\href{https://github.com/CHELSEA234/HiFi_IFDL}{\color[HTML]{0000EE} { Link}}&\cellcolor[HTML]{ffcfd2}  T/I2I                                                                      & \cellcolor[HTML]{ffcfd2} $\sim$600,000                                                    & \cellcolor[HTML]{ffcfd2}1,300,000            &\cellcolor[HTML]{ffcfd2} FFHQ~\cite{karras2019style},      COCO, LSUN  &\cellcolor[HTML]{ffcfd2} DDPM~\cite{ho2020denoising}, GLIDE~\cite{nichol2021glide}, LDM~\cite{rombach2021highresolution},GANs                                                                                                                      \\ \cline{2-11} 
                                     &\cellcolor[HTML]{ffcfd2} DiffForensics\cite{wang2023dire} &\cellcolor[HTML]{ffcfd2} 2023   &\cellcolor[HTML]{ffcfd2}     &\cellcolor[HTML]{ffcfd2} General               &\cellcolor[HTML]{ffcfd2}\href{https://github.com/ZhendongWang6/DIRE}{\color[HTML]{0000EE} { Link}}& \cellcolor[HTML]{ffcfd2} T/I2I                                                                                     & \cellcolor[HTML]{ffcfd2}232,000                                                          &\cellcolor[HTML]{ffcfd2} 232,000              & \cellcolor[HTML]{ffcfd2}LSUN, ImageNet                                                                                                                                                                                           & \cellcolor[HTML]{ffcfd2} LDM~\cite{rombach2021highresolution}, DDPM~\cite{ho2020denoising},     VQDM~\cite{gu2022vector},  ADM~\cite{dhariwal2021diffusion}                                                                                                                                           \\ \cline{2-11} 
                                     & \cellcolor[HTML]{ffcfd2}CocoGlide\cite{guillaro2023trufor}  &\cellcolor[HTML]{ffcfd2} 2023    &\cellcolor[HTML]{ffcfd2}           & \cellcolor[HTML]{ffcfd2}General               &\cellcolor[HTML]{ffcfd2}\href{https://github.com/grip-unina/TruFor}{\color[HTML]{0000EE} { Link}}&\cellcolor[HTML]{ffcfd2}  T2I                                                                                           &\cellcolor[HTML]{ffcfd2} 512                                                              &\cellcolor[HTML]{ffcfd2} 512                  &\cellcolor[HTML]{ffcfd2}COCO                                                   & \cellcolor[HTML]{ffcfd2}GLIDE\cite{nichol2021glide}                                                                 \\ \cline{2-11} 
                                     
                                     &\cellcolor[HTML]{ffcfd2}LSUNDB\cite{ricker2022towards} &\cellcolor[HTML]{ffcfd2} 2023   &\cellcolor[HTML]{ffcfd2}                   &\cellcolor[HTML]{ffcfd2} General               &\cellcolor[HTML]{ffcfd2}\href{ https://zenodo.org/records/7528113 }{\color[HTML]{0000EE} { Link}}&\cellcolor[HTML]{ffcfd2}  T/I2I                                                                                  &\cellcolor[HTML]{ffcfd2} 250,000                                                               &\cellcolor[HTML]{ffcfd2} 250,000               &\cellcolor[HTML]{ffcfd2} LSUN                                                                                               & \cellcolor[HTML]{ffcfd2}DDPM~\cite{ho2020denoising},  LDM~\cite{rombach2021highresolution}, StyleGAN~\cite{karras2019style}                                                                                       \\ \cline{2-11} 
                                     &\cellcolor[HTML]{ffcfd2}UniFake\cite{ojha2023towards}   & \cellcolor[HTML]{ffcfd2}2023        &\cellcolor[HTML]{ffcfd2}            &\cellcolor[HTML]{ffcfd2} General               & \cellcolor[HTML]{ffcfd2}\href{https://drive.google.com/file/d/1FXlGIRh_Ud3cScMgSVDbEWmPDmjcrm1t/view}{\color[HTML]{0000EE} { Link}}& \cellcolor[HTML]{ffcfd2} T2I                                                                                           & \cellcolor[HTML]{ffcfd2}8,000                                                            &\cellcolor[HTML]{ffcfd2} 8,000                &\cellcolor[HTML]{ffcfd2} LAION-400M~\cite{schuhmann2021laion}                                                                                                                                                                                                   &\cellcolor[HTML]{ffcfd2} LDM~\cite{rombach2021highresolution}, GLIDE~\cite{nichol2021glide}                                                                          \\ \cline{2-11} 
                                     &\cellcolor[HTML]{ffcfd2} REGM~\cite{asnani2023reverse}  &\cellcolor[HTML]{ffcfd2} 2023      &\cellcolor[HTML]{ffcfd2}                  & \cellcolor[HTML]{ffcfd2}General               &  \cellcolor[HTML]{ffcfd2}\href{https://drive.google.com/file/d/1bAmC_9aMkWJB_scGvOOWvNeLa9FBoMUr/view}{\color[HTML]{0000EE} { Link}}& \cellcolor[HTML]{ffcfd2} T/I2I                                                                & \cellcolor[HTML]{ffcfd2}-                                                             & \cellcolor[HTML]{ffcfd2}116,000              & \cellcolor[HTML]{ffcfd2}CelebA~\cite{liu2015faceattributes}, LSUN                                                                                             &\cellcolor[HTML]{ffcfd2} 116 publicly available GMs                              \\ \cline{2-11} 
                                     &\cellcolor[HTML]{ffcfd2}DMimage~\cite{corvi2023detection}  & \cellcolor[HTML]{ffcfd2}2022       &\cellcolor[HTML]{ffcfd2}\checkmark                   & \cellcolor[HTML]{ffcfd2}General               &\cellcolor[HTML]{ffcfd2}\href{ https://github.com/grip-unina/DMimageDetection }{\color[HTML]{0000EE} { Link}}& \cellcolor[HTML]{ffcfd2} T2I                                                                          & \cellcolor[HTML]{ffcfd2}200,000                                                          & \cellcolor[HTML]{ffcfd2}200,000              &\cellcolor[HTML]{ffcfd2} COCO, LSUN                                                                                                 &\cellcolor[HTML]{ffcfd2} LDM~\cite{rombach2021highresolution}                                                                                                                                                   \\ \cline{2-11} 
                                     &\cellcolor[HTML]{ffcfd2}AIGCD\cite{zhong2023rich}     & \cellcolor[HTML]{ffcfd2}2023      &\cellcolor[HTML]{ffcfd2}\checkmark             &\cellcolor[HTML]{ffcfd2} General               &\cellcolor[HTML]{ffcfd2}\href{ https://github.com/Ekko-zn/AIGCDetectBenchmark }{\color[HTML]{0000EE} { Link}}& \cellcolor[HTML]{ffcfd2} T/I2I                                                                                         &\cellcolor[HTML]{ffcfd2} 360,000                                                               &\cellcolor[HTML]{ffcfd2} 508,500              &\cellcolor[HTML]{ffcfd2}  LSUN,      COCO, FFHQ~\cite{karras2019style}                                                                                                                                                                         & \cellcolor[HTML]{ffcfd2}SDMs, GANs,ADM~\cite{dhariwal2021diffusion},DALL·E 2,GLIDE\cite{nichol2021glide}                                                  \\ \cline{2-11} 
                                     &\cellcolor[HTML]{ffcfd2}DIF~\cite{sinitsa2023deep}   &\cellcolor[HTML]{ffcfd2} 2023     &\cellcolor[HTML]{ffcfd2}                &\cellcolor[HTML]{ffcfd2} General               &  \cellcolor[HTML]{ffcfd2}\href{https://github.com/sergo2020/dif_pytorch_official}{\color[HTML]{0000EE} { Link}}& \cellcolor[HTML]{ffcfd2} T/I2I                                                                        & \cellcolor[HTML]{ffcfd2}84,300                                                             &\cellcolor[HTML]{ffcfd2} 84,300                  & \cellcolor[HTML]{ffcfd2}LAION-5B~\cite{schuhmann2022laion}                                                                                                                                                                                                & \cellcolor[HTML]{ffcfd2} SDMs, DALL·E 2, GLIDE~\cite{nichol2021glide}, GANs                                                                                                                                                                                                      \\ \cline{2-11} 
                              
                                     & \cellcolor[HTML]{ffcfd2}Fake2M\cite{lu2023seeing}     &  \cellcolor[HTML]{ffcfd2}2023       &\cellcolor[HTML]{ffcfd2}\checkmark           &\cellcolor[HTML]{ffcfd2} General               & \cellcolor[HTML]{ffcfd2}\href{https://github.com/Inf-imagine/Sentry}{\color[HTML]{0000EE} { Link}}&  \cellcolor[HTML]{ffcfd2}  T/I2I                                                                         &  \cellcolor[HTML]{ffcfd2}-                                                            & \cellcolor[HTML]{ffcfd2} 2,300,000            &  \cellcolor[HTML]{ffcfd2}CC3M\cite{sharma2018conceptual}                                                                                                                                                                                                    & \cellcolor[HTML]{ffcfd2} SD-V1.5~\cite{rombach2022high},  IF~\cite{deepfloyd2023if}, StyleGAN3 \\ \cline{2-11} 
                              
                                     & \cellcolor[HTML]{ffcfd2}SID-Set~\cite{huang2024sida}     &  \cellcolor[HTML]{ffcfd2}2024       &\cellcolor[HTML]{ffcfd2}           &\cellcolor[HTML]{ffcfd2} Social Media               & \cellcolor[HTML]{ffcfd2}\href{https://github.com/hzlsaber/SIDA}{\color[HTML]{0000EE} { Link}}&  \cellcolor[HTML]{ffcfd2}  T/I2I                                                                         &  \cellcolor[HTML]{ffcfd2}100,000                                                            & \cellcolor[HTML]{ffcfd2} 200,000            &  \cellcolor[HTML]{ffcfd2}COCO, Flickr30k, MagicBrush                                                                                                                                                                                                    & \cellcolor[HTML]{ffcfd2}FLUX~\cite{fluxai2025}                                             \\ \cline{2-11} 
                              
                                     & \cellcolor[HTML]{ffcfd2}Chameleon~\cite{yan2024sanity}     &  \cellcolor[HTML]{ffcfd2}2024       &\cellcolor[HTML]{ffcfd2}\checkmark           &\cellcolor[HTML]{ffcfd2} Social Media               & \cellcolor[HTML]{ffcfd2}\href{https://github.com/shilinyan99/AIDE}{\color[HTML]{0000EE} { Link}}&  \cellcolor[HTML]{ffcfd2}  T/I2I                                                                         &  \cellcolor[HTML]{ffcfd2}14,863                                                            & \cellcolor[HTML]{ffcfd2}  11,170            &  \cellcolor[HTML]{ffcfd2}Unsplash~\cite{unsplash}                                                                                                                                                                                                    & \cellcolor[HTML]{ffcfd2}GANs, SDMs, DALL·E 2,GLIDE\cite{nichol2021glide}   \\ \cline{2-11} 
                              
                                     & \cellcolor[HTML]{ffcfd2}DF40~\cite{yan2024df40}     &  \cellcolor[HTML]{ffcfd2}2024       &\cellcolor[HTML]{ffcfd2}\checkmark           &\cellcolor[HTML]{ffcfd2} Face              & \cellcolor[HTML]{ffcfd2}\href{https://github.com/YZY-stack/DF40}{\color[HTML]{0000EE} { Link}}&  \cellcolor[HTML]{ffcfd2}  T/I2I                                                                         &  \cellcolor[HTML]{ffcfd2}-                                                           & \cellcolor[HTML]{ffcfd2}$\sim$~1,000,000              &  \cellcolor[HTML]{ffcfd2}FF++, CDF, FFHQ, CelebA & \cellcolor[HTML]{ffcfd2}SDMS, GANs, Midjourney, DDPM  \\ \cline{2-11} 
                              
                                     & \cellcolor[HTML]{ffcfd2}FakeBench~\cite{li2024fakebench}     &  \cellcolor[HTML]{ffcfd2}2024       &\cellcolor[HTML]{ffcfd2}\checkmark           &\cellcolor[HTML]{ffcfd2}General              & \cellcolor[HTML]{ffcfd2}\href{https://github.com/Yixuan423/FakeBench}{\color[HTML]{0000EE} { Link}}&  \cellcolor[HTML]{ffcfd2}  T/I2I                                                                         &  \cellcolor[HTML]{ffcfd2}3,000                                                           & \cellcolor[HTML]{ffcfd2}3,000            &  \cellcolor[HTML]{ffcfd2}  10 Public Datasets                                                                                                                                                   & \cellcolor[HTML]{ffcfd2}10 Generative Models   \\ \cline{2-11} 
                              
                                     & \cellcolor[HTML]{ffcfd2}AI-Face~\cite{lin2024ai}     &  \cellcolor[HTML]{ffcfd2}2024       &\cellcolor[HTML]{ffcfd2}\checkmark           &\cellcolor[HTML]{ffcfd2}Face              & \cellcolor[HTML]{ffcfd2}\href{https://github.com/Purdue-M2/AI-Face-FairnessBench}{\color[HTML]{0000EE} { Link}}&  \cellcolor[HTML]{ffcfd2}  T/I2I                                                                         &  \cellcolor[HTML]{ffcfd2}400,885                                                          & \cellcolor[HTML]{ffcfd2}1,245,660           &  \cellcolor[HTML]{ffcfd2}6 Public datasets                                                                                                                                                   & \cellcolor[HTML]{ffcfd2}SDMs, GANs, Midjourney, IF                                                                                                                                                         \\ \hline
\multirow{7}{*}{\cellcolor[HTML]{FFFFFF} \textbf{Video}}                         & \cellcolor[HTML]{fec89a} WildDeepfake\cite{zi2020wilddeepfake}    & \cellcolor[HTML]{fec89a}2021    &\cellcolor[HTML]{fec89a}        & \cellcolor[HTML]{fec89a}Face                 &  \cellcolor[HTML]{fec89a}\href{https://github.com/OpenTAI/wild-deepfake}{\color[HTML]{0000EE} { Link}}& \cellcolor[HTML]{fec89a}-                                                                             &\cellcolor[HTML]{fec89a}3,805                                                           & \cellcolor[HTML]{fec89a}3,509              & \cellcolor[HTML]{fec89a}Social Media                                                                                                                                                                                                & \cellcolor[HTML]{fec89a}Social Media                                                                                                             \\ \cline{2-11}  & \cellcolor[HTML]{fec89a} DiffHead\cite{stypulkowski2024diffused}    & \cellcolor[HTML]{fec89a}2023    &\cellcolor[HTML]{fec89a}        & \cellcolor[HTML]{fec89a}Face                 &  \cellcolor[HTML]{fec89a}\href{https://drive.google.com/file/d/1zWSqtV7O4WGkgh6WB55b8Mdg2lXXUudH/view}{\color[HTML]{0000EE} { Link}}& \cellcolor[HTML]{fec89a} I.A2V                                                                             &\cellcolor[HTML]{fec89a} -                                                           & \cellcolor[HTML]{fec89a}820              & \cellcolor[HTML]{fec89a}CREMA~\cite{cao2014crema}                                                                                                                                                                                                & \cellcolor[HTML]{fec89a}Diffused Heads: build on DDPM                                                                                                            \\ \cline{2-11}  & \cellcolor[HTML]{fec89a} DVF\cite{song2024learning}    & \cellcolor[HTML]{fec89a}2024    &\cellcolor[HTML]{fec89a}        & \cellcolor[HTML]{fec89a}General                 &  \cellcolor[HTML]{fec89a}\href{https://github.com/SparkleXFantasy/MM-Det}{\color[HTML]{0000EE} { Link}}& \cellcolor[HTML]{fec89a} I/T2V                                                                             &\cellcolor[HTML]{fec89a}2,750                                                          & \cellcolor[HTML]{fec89a}3,938              & \cellcolor[HTML]{fec89a}Internvid~\cite{wang2023internvid}, Youtube-8M~\cite{abu2016youtube}                                                                                                                                                                                                & \cellcolor[HTML]{fec89a}8 Diffusion Models  \\ \cline{2-11} & \cellcolor[HTML]{fec89a} GenVideo\cite{chen2024demamba}    & \cellcolor[HTML]{fec89a}2024    &\cellcolor[HTML]{fec89a}\checkmark        & \cellcolor[HTML]{fec89a}General                 &  \cellcolor[HTML]{fec89a}\href{https://github.com/chenhaoxing/DeMamba}{\color[HTML]{0000EE} { Link}}& \cellcolor[HTML]{fec89a} I/T2V                                                                             &\cellcolor[HTML]{fec89a}1,223,511                                                          & \cellcolor[HTML]{fec89a}1,078,838              & \cellcolor[HTML]{fec89a} \begin{tabular}[c]{@{}c@{}} Kinetics-400~\cite{kay2017kinetics}, Youku-mPLUG~\cite{xu2023youku},\\ MSR-VTT~\cite{xu2016msr}    \end{tabular}                                                                                                                                                                                         & \cellcolor[HTML]{fec89a}20 Generative Models  \\ \cline{2-11}  & \cellcolor[HTML]{fec89a} GenVidBench\cite{ni2025genvidbench}    & \cellcolor[HTML]{fec89a}2025    &\cellcolor[HTML]{fec89a}\checkmark        & \cellcolor[HTML]{fec89a}General                 &  \cellcolor[HTML]{fec89a}\href{https://genvidbench.github.io/}{\color[HTML]{0000EE} { Link}}& \cellcolor[HTML]{fec89a} I/T2V                                                                             &\cellcolor[HTML]{fec89a}33,931                                                         & \cellcolor[HTML]{fec89a}110,400              & \cellcolor[HTML]{fec89a}Vript\cite{yang2024vript}, HD-VG-130M~\cite{wang2025swap}                                                                                                                                                                                  & \cellcolor[HTML]{fec89a}8 Generative Models  \\ \cline{2-11} & \cellcolor[HTML]{fec89a} PDID\cite{walker2024merging}    & \cellcolor[HTML]{fec89a}2024    &\cellcolor[HTML]{fec89a}        & \cellcolor[HTML]{fec89a}Social Media                 &  \cellcolor[HTML]{fec89a}\href{https://airtable.com/appOU03dlKuBdbmty/shrEkrIYINbrcKQ3z/tbleGYjNLn2D4Xfzs}{\color[HTML]{0000EE} { Link}}& \cellcolor[HTML]{fec89a}-                                                                             &\cellcolor[HTML]{fec89a}-                                                         & \cellcolor[HTML]{fec89a}-             & \cellcolor[HTML]{fec89a}Social Media                                                                                                                                                                                               & \cellcolor[HTML]{fec89a}Social Media   \\ \hline
\multirow{3}{*}{\cellcolor[HTML]{FFFFFF} \textbf{Audio}}                 &\cellcolor[HTML]{C3ACD0} In-the-Wild\cite{muller2022does}   & \cellcolor[HTML]{C3ACD0}2022    &\cellcolor[HTML]{C3ACD0}\checkmark        &\cellcolor[HTML]{C3ACD0} Speech                 &  \cellcolor[HTML]{C3ACD0}\href{https://deepfake-total.com/in_the_wild}{\color[HTML]{0000EE} { Link}}&\cellcolor[HTML]{C3ACD0}-                                                                            &\cellcolor[HTML]{C3ACD0}20.7 hours                                                             &\cellcolor[HTML]{C3ACD0}17.2 hours               & \cellcolor[HTML]{C3ACD0}   Social Media, Video Streaming Platforms                                                                                                                                                                                            & \cellcolor[HTML]{C3ACD0} Social Media, Video Streaming Platforms   \\    \cline{2-11}      &\cellcolor[HTML]{C3ACD0} LibriSeVoc\cite{sun2023ai}   & \cellcolor[HTML]{C3ACD0}2023    &\cellcolor[HTML]{C3ACD0}        &\cellcolor[HTML]{C3ACD0} Speech                 &  \cellcolor[HTML]{C3ACD0}\href{https://github.com/csun22/Synthetic-Voice-Detection-Vocoder-Artifacts}{\color[HTML]{0000EE} { Link}}&\cellcolor[HTML]{C3ACD0}  T2A                                                                            &\cellcolor[HTML]{C3ACD0} 13,201                                                             &\cellcolor[HTML]{C3ACD0} 79,206               & \cellcolor[HTML]{C3ACD0}LibriTTS\cite{zen2019libritts}                                                                                                                                                                                                   & \cellcolor[HTML]{C3ACD0}DiffWave~\cite{kong2020diffwave}, WaveNet~\cite{oord2016wavenet}                                                                                                      \\ \cline{2-11}  &\cellcolor[HTML]{C3ACD0} SONAR\cite{li2024sonar}   & \cellcolor[HTML]{C3ACD0}2024    &\cellcolor[HTML]{C3ACD0}\checkmark        &\cellcolor[HTML]{C3ACD0} Speech                 &  \cellcolor[HTML]{C3ACD0}\href{https://github.com/Jessegator/SONAR}{\color[HTML]{0000EE} { Link}}&\cellcolor[HTML]{C3ACD0}  T2A                                                                            &\cellcolor[HTML]{C3ACD0}-                                                             &\cellcolor[HTML]{C3ACD0}2,274               & \cellcolor[HTML]{C3ACD0}LibriTTS\cite{zen2019libritts}                                                                                                                                                                                                   & \cellcolor[HTML]{C3ACD0}OpenAI~\cite{openai2025texttospeech}, Seed-TTS~\cite{anastassiou2024seed}, AudioGen~\cite{kreuk2022audiogen}  \\ \cline{2-11}  &\cellcolor[HTML]{C3ACD0} ASVspoof 2024\cite{wang2024asvspoof}   & \cellcolor[HTML]{C3ACD0}2024    &\cellcolor[HTML]{C3ACD0}\checkmark    &\cellcolor[HTML]{C3ACD0} Speech                 &  \cellcolor[HTML]{C3ACD0}\href{https://www.asvspoof.org/workshop2024}{\color[HTML]{0000EE} { Link}}&\cellcolor[HTML]{C3ACD0}  T/A2A                                                                            &\cellcolor[HTML]{C3ACD0}$\sim$289,527                                                             &\cellcolor[HTML]{C3ACD0}$\sim$1,211,186               & \cellcolor[HTML]{C3ACD0}MLS-English~\cite{pratap2020mls}                                                                                                                                                                                                 & \cellcolor[HTML]{C3ACD0}32 Manipulation Methods \\  \hline

\multirow{12}{*}{ \textbf{ \begin{tabular}[c]{@{}c@{}} Multi-\\ modal \end{tabular}}} & \cellcolor[HTML]{f3d5b5}$DGM^4$ \cite{shao2023detecting}        & \cellcolor[HTML]{f3d5b5}2023  &\cellcolor[HTML]{f3d5b5}        &\cellcolor[HTML]{f3d5b5}News                &\cellcolor[HTML]{f3d5b5}\href{ https://github.com/rshaojimmy/MultiModal-DeepFake }{\color[HTML]{0000EE} { Link}}       & \cellcolor[HTML]{f3d5b5}      T/I2T                                                                             &\cellcolor[HTML]{f3d5b5}77,426                                                          & \cellcolor[HTML]{f3d5b5}   152,574            &\cellcolor[HTML]{f3d5b5} Visual News~\cite{liu2020visual}   & \cellcolor[HTML]{f3d5b5}  B-GST~\cite{sudhakar2019transforming}, StyleCLIP~\cite{patashnik2021styleclip},    HFGI~\cite{wang2022high}                                                                                                                                                                                                                                     \\ \cline{2-11} 
                                     &\cellcolor[HTML]{f3d5b5}  COCOFake\cite{amoroso2024parents}  & \cellcolor[HTML]{f3d5b5}2023        &\cellcolor[HTML]{f3d5b5}           &\cellcolor[HTML]{f3d5b5} General              &\cellcolor[HTML]{f3d5b5}\href{https://github.com/aimagelab/ELSA_COCO-Fake}{\color[HTML]{0000EE} { Link}}& \cellcolor[HTML]{f3d5b5} T/I2T                                                                                  & \cellcolor[HTML]{f3d5b5}  113,287                                                         &\cellcolor[HTML]{f3d5b5} 566,435             &  \cellcolor[HTML]{f3d5b5}    COCO                                                                               &\cellcolor[HTML]{f3d5b5} SDMs          \\ \cline{2-11}  &\cellcolor[HTML]{f3d5b5}AV-Deepfake1M\cite{cai2024av}  & \cellcolor[HTML]{f3d5b5}2023     &\cellcolor[HTML]{f3d5b5}\checkmark               & \cellcolor[HTML]{f3d5b5}Face             &\cellcolor[HTML]{f3d5b5}\href{https://github.com/ControlNet/AV-Deepfake1M}{\color[HTML]{0000EE} { Link}}         &   \cellcolor[HTML]{f3d5b5}\begin{tabular}[c]{@{}c@{}} T2A \\I2I    \end{tabular}                                                                             & \cellcolor[HTML]{f3d5b5}286,721                                                      & \cellcolor[HTML]{f3d5b5}860,039            &      \cellcolor[HTML]{f3d5b5}Voxceleb2~\cite{chung2018voxceleb2},                                                                  & \cellcolor[HTML]{f3d5b5}VITS~\cite{kim2021conditional}, YourTTS~\cite{casanova2022yourtts}, TalkLip~\cite{wang2023seeing}         \\ \cline{2-11} & \cellcolor[HTML]{f3d5b5}$D^3$~\cite{baraldi2024contrasting}     &  \cellcolor[HTML]{f3d5b5}2024       &\cellcolor[HTML]{f3d5b5}           &\cellcolor[HTML]{f3d5b5}General              & \cellcolor[HTML]{f3d5b5}\href{https://github.com/aimagelab/CoDE}{\color[HTML]{0000EE} { Link}}&  \cellcolor[HTML]{f3d5b5}T2I                                                                         &  \cellcolor[HTML]{f3d5b5}$\sim$2,300,000                                                           & \cellcolor[HTML]{f3d5b5}$\sim$9,200,000            &  \cellcolor[HTML]{f3d5b5}LAION-400M                                                                                                                                                   & \cellcolor[HTML]{f3d5b5}SDMs, IF   \\ \cline{2-11}  &  \cellcolor[HTML]{f3d5b5}$M^3A$\cite{xu2024m3a}  & \cellcolor[HTML]{f3d5b5}2024     &\cellcolor[HTML]{f3d5b5}-               & \cellcolor[HTML]{f3d5b5}News            &\cellcolor[HTML]{f3d5b5}-         &   \cellcolor[HTML]{f3d5b5}\begin{tabular}[c]{@{}c@{}} T2T/I/V/A \\T/I/V/A2T    \end{tabular}                                                                             & \cellcolor[HTML]{f3d5b5}708,425                                                         & \cellcolor[HTML]{f3d5b5}6,566,386            &      \cellcolor[HTML]{f3d5b5}60 News Outlets                                                                               & \cellcolor[HTML]{f3d5b5}LLaMA2, GPT-4, GLIDE~\cite{nichol2021glide}, SD, Tango~\cite{ghosal2023text}          \\ \cline{2-11} &  \cellcolor[HTML]{f3d5b5}LOKI\cite{ye2024loki}  & \cellcolor[HTML]{f3d5b5}2024     &\cellcolor[HTML]{f3d5b5}\checkmark               & \cellcolor[HTML]{f3d5b5}General             &\cellcolor[HTML]{f3d5b5}\href{ https://opendatalab.github.io/LOKI/ }{\color[HTML]{0000EE} { Link}}         &   \cellcolor[HTML]{f3d5b5}\begin{tabular}[c]{@{}c@{}} T2T/I/V/A \\T/I/V/A2T    \end{tabular}                                                                             & \cellcolor[HTML]{f3d5b5}$\sim$9,000                                                         & \cellcolor[HTML]{f3d5b5}$\sim$9,000            &      \cellcolor[HTML]{f3d5b5}   21 Public Datasets                                                                           & \cellcolor[HTML]{f3d5b5}43 Generative Models  \\ \cline{2-11} &  \cellcolor[HTML]{f3d5b5}MMFakeBench\cite{liu2024mmfakebench}  & \cellcolor[HTML]{f3d5b5}2024     &\cellcolor[HTML]{f3d5b5}\checkmark               & \cellcolor[HTML]{f3d5b5}Social Media             &\cellcolor[HTML]{f3d5b5}\href{https://liuxuannan.github.io/MMFakeBench.github.io/}{\color[HTML]{0000EE} { Link}}         &   \cellcolor[HTML]{f3d5b5}\begin{tabular}[c]{@{}c@{}} T2T/I \\I2I    \end{tabular}                                                                             & \cellcolor[HTML]{f3d5b5}-                                                      & \cellcolor[HTML]{f3d5b5}$\sim$11,000            &      \cellcolor[HTML]{f3d5b5}\begin{tabular}[c]{@{}c@{}} MS-COCO, VisualNews,\\Reddit,FEVER~\cite{Thorne18Fever}\end{tabular}                                                                   & \cellcolor[HTML]{f3d5b5}GPT-3.5, SD-XL,  DALL·E 3, Midjourney         \\ \cline{2-11} &  \cellcolor[HTML]{f3d5b5}Deepfake-Eval\cite{chandra2025deepfake}  & \cellcolor[HTML]{f3d5b5}2024     &\cellcolor[HTML]{f3d5b5}\checkmark               & \cellcolor[HTML]{f3d5b5}Social Media             &\cellcolor[HTML]{f3d5b5}\href{https://github.com/nuriachandra/Deepfake-Eval-2024}{\color[HTML]{0000EE} { Link}}         &   \cellcolor[HTML]{f3d5b5}\begin{tabular}[c]{@{}c@{}} T2I/A/V \\I2I/V    \end{tabular}                                                                             & \cellcolor[HTML]{f3d5b5}3,390                                                      & \cellcolor[HTML]{f3d5b5}2,441            &      \cellcolor[HTML]{f3d5b5}Social Media                                                                   & \cellcolor[HTML]{f3d5b5}Social Media         \\ \cline{2-11} &  \cellcolor[HTML]{f3d5b5}ILLUSION\cite{thakralillusion}  & \cellcolor[HTML]{f3d5b5}2025     &\cellcolor[HTML]{f3d5b5}\checkmark               & \cellcolor[HTML]{f3d5b5}General             &\cellcolor[HTML]{f3d5b5}\href{https://www.iab-rubric.org/illusion-database}{\color[HTML]{0000EE} { Link}}         &   \cellcolor[HTML]{f3d5b5}\begin{tabular}[c]{@{}c@{}} T2A/I \\I2I    \end{tabular}                                                                             & \cellcolor[HTML]{f3d5b5}139,740                                                      & \cellcolor[HTML]{f3d5b5}1,232,246            &      \cellcolor[HTML]{f3d5b5}\begin{tabular}[c]{@{}c@{}}CelebV-Text~\cite{yu2023celebv}, COCO, \\MusicCaps~\cite{agostinelli2023musiclm}, Social Media\end{tabular}                                                                   & \cellcolor[HTML]{f3d5b5} 28 Generative Methods      \\ \cline{2-11} 
                                     
                                      \hline
\end{tabular}%
 }
\vspace{-2mm} 
\label{tab:datasets}
\end{table*}

\section{Generation}\label{sec:LAIMs_generated_Multimedia}
\noindent
In this section, we provide an overview of large generative AI models, their generation mechanisms, and the type of content they generate. 

\smallskip
\noindent
$\clubsuit$ \textbf{Text}.
Machine-generated text, primarily driven by the advent of LLMs, is increasingly permeating many aspects of our daily lives. The exceptional proficiency of LLMs in understanding, following, and complex reasoning~\cite{yang2023harnessing} has established their dominance in text creation.  Recently, we have witnessed a surge of LLMs including 
OpenAI GPT-4~\cite{achiam2023gpt},  Google Gemini~\cite{team2023gemini}, Meta LLaMA2~\cite{touvron2023llama2}
as well as the remarkable performance they have achieved in various tasks, such as News~\cite{zellers2019defending, uchendu2021turingbench}, Question Answering~\cite{fan2019eli5},  Biomedicine~\cite{wan2023med}, Code Generation~\cite{zheng2023codegeex}, Tweets~\cite{fagni2021tweepfake}, and Scientific Writing~\cite{rosati-2022-synscipass}, see Fig.~\ref{fig:generration_overview} a). More details can be found in \cite{hadi2023large, zhao2023survey}.


\raisebox{-0.3ex}{\scalebox{1.3}{\ding{192}}} \textbf{LAIM Generated Datasets (Recommended)}. The datasets for LLM-generated text detection are listed in Table~\ref{tab:datasets}. Recent efforts have introduced large-scale and challenging benchmarks to evaluate detector robustness under realistic and adversarial settings. For instance, RAID~\cite{dugan-etal-2024-raid} is the largest and most comprehensive benchmark to date, containing over 6 million texts generated by 11 LLMs across 8 domains, 4 decoding strategies, and 11 types of adversarial attacks. It supports fine-grained evaluation of generalization and robustness by simulating real-world perturbations such as paraphrasing, machine translation, and sampling noise.
DetectRL~\cite{wu2024detectrl} focuses on real-world detection by integrating reinforcement learning strategies and human-in-the-loop edits to simulate text that mimics post-edited or polished LLM outputs. MAGE~\cite{li2023deepfake} presents a diverse benchmark combining 27 LLMs and 7 writing tasks, including both creative and factual domains, enabling a broad analysis of generalization performance across model families and genres.

\underline{Social Media.}
F3~\cite{lucas-etal-2023-fighting} introduces a novel dataset to study both generation and detection of disinformation across news articles and social posts, leveraging perturbation- and paraphrase-based generation with models like GPT-3.5. Similarly, MultiSocial~\cite{macko2024multisocial} fills a critical gap by offering the first multilingual, multi-platform benchmark of social media texts, covering 22 languages and 5 platforms. Its focus on informal, short-form user-generated content makes it particularly valuable for evaluating cross-lingual and domain-specific detection robustness.

\raisebox{-0.3ex}{\scalebox{1.3}{\ding{193}}}  \textbf{Non-LAIM Generated Datasets}. 
AA~\cite{uchendu2020authorship}, TweepFake~\cite{fagni2021tweepfake}, and GPT-2 Output~\cite{solaiman2019release} are widely used datasets produced by earlier language models (\eg, n-gram models, RNNs, LSTMs, or smaller transformer models like GPT-2). These datasets typically include machine-generated text that exhibits limited fluency, shallow coherence, and frequent repetition, especially in longer sequences. OpenAI’s analysis of GPT-2 outputs revealed notable distributional artifacts, such as the underuse of proper nouns and overuse of pronouns and generic expressions, which early detectors could effectively exploit~\cite{solaiman2019release}. In contrast, text generated by LLMs, as seen in datasets like RAID~\cite{dugan-etal-2024-raid} or DetectRL~\cite{wu2024detectrl}, is significantly more fluent, coherent, and contextually accurate.

\smallskip
\noindent
$\clubsuit$ \textbf{Image}. In the challenging task of image synthesis, diffusion models (DMs) \cite{ho2020denoising, song2019generative, song2020score} have emerged as the new state-of-the-art family of deep generative models. The image generation process in DMs usually contains two processes \cite{ho2020denoising}: a forward process that progressively destroys data by adding noise and a reverse process that learns to generate new data by denoising. More details can be found in \cite{li2023multimodal, yang2023diffusionsurvey}.
Current research on diffusion models is mostly based on three predominant formulations: denoising diffusion probabilistic models (DDPMs)~\cite{ho2020denoising}, score-based generative models (SGMs)~\cite{song2019generative}, and stochastic differential equations (Score SDEs)~\cite{song2020score}. Building upon them, more advanced models have emerged in image generation, 
including OpenAI DALL·E 3~\cite{betker2023improving}, Stable Diffusion 3.5~\cite{sd_3}, Google Imagen 3~\cite{Imagen_3}, Midjourney 7~\cite{midjourney}, Amazon Titan Image Generator 2~\cite{amazon_titan}, and Meta Emu Edit~\cite{sheynin2023emu}, see Fig.~\ref{fig:generration_overview} b).

\raisebox{-0.3ex}{\scalebox{1.3}{\ding{192}}} \textbf{LAIM Generated Datasets (Recommended)}. There are several million-scale LAIM-generated image detection datasets, see Table~\ref{tab:datasets}. Among them, GenImage\cite{zhu2023genimage} offers over 2.6M real and fake images across 1000 ImageNet categories, generated by both GANs and diffusion models. It emphasizes diverse content and proposes tasks such as cross-generator and degraded-image detection to simulate real-world challenges. DiffusionDB\cite{wang2022diffusiondb}, in contrast, uniquely pairs 14M images with 1.8M user-written prompts, revealing generation patterns and biases in human-guided diffusion-based generation, though it is limited to Stable Diffusion outputs.
AI-Face~\cite{lin2024ai} introduces the first million-scale demographically annotated dataset of AI-generated faces, incorporating 37 generation methods (deepfakes, GANs, and DMs) and annotated gender, age, and skin tone. It also provides a comprehensive fairness benchmark for evaluating detection disparities across demographic groups. DF40\cite{yan2024df40} pushes realism and diversity further by including 40 modern face synthesis techniques spanning face-swapping, reenactment, editing, and full-synthesis, reflecting the heterogeneity of real-world manipulations and bridging the gap between academic and practical detection scenarios.

\underline{Social Media.}
SID-Set~\cite{huang2024sida} is a comprehensive dataset with 300K real, synthetic, and tampered images annotated for detection, localization, and explanation. It emphasizes scene complexity and visual plausibility, targeting real-world misinformation scenarios. Chameleon~\cite{yan2024sanity} complements this by curating images that consistently fool human annotators, representing high-resolution, semantically rich content from categories such as humans, animals, and objects. Detection models struggle significantly on this benchmark, highlighting current limitations.

\raisebox{-0.3ex}{\scalebox{1.3}{\ding{193}}} \textbf{Non-LAIM Generated Datasets}. FaceForensics++ (FF++) \cite{rossler2019faceforensics++}, DeepfakeDetection (DFD) \cite{googledeepfakes2019}, Deepfake Detection Challenge (DFDC) \cite{deepfakedetection2021}, and Celeb-DF \cite{li2020celebdf} are four most representative deepfake video datasets. Although originally constructed as video datasets, they are often used for frame-level detection; therefore, we also categorize them under image-based datasets. They are created by early manipulation methods, such as graphics-based~\cite{FaceSwap2018, thies2016face2face}, autoencoder-based~\cite{fakeapp}, and GAN-based~\cite{thies2019deferred}. These methods typically rely on paired source-target videos, facial tracking, and region blending to produce partially manipulated videos that often leave visible artifacts, such as blending boundaries around the face, color mismatches between the manipulated region and surrounding skin, temporal flickering across frames, and unnatural mouth or eye movements. In contrast, diffusion models generate content entirely from scratch, using text or image prompts. This enables highly customized and diverse outputs in terms of style, identity, and context. Unlike traditional deepfake methods, diffusion models do not require explicit source-target video pairs and often produce artifact-free content, making detection significantly more challenging.

\smallskip
\noindent
$\clubsuit$ \textbf{Video}. In the pursuit of high-quality video generation, recent research has also turned to diffusion models. Early work~\cite{harvey2022flexible, yang2023diffusion} are based on DDPM~\cite{ho2020denoising} for video generation. Current research extends text-to-image (T2I) diffusion models to text-to-video (T2V) generation.  
Early T2V efforts include Meta’s Make-A-Video~\cite{singer2022make}, which extends a diffusion-based T2I model using spatiotemporally factorized diffusion, and Emu Video~\cite{girdhar2023emu}, a simplified two-stage successor. Google’s Imagen Video~\cite{ho2022imagen}, Stability AI’s Stable Video Diffusion~\cite{blattmann2023stable}, Runway’s GEN-2~\cite{runway_gen2}, and Pika~\cite{pika} also contributed video generation pipelines.
More recently, the field has advanced rapidly with powerful models such as OpenAI’s Sora~\cite{openai2024sora}, Google DeepMind’s Veo2~\cite{deepmind2025veo2}.
as shown in Fig.~\ref{fig:generration_overview} c).

\raisebox{-0.3ex}{\scalebox{1.3}{\ding{192}}} \textbf{LAIM Generated Datasets (Recommended)}. Though building upon the success of T2I generation, T2V generation requires temporally smooth and realistic motion that matches the text, besides high-quality visual content, making it still in the nascent stage compared with image generation. Furthermore, video generation is much more computationally costly than image synthesis. 
Recent work has begun to contribute to LAIM-generated video datasets, see Table~\ref{tab:datasets}. DiffHead~\cite{stypulkowski2024diffused} provides high-quality talking-head videos generated from a single image and audio using a diffusion model. While domain-specific, its temporal smoothness and realistic expressions make it well-suited for evaluating detectors on subtle artifacts like lip-sync mismatch and unnatural facial motion. On the other hand, DVF~\cite{song2024learning} provides a broader testing ground for general-purpose video detection. It consists of fake videos generated by eight different diffusion models, including both open-source and commercial systems. GenVideo~\cite{chen2024demamba} and GenVidBench~\cite{ni2025genvidbench} represent the latest efforts toward scaling and realism. GenVideo provides an extensive corpus of over 2.3 million videos, collected from diverse real-world sources and spanning 20 generative models. Its emphasis on massive diversity and model heterogeneity makes it valuable for studying broad generalization across unseen generators. GenVidBench focuses on constructing a challenging benchmark for AI-generated video detection, emphasizing cross-source and cross-generator generalization.

\underline{Social Media.}
Deepfake videos are increasingly shared across platforms like YouTube, Twitter, and TikTok, highlighting the need for datasets that reflect the real-world complexity of such content. Unlike lab-created datasets—produced under controlled conditions with limited diversity—videos circulating online exhibit greater variability in quality, compression, and context. These differences are critical, as detectors trained on lab data often fail to generalize to the noisy, unpredictable nature of in-the-wild deepfakes seen on social platforms. WildDeepfake~\cite{zi2020wilddeepfake} collects over 700 deepfake videos directly from the internet, capturing a wide range of social media-style content with diverse backgrounds, multiple speakers, varying resolutions, and real-world compression effects, making it an effective benchmark for evaluating detector robustness in open-world scenarios. Complementing this, the Political Deepfakes Incidents Database (PDID)~\cite{walker2024merging} compiles verified political deepfakes that have circulated widely on social platforms, enabling the study of how deepfakes influence discourse, political trust, and public perception. It supports detection research that goes beyond technical accuracy, addressing detection failures and social impacts.

\raisebox{-0.3ex}{\scalebox{1.3}{\ding{193}}} \textbf{Non-LAIM Generated Datasets}.  FaceForensics++ (FF++) \cite{rossler2019faceforensics++}, DeepfakeDetection (DFD) \cite{googledeepfakes2019}, Deepfake Detection Challenge (DFDC) \cite{deepfakedetection2021}, and Celeb-DF \cite{li2020celebdf} are four most representative deepfake video datasets. Since they have also been used for image-level detection and were previously discussed, we omit their detailed discussion here to avoid redundancy.

\smallskip
\noindent
$\clubsuit$ \textbf{Audio}. Most audio synthesis by diffusion models focuses on text-to-speech (TTS) tasks. One line of work~\cite{jeong2021diff, chen2020wavegrad, kong2020diffwave} first generates acoustic features, e.g., mel-spectrogram, and then outputs waveform with a vocoder. Another branch of work~\cite{chen2021wavegrad, shi2023iton} attempts to solve the TTS task in an end-to-end manner, as shown in Fig.~\ref{fig:generration_overview} d). Diff-TTS~\cite{jeong2021diff} is the first work that applies DDPM~\cite{ho2020denoising} to mel-spectrogram generation. It transforms the noise signal into a mel-spectrogram via diffusion time steps. \cite {chen2020wavegrad, kong2020diffwave} apply diffusing models to vocoder for generating waveform based on mel-spectrogram. Instead of treating acoustic modeling and vocoder modeling as independent processes, ~\cite{chen2021wavegrad, shi2023iton} generate audio from text without acoustic features as explicit representation.


\raisebox{-0.3ex}{\scalebox{1.3}{\ding{192}}} \textbf{LAIM Generated Datasets (Recommended)}. Due to comparatively less research attention on the detection of fake audio, there are less LAIM-generated audio datasets compared to text and image. As shown in  Table~\ref{tab:datasets}, ASVspoof 2024~\cite{wang2024asvspoof} introduces the most comprehensive benchmark to date, incorporating crowdsourced bona fide and spoofed speech from over 4,000 speakers and covering 32 manipulation methods, including advanced TTS ZMM-TTS~\cite{gong2024zmm}, voice conversion (VC) DiffVC~\cite{popov2021diffusion}, and adversarial audio filtered through common codecs. This makes it particularly suitable for evaluating detection robustness in telephony and speaker verification scenarios. SONAR~\cite{li2024sonar} focuses on evaluating generalization in audio deepfake detection, compiling over 2,200 AI-generated samples from nine state-of-the-art TTS models, including commercial systems like OpenAI~\cite{openai2025texttospeech} and Seed-TTS~\cite{anastassiou2024seed}, which pose significant challenges for current detectors due to their high fidelity and proprietary training. Finally, LibriSeVoc~\cite{sun2023ai} takes a focused approach by isolating neural vocoder artifacts through self-vocoding of real speech using six different vocoders, including DiffWave~\cite{kong2020diffwave} and WaveGrad~\cite{chen2020wavegrad}, enabling fine-grained artifact-based detection via a multi-task learning strategy.

\underline{Social Media.}
Müller et al.~\cite{muller2022does} introduce an in-the-wild audio deepfake dataset to reflect the challenges of detection in real-world, social media contexts. Unlike lab-controlled datasets like ASVspoof, their collection includes 37.9 hours of audio from 58 celebrities and politicians, sourced from social platforms and matched for speaker style, emotion, and noise. This dataset captures typical social media artifacts and content absurdity, making it highly realistic. Evaluation shows that models trained on ASVspoof 2019~\cite{wang2020asvspoof} degrade by about 200\% to 1000\% EER when tested on this dataset, highlighting the urgent need for robust, socially grounded benchmarks for audio deepfake detection.

\raisebox{-0.3ex}{\scalebox{1.3}{\ding{193}}} \textbf{Non-LAIM Generated Datasets}. Before the rise of LAIMs (\eg, diffusion models and LLMs), a number of influential audio datasets (\eg, WaveFake~\cite{frank2021wavefake}, ASVspoof 2019~\cite{wang2020asvspoof}, and ASVspoof 2021~\cite{liu2023asvspoof}) were developed using traditional generative techniques such as GAN-based vocoders, like MelGAN~\cite{kumar2019melgan}, HiFiGAN~\cite{kong2020hifi}. 
Traditional spoofed speech is typically created from limited speaker data using purpose-built models with shallow linguistic understanding and observable signal-level artifacts. In contrast, LAIM-generated audio—such as from OpenAI~\cite{openai2025texttospeech}, VALL-E~\cite{wang2023neural}, or Voicebox~\cite{le2023voicebox}—draws on massive multilingual corpora and produces highly coherent, context-aware, and personalized speech, often free of detectable synthesis artifacts. As such, detectors trained on non-LAIM data often fail to generalize to LAIM scenarios, underscoring the need for new benchmarks and detection methods specifically tailored to the emerging class of foundation model-based audio generation.
WaveFake\cite{frank2021wavefake} comprises over 117,000 audio clips synthesized from six neural vocoders (\eg, MelGAN, HiFi-GAN), focusing on signal-level differences across architectures. Similarly, ASVspoof 2019\cite{wang2020asvspoof} covers logical and physical access scenarios using VC and TTS techniques including Tacotron2~\cite{shen2018natural} and WaveNet~\cite{oord2016wavenet}. The ASVspoof 2021~\cite{liu2023asvspoof} marked a progression toward realism by introducing codec-transcoded, compressed, and replayed speech in semi-wild conditions, but still relied on predefined spoofing algorithms, not modern autoregressive or diffusion-based LLM architectures.

\smallskip
\noindent
$\clubsuit$ \textbf{Multimodal}. Multimodal learning refers to an embodied learning situation that includes learning multiple modalities
such as text, image, video, and audio~\cite{khalid2021evaluation}. From a generation perspective, visual generation tasks, such as text-to-image and text-to-video, are regarded as multimodal generation. The generative models for these tasks are trained to learn visual representations and language understanding for visual generation. From the detection aspect, detectors that learn multiple modalities for forgery detection are categorized into multimodal.
In this context, we define "multimodal generation" from a detection perspective, referring to frameworks capable of creating multimodal output.  Multimodal generation process normally contains foundation models as encoders (\eg, CLIP~\cite{radford2021learning}, ViT~\cite{dosovitskiy2020image} ) and decoders (\eg, Stable Diffusion~\cite{rombach2021highresolution} ), and a LLM for taking  language-like representations from encoders for semantic understanding and produces modality signal tokens for guiding content output, 
see Fig.~\ref{fig:generration_overview} e).

Most recent work, such as OpenAI GPT-4o~\cite{openai2024gpt4o},  Google Gemini 2.5~\cite{team2023gemini}, Meta LLaMA4~\cite{metaai2025llama4}, Claude 3.5 Sonnet~\cite{anthropic2024claude35sonnet}, DeepSeek V3~\cite{liu2024deepseek} , demonstrates rapid progress in unified multimodal modeling. These models are capable of processing and reasoning over diverse modalities with a single framework.

\raisebox{-0.3ex}{\scalebox{1.3}{\ding{192}}} \textbf{LAIM Generated Datasets (Recommended)}. The multimodal datasets listed in Table~\ref{tab:datasets} capture diverse forms of synthetic media, spanning text, image, video, and audio. Early datasets such as $DGM^4$\cite{shao2023detecting} and COCOFake\cite{amoroso2024parents} focused on image-text pair manipulations. $DGM^4$ introduces the task of detecting and grounding manipulations in both images and texts simultaneously, providing fine-grained annotations (e.g., bounding boxes, token-level edits) and simulating realistic multimodal misinformation scenarios such as fake news. COCOFake, in contrast, leverages COCO captions and Stable Diffusion to create over 1.2M text-guided fake images, highlighting both semantic and perceptual mismatches in multimodal generation.
More recently, $M^3A$\cite{xu2024m3a} and LOKI\cite{ye2024loki} have expanded multimodal detection benchmarks to unprecedented scale and modality diversity. $M^3A$ contains 6.5 million manipulated news samples across text, image, audio, and video, sourced from 60 reputable outlets and generated using diverse strategies like named entity manipulation and modality mismatching. It supports a wide range of tasks including deepfake detection, out-of-context analysis, and fact-checking. LOKI evaluates synthetic detection across 5 modalities and 26 subcategories, with rich, fine-grained anomaly annotations and explainability-focused benchmarks. It includes not only binary classification, but also reasoning-based tasks such as abnormal detail selection and explanation.

\underline{Social Media.}
MMFakeBench~\cite{liu2024mmfakebench} advances the field with the first benchmark for mixed-source multimodal misinformation, combining three key forgery types: textual veracity distortion, visual veracity distortion, and cross-modal inconsistency. It contains 11K curated samples across 12 subcategories, revealing how current models struggle with realistic, multi-source misinformation. Deepfake-Eval~\cite{chandra2025deepfake} introduces an in-the-wild multimodal benchmark collected from social media and user submissions, covering video, audio, and image modalities, with over 2,400 generated samples and 3,300 real samples.

\raisebox{-0.3ex}{\scalebox{1.3}{\ding{193}}} \textbf{Non-LAIM Generated Datasets}. In addition to LAIM-generated multimodal benchmarks, several earlier representative datasets were developed using non-foundational, task-specific generation methods. Notably, FakeAVCeleb\cite{khalid2021fakeavceleb} introduces a multimodal benchmark with paired deepfake videos and cloned audios, generated using models like FSGAN~\cite{nirkin2019fsgan}, Faceswap~\cite{korshunova2017fast}, and SV2TTS~\cite{jia2018transfer}. While it enables fine-grained evaluation across audio-video consistency, its generation methods rely on traditional GANs and vocoders, lacking the expressiveness and semantic generality of diffusion or autoregressive models. NewsCLIPpings\cite{luo2021newsclippings} takes a different approach, focusing on out-of-context multimodal misinformation, where real images are automatically mismatched with real captions using CLIP-based semantic similarity. In contrast, recent LAIM-based datasets introduce greater content diversity, generative flexibility, and semantic complexity, making detection fundamentally more challenging and motivating the need for LAIM-focused detection research.

\section{Detection}
\label{sec:detection}

\definecolor{5anodecolor}{HTML}{C3ACD0}
\definecolor{tnodecolor}{HTML}{C8E4B2}
\definecolor{basiccolor}{HTML}{CAEDFF}
\definecolor{5inodeccolor}{HTML}{ffcfd2}
\definecolor{5vnodeccolor}{HTML}{fec89a}
\definecolor{5mmnodeccolor}{HTML}{f3d5b5}
\usetikzlibrary{shadows}
\begin{figure*}
    \centering

\tikzset{
    basic/.style  = {draw, fill=basiccolor,text width=2cm, align=center, font=\sffamily, rectangle},
    rotated/.style = {basic, rotate=90},
    root/.style   = {basic, rounded corners=2pt, thin, align=center, fill=green!30},
    onode/.style = {basic, thin, rounded corners=2pt, align=center, fill=green!60,text width=3cm,},
    tnode/.style = {basic, thin, align=left, fill=pink!60, text width=5em, align=center},
    nnode/.style = {basic, thin, align=left, fill=yellow!30, text width=16em, align=center},
    4tnode/.style = {basic, thin, align=left, fill=tnodecolor, text width=16em, align=center},
    4mmnode/.style = {basic, thin, align=left, fill=5mmnodeccolor, text width=16em, align=center},
    5inode/.style = { thin, align=left,font=\sffamily\small, fill=5inodeccolor, text width=22em, align=center, },
    5vnode/.style = { thin, align=left,font=\sffamily\small, fill=5vnodeccolor, text width=22em, align=center, },
    5anode/.style = { thin, align=left,font=\sffamily\small, fill=5anodecolor, text width=22em, align=center, },
    5mmnode/.style = { thin, align=left, font=\sffamily\small, fill=5mmnodeccolor, text width=22em, align=center, },
    5tnode/.style = { thin, align=left, font=\sffamily\small, fill=tnodecolor, text width=22em, align=center, },
    4inode/.style = {basic, thin, align=left, fill=5inodeccolor, text width=16em, align=center},
    5tmnode/.style = { thin, align=left, fill=green!30, text width=22em, align=center},
    textnode/.style = {basic, thin, align=left, fill=orange!60, text width=12em, align=center},
    xnode/.style = {basic, thin, rounded corners=2pt, align=center, fill=blue!20,text width=1cm,},
    wnode/.style = {basic, thin, align=left, fill=pink!10!blue!80!red!10, text width=6.5em},
    edge from parent/.style={draw=black, edge from parent fork right}

}
\resizebox{\textwidth}{!}{%
\begin{forest} for tree={
    l = 0.001cm,
    s sep=0.04cm,
    grow=east,
    growth parent anchor=west,
    parent anchor=east,
    child anchor=west,
    edge path={\noexpand\path[\forestoption{edge},->, >={latex}] 
         (!u.parent anchor) -- +(10pt,0pt) |-  (.child anchor) 
         \forestoption{edge label};}
}
[{Detection\\(\S \ref{sec:detection})}, basic,  l sep=5mm,
    [Multi-\\modal (\S \ref{sec:Multimodal_det}), xnode,  l sep=5mm,
        [Beyond \\(\S \ref{sec:multimodal_beyond})\, , tnode, l sep=5mm,
        [$\clubsuit$ Empirical Study\, , nnode,l       sep=66.0mm,
        [{Coccomini et al.~\cite{coccomini2024detecting},VERITE~\cite{papadopoulos2024verite}, Clipping~\cite{khan2024clipping}, Jia et al.~\cite{Jia_2024_CVPR}} \, , 5mmnode]]
        [$\clubsuit$ Localization\, , nnode,l sep=5mm,
        [{{\scalebox{1.3}{\ding{194}}} MLLM-based} \, , 4mmnode,l sep=5mm,
        [{FakeShield~\cite{xu2024fakeshield}, ForgeryGPT~\cite{liu2024forgerygpt}} \, , 5mmnode]]
        [{{\scalebox{1.3}{\ding{193}}} Frequency-based} \, , 4mmnode,l sep=5mm,
        [{UFAFormer~\cite{liu2023unified}} \, , 5mmnode]]
        [{{\scalebox{1.3}{\ding{192}}} Spatial-based} \, , 4mmnode,l sep=5mm,
        [{HAMMER~\cite{shao2023detecting}, HAMMER++~\cite{shao2023detecting}, Wang et al.~\cite{wang2023exploiting}} \, , 5mmnode]]]
        [$\clubsuit$ Interpretability\, , nnode,l       sep=66.0mm,
        [{Xu et al.~\cite{xu2023combating}, FFAA~\cite{huang2024ffaa}, X²-DFD~\cite{chen2024textit}} \, , 5mmnode]]
        [$\clubsuit$ Generalization\, , nnode,l sep=5mm,
        [{{\scalebox{1.3}{\ding{193}}} Contrastive Learning} \, , 4mmnode,l sep=5mm,
        [{LASTED~\cite{wu2023generalizable}} \, , 5mmnode]]
        [{{\scalebox{1.3}{\ding{192}}} Prompt Tuning} \, , 4mmnode,l sep=5mm,
        [{AntifakePrompt~\cite{chang2023antifakeprompt}} \, , 5mmnode]]
        ]
        [$\clubsuit$ Attribution\, , nnode,l       sep=66.0mm,
        [{De-Fake~\cite{sha2023fake}, FIDAVL~\cite{keita2025fidavl}} \, , 5mmnode]]]
        [Pure \\(\S \ref{sec:multimodal_pure})\, , tnode,l sep=5mm,
        [$\clubsuit$ Text-image Inconsistency\, , nnode,l       sep=66.0mm,
        [{DIDAN~\cite{tan2020detecting}, D-TIIL~\cite{huang2023exposing}} \, , 5mmnode]]
        [$\clubsuit$ Prompt-guided\, , nnode,l       sep=66.0mm,
        [{Amoroso et al.~\cite{amoroso2024parents}, MM-Det~\cite{song2024learning}, Human Action CLIPS~\cite{bohacek2024human}} \, , 5mmnode]]
        ]
        ]
    [Audio\\(\S \ref{sec:Audio_det}), xnode,  l sep=5mm,
        [Beyond \\(\S \ref{sec:audio_beyond})\, , tnode, l sep=5mm,
        [$\clubsuit$ Generalization\, , nnode,l       sep=66.0mm,
        [{Ren et al.~\cite{ren2025improving}} \, , 5anode]]]
        [Pure \\(\S \ref{sec:audio_pure})\, , tnode, l sep=5mm,
        [$\clubsuit$ Vocoder-based\, , nnode,l       sep=66.0mm,
        [{DetectVocoder~\cite{sun2023ai}} \, , 5anode]]]
        ]
    [Video\\(\S \ref{sec:Video_det}), xnode,  l sep=5mm,
        [Beyond \\(\S \ref{sec:video_beyond})\, , tnode, l sep=5mm,
        [$\clubsuit$ Empirical Study\, , nnode,l       sep=66.0mm,
        [{Vahdati et al.~\cite{vahdati2024beyond}} \, , 5vnode]]
        [$\clubsuit$ Generalization\, , nnode,l       sep=66.0mm,
        [{RevisitVideo~\cite{kamat2023revisiting}} \, , 5vnode]]]  
        [Pure \\(\S \ref{sec:video_pure})\, , tnode, l sep=5mm,
        [$\clubsuit$ Spatial\&Temporal-based Methods\, , nnode,l       sep=66.0mm,
        [{DuB3D~\cite{ji2024distinguish}, He et al.~\cite{he2024exposing}} \, , 5vnode]]]  
        ]
    [Image\\(\S \ref{sec:Image_det}), xnode,  l sep=5mm,
        [Beyond \\(\S \ref{sec:image_beyond})\, , tnode,l sep=5mm,
        [$\clubsuit$ Empirical Study\, , nnode,l       sep=66.0mm,
        [{Corvi et al.~\cite{corvi2023detection, corvi2023intriguing}, Ricker et al.~\cite{ricker2022towards}, Cocchi et al.~\cite{cocchi2023unveiling}, Papa et al.~\cite{papa2023use}, Porcile et al.~\cite{porcile2023finding}, Mandelli et al.~\cite{mandelli2022forensic}, Carriere et al.~\cite{carriere2023beyond}} \, , 5inode]]
        [$\clubsuit$ Robustness\, , nnode,l sep=5mm,
        [{{\scalebox{1.3}{\ding{193}}}Post-Processing Robustness} \, , 4inode,l sep=5mm,
        [{GLFF~\cite{ju2023glff}, Xu et al.~\cite{xu2023exposing}, Local Statistics~\cite{jer2023local}, Yang~\cite{yang2025all}} \, , 5inode]]
        [{{\scalebox{1.3}{\ding{192}}}Adversarial Attack Robustness} \, , 4inode,l sep=5mm,
        [{D4~\cite{Hooda2024D4}, De Rosa et al.~\cite{de2024exploring}} \, , 5inode]]]
        [$\clubsuit$ Localization\, , nnode,l sep=5mm,
        [{{\scalebox{1.3}{\ding{193}}}Weakly-supervised} \, , 4inode,l sep=5mm,
        [{Tantaru et al.~\cite{tantaru2023weakly}} \, , 5inode]]
        [{{\scalebox{1.3}{\ding{192}}}Fully-supervised} \, , 4inode,l sep=5mm,
        [{IFDL~\cite{guo2023hierarchical}, PAL~\cite{zhang2023perceptual}, TruFor~\cite{guillaro2023trufor}, UnionFormer~\cite{Li_2024_CVPR}} \, , 5inode]]]
        [$\clubsuit$ Interpretability\, , nnode,l       sep=66.0mm,
        [{Aghasanli et al.~\cite{aghasanli2023interpretable}, Color Statistics~\cite{uhlenbrock2024did}} \, , 5inode]]
        [$\clubsuit$ Generalization\, , nnode,l       sep=66.0mm,
        [{Online Detection~\cite{epstein2023online}, Ojha et al.~\cite{ojha2023towards}, Cozzolino et al.~\cite{cozzolino2023raising}, LSDA~\cite{yan2023transcending}, Fingerprintnet~\cite{jeong2022fingerprintnet}, FACTOR~\cite{reiss2023detecting}, Dogoulis et al.~\cite{dogoulis2023improving}, DNF~\cite{zhang2023diffusion} CoDE~\cite{baraldi2024contrasting}, HRR~\cite{yuan2025hrr}, AIDE~\cite{yan2024sanity}, SUR-LID~\cite{cheng2024stacking}, B-Free~\cite{guillaro2024bias}, Breaking~\cite{zheng2024breaking}, Chen~\cite{chen2025dual}} \, , 5inode]]
        [$\clubsuit$ Attribution and Model Parsing\, , nnode,l sep=5mm,
        [{{\scalebox{1.3}{\ding{193}}}Model Parsing} \, , 4inode,l sep=5mm,
        [{Asnani et al.~\cite{asnani2023reverse}} \, , 5inode]]
        [{{\scalebox{1.3}{\ding{192}}}Attribution} \, , 4inode,l sep=5mm,
        [{Guarnera et al.~\cite{guarnera2024level}} \, , 5inode]]
        ]] 
        [Pure \\(\S \ref{sec:image_pure})\, , tnode, l sep=5mm,
        [$\clubsuit$ Distribution-based Methods\, , nnode,l       sep=66.0mm,
        [{ZED~~\cite{cozzolino2024zero}} \, , 5inode]]
        [$\clubsuit$ Frequency-based Methods\, , nnode,l       sep=66.0mm,
        [{Wavelet-Packets~\cite{wolter2022wavelet}, AUSOME~\cite{poredi2023ausome}, Xi et al.~\cite{xi2023ai}, Synthbuster~\cite{bammey2023synthbuster}, Lanzino et al. \cite{Lanzino_2024_CVPR}} \, , 5inode]]
        [$\clubsuit$ Spatial-based Methods\, , nnode,l       sep=66.0mm,
        [{AIGCD~\cite{zhong2023rich}, Nguyen et al.~\cite{nguyen2023unmasking}, multiLID~\cite {lorenz2023detecting}} \, , 5inode]]
        [$\clubsuit$ Diffuser Fingerprints \, , nnode,l       sep=66.0mm,
        [{DIF~\cite{sinitsa2023deep}, DIRE~\cite{wang2023dire}, SeDID~\cite{ma2023exposing},LaRE²~\cite{Luo_2024_CVPR}, Rajan et al.~\cite{rajan2025aligned}, Brokman et al.~\cite{brokman2025manifold}} \, , 5inode]]
        [$\clubsuit$ Physical/Physiological\, , nnode,l       sep=66.0mm,
        [{Borji~\cite{borji2023qualitative}, Perspective~\cite{farid2022perspective}, Lighting~\cite{farid2022lighting}} \, , 5inode]]
        [$\clubsuit$ Watermarking\, , nnode,l       sep=66.0mm,
        [{WaDiff~\cite{min2024watermark}, Stable Signature~\cite{fernandez2023stable}, Tree-Ring~\cite{wen2023tree}, Zhu et al.~\cite{zhu2024watermark},  Gaussian Shading~\cite{yang2024gaussian}} \, , 5inode]]
         ]
         ]
    [Text\\(\S \ref{sec:Text_det}), xnode,  l sep=5mm,
        [Beyond \\(\S \ref{sec:text_beyond})\, , tnode, l sep=5mm,          
        [$\clubsuit$ Empirical Study\, , nnode,l sep=66.0mm,
        [{ChatLog~\cite{tu2023chatlog}, Zero-Shot~\cite{pu2023zero}, Training-based~\cite{xu2023generalization}, Sarvazyan et al.~\cite{sarvazyan2023supervised}, MAGE~\cite{li2023deepfake}, Wahle et al.~\cite{wahle2022large}, Becker et al.~\cite{becker2023paraphrase}, MGTBench~\cite{he2024mgtbench}, Guo et al.~\cite{guo2023close},Chen et al.~\cite{chen2023can},Antoun et al.~\cite{antoun2023text}}\, , 5tnode]]
        [$\clubsuit$ Robustness\, , nnode,l sep=5mm,
        [{\scalebox{1.3}{\ding{193}}} LAIM-Polished Robustness \,, 4tnode,l sep=5mm,
        [{Yang et al.~\cite{yang2023chatgpt},} \, , 5tnode]]
        [{\scalebox{1.3}{\ding{192}}} Adversarial Attack Robustness \,, 4tnode,l sep=5mm,
        [{Shi et al.~\cite{shi2024red}, RADAR~\cite{hu2023radar}, OUTFOX~\cite{koike2023outfox}, J-Guard~\cite{kumarage2023j}} \, , 5tnode]]]
        [$\clubsuit$ Interpretability\, , nnode,l sep=66.0mm,
         [{DNA-GPT~\cite{yang2023dna},Kirchenbauer et al.~\cite{kirchenbauer2023watermark}, Mitrovic et al.~\cite{mitrovic2023chatgpt}, Check-GPT~\cite{liu2024detectability},Yang et al.~\cite{yang2023chatgpt}} \, , 5tnode]]
        [$\clubsuit$ Generalization\, , nnode,l sep=66.0mm,
        [{Ghostbuster~\cite{verma2023ghostbuster},Conda~\cite{bhattacharjee2023conda},Text Fluoroscopy~\cite{yu2024text}, DeTeCtive\cite{guo2024detective}, PHD~~\cite{tulchinskii2023intrinsic} } \, , 5tnode]]
        [$\clubsuit$ Attribution\, , nnode,l sep=66.0mm,
        [{Linguistic Model~\cite{uchendu2020authorship}, GPT-who~\cite{venkatraman2023gpt},LLMDet~\cite{wu2023llmdet}, Few-Shot~\cite{Soto2024few-shot}, Sniffer~\cite{li2023origin},TuringBench~\cite{uchendu-etal-2021-turingbench-benchmark}, Contra-X~\cite{ai-etal-2022-whodunit}, XLNet-FT~\cite{munir2021through}, TopFormer~\cite{uchendu2024topformer}} \, , 5tnode]]
        [$\clubsuit$ Efficiency\, , nnode,l sep=66.0mm,
        [{Efficient-DetectGPT~\cite{deng2023efficient}, Fast-DetectGPT~\cite{bao2023fast}, DetectLLM~\cite{su2023detectllm}, SeqXGPT~\cite{wang2023seqxgpt}, Glimpse~\cite{bao2024glimpse} } \, , 5tnode]]] 
        [Pure \\(\S \ref{sec:text_pure})\, , tnode, l sep=5mm,
        [$\clubsuit$ Hard-explainable Methods\, , nnode,l sep=66.0mm,
        [{HowkGPT~\cite{vasilatos2023howkgpt}, GPTZero~\cite{GPTZero}, Binoculars~\cite{hans2024spotting},DetectGPT~\cite{mitchell2023detectgpt},MPU~\cite{tian2023multiscale} } \, , 5tnode]]
        [$\clubsuit$ Easy-explainable Methods\, , nnode,l sep=5mm,
         [{\scalebox{1.3}{\ding{193}}} Non-watermarking \,, 4tnode,l sep=5mm,
         [{Pu et al.~\cite{pu2022unraveling},Stylometric Signals~\cite{kumarage2023stylometric}, Coherence Consistency~\cite{liu2022coco}, Zhu et al.~\cite{zhu2023beat}, GECScore~\cite{wu2024wrote}} \, , 5tnode]]
         [{\scalebox{1.3}{\ding{192}}} Watermarking \,, 4tnode,l sep=5mm,
         [{Distillation-Resistant~\cite{zhao-etal-2022-distillation}, Fernandez et al.~\cite{fernandez2023three}, Robust Multi-bit~\cite{yoo2023robust}, Christ et al.~\cite{christ2024undetectable}, Kuditipudi et al.~\cite{kuditipudi2023robust}, Unigram Watermark~\cite{zhao2023provable}, Private Watermark~\cite{liu2023unforgeable}} \, , 5tnode]]
       ]]
         ] ]
\end{forest}
}

    \caption{\small Taxonomy of the literature on detecting multimedia generated by Large AI Models (LAIMs).}
    \label{fig:detection_overview}
\end{figure*}
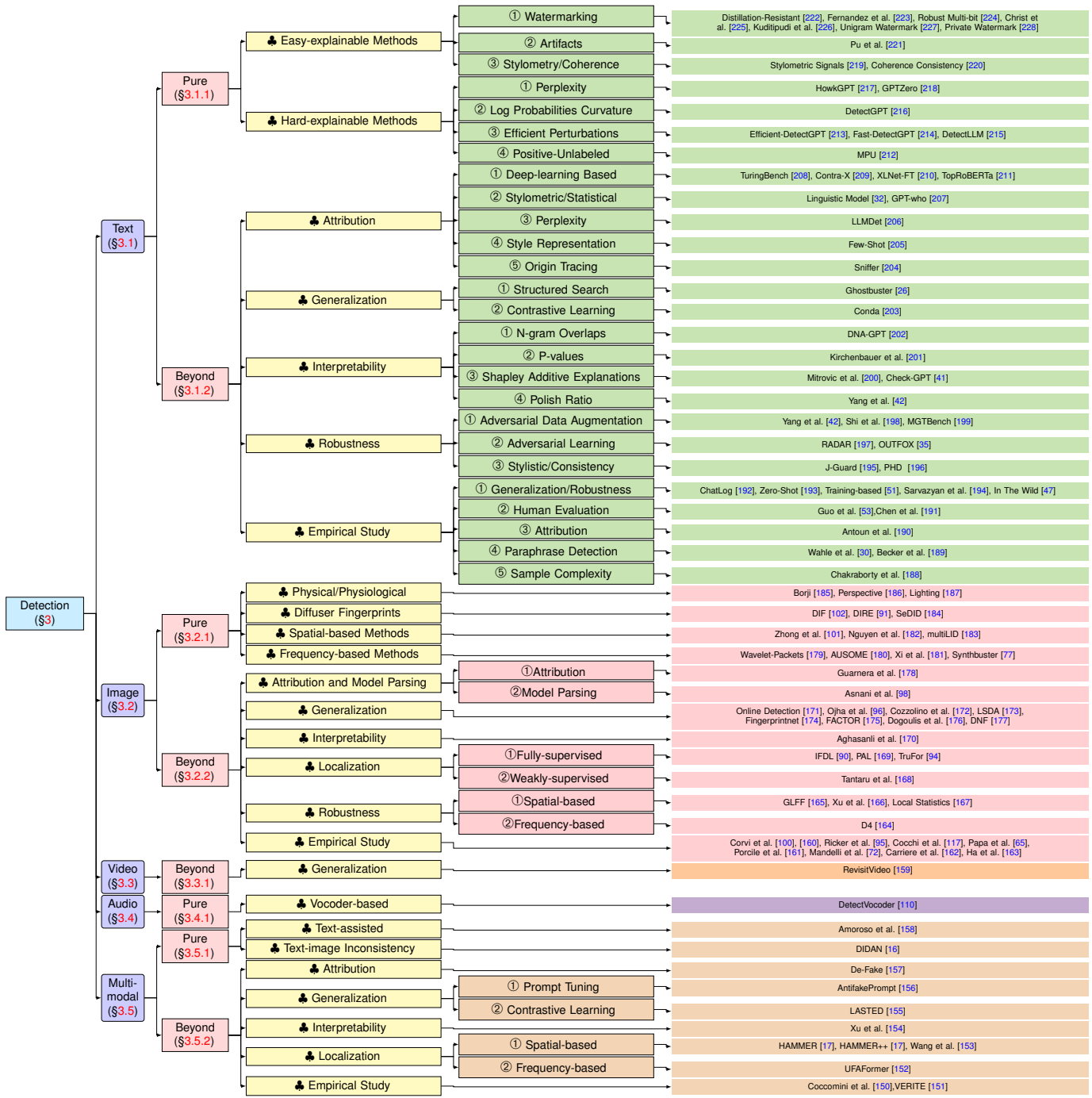
\noindent
In this section, we provide a novel taxonomy of detection methods on LAIM-generated multimedia. An overview of the structure for this section is provided in Fig. \ref{fig:detection_overview}. Specifically, we identify the functionality of these detectors and organize them into two categories for each data modality: 
\begin{enumerate}[leftmargin=*]
    \item \textbf{Pure Detection}. The detection methods in this category only aim to improve detection performance.
    \item \textbf{Beyond Detection}. It endows detectors extra characteristics (\eg, generalizability, robustness, interpretability) while keeping accurate and effective detection ability.
\end{enumerate}

\subsection{Text}\label{sec:Text_det}

There are existing surveys providing a detailed overview of current detection strategies and benchmarks~\cite{wu2023survey, yang2023survey, ghosal2023towards}. Specifically, 
\cite{yang2023survey} categorizes the detection methods into three groups: training-based, zero-shot, and watermarking. Works in each group are further distinguished according to their detection scenarios such as black-box 
and white-box. 
The authors in \cite{wu2023survey} categorize the detection of LLM-generated content based on the techniques applied by detectors (\eg, adversarial learning, watermarking methods, and human-assisted methods) and the dependency of detectors on training (\eg, zero-shot detectors, fine-tuned detectors). 
{\cite{ghosal2023towards}} divides the existing literature into two parts: methods designed to detect LLM-generated text (\eg, watermarking-based, fine-tune-based, and zero-shot) and methods designed to evade detection (\eg, paraphrasing attacks, spoofing attacks). 
Although these organizational strategies apply to recent detection methods, their taxonomy may not be sufficient for adapting to new
or evolving detection techniques. 
To this end, we provide a novel taxonomy based on Pure Detection and Beyond Detection.  

\begin{figure*}[t]
    \centering
    \includegraphics[width=1\textwidth]{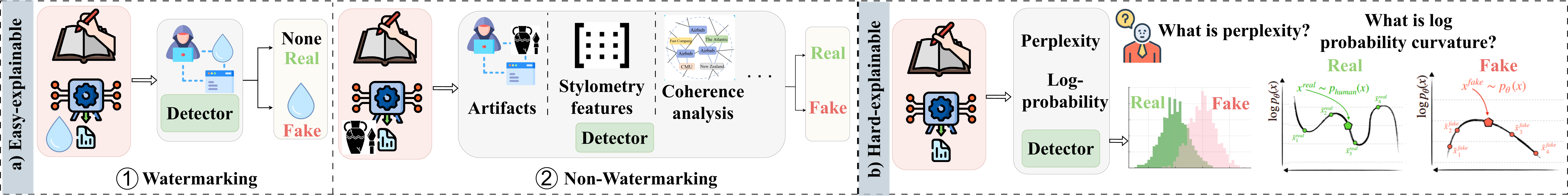}
    \vspace{-6mm}
    \caption{\small Illustrations of pure detection methodologies for LAIM-generated text.}
    \label{fig:text_pure_overview}
    \vspace{-4mm}
\end{figure*}

\subsubsection{Pure Detection}\label{sec:text_pure}
We categorize the pure detection methodologies based on their comprehensibility toward the general populace,  which can be identified as \textbf{Easy-explainable} and \textbf{Hard-explainable} methods, as shown in Fig.~\ref{fig:text_pure_overview}.

\smallskip
\noindent
$\clubsuit$ \textbf{Easy-explainable Methods.} These approaches ensure that humans can straightforwardly understand the principles behind the detection technology. 
Such methods prioritize clarity and accessibility, making the technology approachable for non-specialist users.

\raisebox{-0.3ex}{\scalebox{1.3}{\ding{192}}} \textbf{Watermarking.} One intuitive direction is text watermarking, where detectable patterns are algorithmically embedded in generated content. While most watermarking methods require a secret key for detection~\cite{zhao-etal-2022-distillation, fernandez2023three, yoo2023robust, christ2024undetectable, kuditipudi2023robust, zhao2023provable}, Liu et al.~\cite{liu2023unforgeable} address this limitation by introducing a keyless approach using shared token embeddings between the generator and detector.

\raisebox{-0.3ex}{\scalebox{1.3}{\ding{193}}} \textbf{Non-watermarking.} Another line of work explores textual artifacts—subtle irregularities in token usage or structure left by language models. Pu et al.~\cite{pu2022unraveling} reveal that artifacts tend to appear in token co-occurrence patterns, especially within the head of the vocabulary and in shallow features like high-frequency content words or higher-order N-grams.

Several studies also rely on stylistic or coherence cues to differentiate human from LLM-generated text. Kumarage et al.\cite{kumarage2023stylometric} apply stylometric analysis to quantify differences in writing style, while Liu et al.\cite{liu2022coco} assess entity-level coherence to expose inconsistency patterns more common in machine-generated text.

Finally, Zhu et al.~\cite{zhu2023beat} propose a zero-shot, black-box method that queries ChatGPT to revise input text. The method assumes ChatGPT makes fewer edits to AI-generated text and detects generation by measuring similarity between the original and revised versions—without needing model logits or additional training. Wu et al.~\cite{wu2024wrote} propose GECScore, a zero-shot, black-box method that detects LLM-generated text by measuring similarity between the input and its grammar-corrected version. Based on the assumption that LLM text contains fewer grammatical errors, this approach offers a simple detection signal.

\smallskip
\noindent
$\clubsuit$ \textbf{Hard-explainable Methods.} These methods involve detection techniques and analytical processes that may not be as easily comprehensible to most people but are quite accessible to researchers in the field. 
Several hard-explainable methods leverage statistical signals or weak supervision to detect LLM-generated text. Perplexity, a standard metric in language modeling, is widely used to distinguish between human- and machine-generated content. For instance, \cite{vasilatos2023howkgpt} computes perplexity scores to separate student-written from ChatGPT-generated homework submissions, while GPTZero~\cite{GPTZero} combines perplexity and variance in sentence-level likelihood to assess the likelihood of text being AI-generated. Hans et al.\cite{hans2024spotting} further propose Binoculars, a zero-shot detector that uses the ratio of perplexity to cross-perplexity between two LLMs, offering a strong, training-free signal for detection across domains. Similarly, log probability curvature has been explored in DetectGPT\cite{mitchell2023detectgpt}, which observes that LLM-generated text exhibits more negative curvature in the log probability landscape. By perturbing the input and analyzing changes in curvature, DetectGPT achieves promising zero-shot results. Beyond statistical modeling, Positive-Unlabeled (PU) training has also been applied to tackle label uncertainty. Tian et al.~\cite{tian2023multiscale} argue that short generated texts often resemble human-written content closely and thus are better treated as unlabeled. They formulate LLM-generated text detection as a PU learning problem and propose a multiscale PU framework that maintains detection accuracy across text lengths.
\begin{figure*}[t]
    \centering
    \includegraphics[width=1\textwidth]{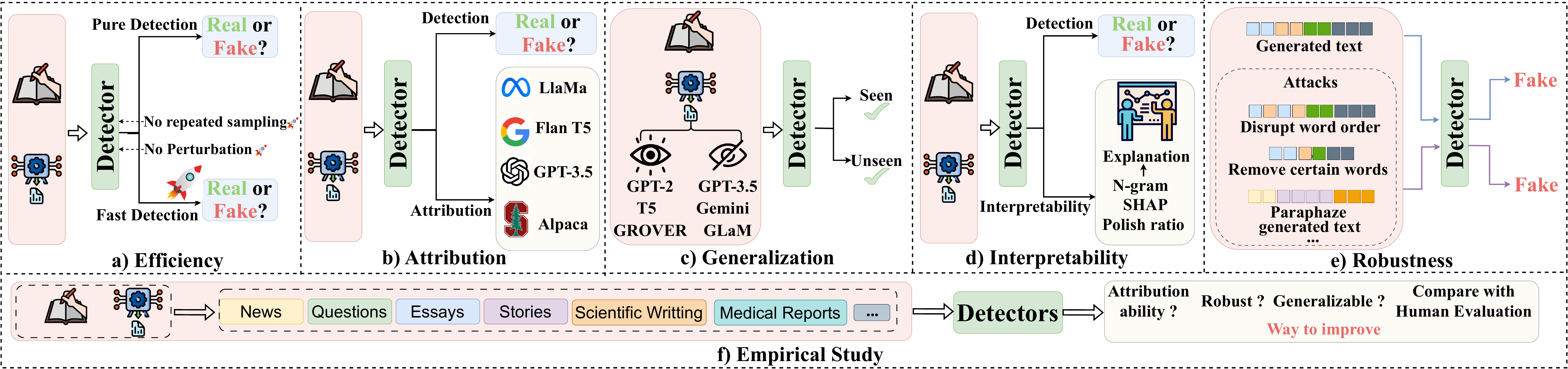}
    \vspace{-6mm}
    \caption{\small Illustrations of beyond detection methodologies for LAIM-generated text.}
    \label{fig:text_beyond_overview}
    \vspace{-4mm}
\end{figure*}

\subsubsection{Beyond Detection}\label{sec:text_beyond}
Detection Methods go beyond distinguishing between human and machine-generated content can be organized (see Fig.~\ref{fig:text_beyond_overview}) as follows:  

\smallskip
\noindent
$\clubsuit$ \textbf{Efficiency.}
Recognizing the use of pure random perturbations in DetectGPT~\cite{mitchell2023detectgpt} requires intensive computational cost,
a series of works adopt various techniques to improve computational efficiency and enhance detection accuracy simultaneously, as shown in Fig.~\ref{fig:text_beyond_overview} a). Specifically, \cite{deng2023efficient} achieves similar performance with up to \textit{two} times fewer queries than DetectGPT with a Bayesian surrogate model by selecting typical samples based on Bayesian uncertainty and interpolating scores from typical samples to other ones, making the perturbation process more focused and less resource intensive. Bao et al.~\cite{bao2023fast} increase the detection speed by \textit{340} times by substituting DetectGPT’s perturbation step with a more efficient sampling step via conditional probability curvature. DetectLLM \cite{su2023detectllm}, another recent contribution, employs normalized perturbed log-rank for text detection generated by LLMs, asserting a lower susceptibility to the perturbation model and the number of perturbations compared to DetectGPT. SeqXGPT~\cite{wang2023seqxgpt} enhances the efficiency of log probability-based detection by operating at the sentence level without requiring costly perturbations. It leverages token-level log probabilities from multiple white-box LLMs and processes them using lightweight convolutional and self-attention networks to capture nuanced statistical patterns that effectively separate human-written from machine-generated sentences. Glimpse~\cite{bao2024glimpse} significantly improves runtime and cost-efficiency in zero-shot detection by extending log-probability curvature methods to proprietary LLMs. It estimates full token distributions from limited top-K API outputs, enabling accurate detection with substantially fewer API calls and reduced computational overhead, even for powerful models like GPT-4.

\smallskip
\noindent
$\clubsuit$ \textbf{Attribution}. Determining which specific model may generate the test content, see Fig.~\ref{fig:text_beyond_overview} b).
Many recent approaches employ deep learning models to capture semantic and structural signals embedded in the text. For instance, Munir et al.\cite{munir2021through} highlight that synthetic texts carry subtle, consistent traces from their originating models, which can be leveraged for attribution. TopRoBERTa\cite{uchendu2024topformer} enhances attribution performance by integrating topological data analysis (TDA) with RoBERTa, enriching contextual semantic and syntactic representations with structural insights. Other methods rely on explicitly defined statistical and linguistic features. Uchendu et al.\cite{uchendu2020authorship} use a Random Forest classifier trained on stylometric indicators such as entropy and readability. GPT-who\cite{venkatraman2023gpt} introduces Uniform Information Density (UID)-based psycholinguistic features to distinguish human from model-generated text, while Wu et al.\cite{wu2023llmdet} utilize proxy perplexity scores for attributing text to models like LLaMA and OPT\cite{liu2021opt}. Soto et al.\cite{Soto2024few-shot} take a few-shot approach, estimating style representations solely from human-authored samples to avoid dependence on generated data. Moving beyond attribution to a single model, Sniffer\cite{li2023origin} infers generative lineage by analyzing statistical discrepancies, enabling traceability across models, for example, linking Alpaca~\cite{alpaca} back to ChatGPT and LLaMA.

\smallskip
\noindent
$\clubsuit$ \textbf{Generalization.} Developing detectors with generalizability that can detect texts generated by generators never seen before, as shown in Fig.~\ref{fig:text_beyond_overview} c).
\cite{verma2023ghostbuster}, which first uses weaker language models to extract structured features from texts through systematic search, and then trains a linear classifier on these selected features to detect AI-generated content. Bhattacharjee et al.~\cite{bhattacharjee2023conda} propose a contrastive domain adaptation framework, combining traditional domain adaptation techniques with contrastive learning to learn domain-invariant representations, thereby improving the generalization ability of detectors. Text Fluoroscopy~\cite{yu2024text} enhances generalization by leveraging intrinsic features from intermediate layers of language models, selecting the most distinctive layer to enable improved cross-domain detection. DeTeCtive~\cite{guo2024detective} further advances this line of work by introducing a multilevel contrastive learning approach that treats detection as a style discrimination problem. It encodes fine-grained stylistic features using auxiliary contrastive objectives, achieving strong generalization to out-of-distribution generators. Notably, it supports training-free incremental adaptation, enabling detectors to incorporate new texts during inference without retraining. Tulchinskii et al.~\cite{tulchinskii2023intrinsic} propose a generator-agnostic method based on intrinsic dimensionality (ID) estimation of sentence embeddings. Without any training, their method generalizes across unseen domains, generators, and languages. 



\smallskip
\noindent
$\clubsuit$ \textbf{Interpretability.} Exploring interpretable detectors that can provide explanations for detection results, see Fig.~\ref{fig:text_beyond_overview} d).
DNA-GPT\cite{yang2023dna} introduces a training-free approach that identifies GPT-generated text by analyzing distinct text continuation patterns between human and AI-generated content, offering evidence based on nontrivial N-gram overlaps to support explainable detection results. Kirchenbauer et al.\cite{kirchenbauer2023watermark} propose a statistical test for detecting watermarks in text, utilizing interpretable p-values to quantify the certainty that the detected pattern (\eg, watermark) is not due to random chance. To gain insight into model reasoning, Mitrovic et al.\cite{mitrovic2023chatgpt} fine-tune a transformer-based model and employ Shapley Additive Explanations (SHAP)~\cite{lundberg2017unified} to extract explanations of the model’s decisions, aiming to uncover the model’s reasoning. Similarly, Liu et al.\cite{liu2024detectability} utilize Shapley Values to compare interpretations derived from word-level and sentence-level results. Addressing the challenge of detecting texts refined at a granular level, such as ChatGPT-polished texts, Yang et al.\cite{yang2023chatgpt} introduce a novel dataset termed HPPT (ChatGPT-polished academic abstracts) and propose the "Polish Ratio" method, which measures ChatGPT’s involvement in text based on editing distance, providing a more reasonable explanation for the detection outcome

\smallskip
\noindent
$\clubsuit$ \textbf{Robustness.} Developing detectors that can handle different attacks, see Fig.~\ref{fig:text_beyond_overview} e).
In light of the vulnerability of detectors to different attacks and robustness issues, a significant body of research has been dedicated to utilizing adversarial learning as a mitigation strategy.

\raisebox{-0.3ex}{\scalebox{1.3}{\ding{192}}} \textbf{Adversarial Attack Robustness}. A growing body of work has focused on improving the robustness of AI-generated text detectors against adversarial perturbations, including paraphrasing, synonym substitution, token-level manipulations, and stylistic shifts.
RADAR~\cite{hu2023radar} introduces a detector trained in tandem with a paraphrasing adversary via adversarial learning, significantly improving detection robustness under paraphrased attacks. Similarly, OUTFOX~\cite{koike2023outfox} advances this idea by co-evolving a detector and an attacker through in-context learning, enabling the detector to adapt to increasingly sophisticated adversarial examples. Red teaming approaches~\cite{shi2024red} further stress-test detection systems by using LLMs to generate synonym-based or instruction-prompted adversarial variants, revealing that even strong detectors can be bypassed when facing semantically equivalent but stylistically modified text. J-Guard~\cite{kumarage2023j}, in contrast, enhances adversarial robustness by encoding journalism-specific stylistic features, such as AP-style conventions, which are less susceptible to word-level transformations. Together, these works demonstrate that models explicitly trained to handle adversarial perturbations, particularly through dynamic attacker-detector interactions or domain-guided cues, are better equipped to maintain detection reliability in real-world settings where malicious evasion is increasingly likely.

\raisebox{-0.3ex}{\scalebox{1.3}{\ding{193}}} \textbf{LAIM-Polished Robustness}. Yang et al.~\cite{yang2023chatgpt} address the challenge of detecting partially AI-influenced text, such as human-written content polished by ChatGPT. They introduce the HPPT dataset and propose the Polish Ratio to quantify ChatGPT’s editing impact, showing improved robustness in identifying collaborative writing scenarios.

\smallskip
\noindent
$\clubsuit$ \textbf{Empirical Study}.
The empirical studies are crucial for advancing our understanding and capabilities in detecting LLM-generated texts, as shown in Fig.~\ref{fig:text_beyond_overview} f). 
Recent work has examined the generalization and robustness of detection models, revealing notable challenges and insights. Tu et al.\cite{tu2023chatlog} observe a decline in RoBERTa-based detectors' effectiveness over time through a month-long evaluation of ChatGPT’s long-form answers. Similarly, Pu et al.\cite{pu2023zero} show that detectors trained on medium-sized LLMs can generalize in zero-shot settings to larger models. Xu et al.\cite{xu2023generalization} further find that detectors often overfit to dataset-specific artifacts due to insufficient data curation, yet are still capable of learning transferable features across domains and tasks. Sarvazyan et al.\cite{sarvazyan2023supervised} report that in-domain fine-tuned detectors perform poorly when tested on outputs from different models, and MAGE\cite{li2023deepfake} confirms this by evaluating 27 LLMs across 10 diverse datasets, showing that detectors frequently collapse under out-of-distribution settings. He et al.\cite{he2024mgtbench} present MGTBench, benchmarking 13 detectors on six LLMs and three datasets, and find that while model-based methods generalize better across datasets, metric-based ones transfer more effectively across models. They also highlight that even small amounts of training data suffice for competitive performance, though domain shifts still cause significant degradation. Other studies focus on paraphrase detection, highlighting threats to academic integrity. 

Wahle et al.\cite{wahle2022large} demonstrate that humans struggle to detect paraphrased GPT-3 outputs, with only 53\% accuracy, while also rating the paraphrases as highly fluent and coherent. Becker et al.\cite{becker2023paraphrase} show that human-written paraphrases tend to be more diverse and difficult than machine-generated ones. Human evaluation further reveals consistent differences in writing style and content. Guo et al.\cite{guo2023close} find that ChatGPT responses are more objective, formal, and detailed than human-written answers, while Chen et al.\cite{chen2023can} discover that LLM-generated misinformation can be more difficult to detect than human-written misinformation, for both humans and LLM-based detectors like GPT-4 and LLaMA2. 

In terms of attribution, Antoun et al.\cite{antoun2023text} conduct a large-scale study involving 50 LLMs and find an inverse correlation between classifier accuracy and model size—larger models are harder to attribute, especially when the classifier is trained on smaller models. Other studies propose representation- and feature-based attribution techniques. Munir et al.\cite{munir2021through} identify subtle textual traces inherited from source models, while TopRoBERTa\cite{uchendu2024topformer} incorporates topological data analysis to enhance RoBERTa’s ability to capture both semantic and structural patterns. Stylometric and psycholinguistic features are also leveraged: Uchendu et al.\cite{uchendu2020authorship} use entropy and readability scores; GPT-who\cite{venkatraman2023gpt} adopts Uniform Information Density; and Wu et al.\cite{wu2023llmdet} apply proxy perplexity scores to attribute texts to models like LLaMA and OPT\cite{liu2021opt}. Soto et al.\cite{Soto2024few-shot} propose a few-shot method that learns style features only from human-authored texts, avoiding reliance on generated samples. Finally, Sniffer\cite{li2023origin} identifies model lineage by analyzing statistical discrepancies between model outputs, successfully linking derived models like Alpaca~\cite{alpaca} to their predecessors such as ChatGPT and LLaMA.

\subsubsection{Analysis} 
\noindent
$\clubsuit$ \textbf{Insights}.
Despite significant advances, most detection methods for LLM-generated text converge on a shared insight: LLMs produce outputs with detectable statistical regularities due to their decoding strategies and optimization objectives. These include over-regular syntax, repetitive phrasing, low grammatical variability, and coherent entity chains~\cite{pu2022unraveling, GPTZero, wu2024wrote, liu2022coco}. Detection methods either exploit these patterns through white-box signals, such as token-level perplexity~\cite{GPTZero}, log-probability curvature~\cite{mitchell2023detectgpt, bao2023fast}, and cross-model comparisons~\cite{wang2023seqxgpt, su2023detectllm}, or rely on black-box cues, including revision similarity and grammar correction~\cite{wu2024wrote, zhu2023beat}. While white-box approaches offer higher precision, they require access to model internals or APIs~\cite{bao2024glimpse}, whereas black-box techniques are more flexible but often less robust. Several methods demonstrate strong zero-shot generalization by leveraging generation-intrinsic features rather than dataset-specific cues~\cite{hans2024spotting, bao2024glimpse}. However, trade-offs persist between interpretability and performance: interpretable methods like watermarking or edit-based signals are easy to explain but fragile, while statistical detectors offer higher accuracy at the cost of opacity~\cite{zhao2023provable, kumarage2023stylometric}. Notably, short-form content remains especially difficult to detect due to limited context, prompting recent interest in Positive-Unlabeled frameworks and style-based adaptation~\cite{tian2023multiscale, Soto2024few-shot}. Future detection systems may benefit from hybrid designs that combine white-box precision, black-box accessibility, and interpretability, while remaining resilient across languages, formats, and model variants.

\smallskip
\noindent
$\clubsuit$ \textbf{Recommendations.}
In light of the above analysis, several general strategies can be recommended for improving LAIM-generated text detection systems. To address the reliance on model internals, hybrid frameworks that combine white-box precision with black-box accessibility may offer a balance between accuracy and deployability. For instance, approximating internal signals (\eg, via top-K outputs or surrogate modeling) can expand the applicability of otherwise restricted methods. To enhance zero-shot generalization, it is beneficial to prioritize detection cues rooted in generation-intrinsic behaviors, such as token-level statistical regularities, rather than model-specific representations. For improved interpretability without sacrificing performance, a layered design that combines transparent surface-level heuristics with deeper statistical or neural backends can support both expert and non-expert use cases. Lastly, the persistent challenge of short form content calls for the integration of uncertainty-aware or semi-supervised learning techniques, such as Positive-Unlabeled modeling or few-shot adaptation, which can help maintain robustness under sparse or noisy inputs.

\smallskip
\noindent
$\clubsuit$ \textbf{Social Media.}
The widespread availability of LLMs has raised significant concerns about AI-generated content on social media platforms. Early works emphasized the challenges and initiated systematic efforts to detect and quantify this phenomenon.
In late 2023, Lucas et al.~\cite{lucas-etal-2023-fighting} proposed the F3 framework, leveraging GPT-3.5's generative capabilities not only to create synthetic disinformation but also to detect it using in-context, zero-shot reasoning techniques. Their findings highlighted that modern LLMs, if properly prompted, can outperform traditional fine-tuned detectors in identifying disinformation posts on platforms like Twitter and news outlets.
Around the same time, Cui et al.~\cite{cui2023said} introduced the SAID benchmark, one of the first efforts to collect real-world AI-generated texts from social media platforms like Zhihu~\cite{zhihu2025} and Quora~\cite{quora2025}. SAID emphasized that detection in real-world settings is substantially harder than in simulated environments, and that user context information can greatly improve AI detection accuracy.

To address the multilingual and multi-platform gap, Macko et al.\cite{macko2024multisocial} introduced MultiSocial, a benchmark covering 22 languages and five platforms (\eg, Twitter, Telegram, Gab), showing that fine-tuned detectors outperform zero-shot methods on short, informal texts, though platform and language differences still pose challenges. Meanwhile, Sun et al.\cite{sun2024we} conducted a large-scale study on Medium, Reddit, and Quora using the SM-D dataset, revealing a sharp rise in AI-generated content after late 2022, especially on Medium and Quora—and noting that such posts are often more objective, technical, and standardized than human-written ones.

Kumarage et al.~\cite{kumarage2023stylometric} tackled the problem of detecting AI-generated tweets at a timeline level by developing a stylometric change point detection framework. Rather than detecting individual tweets, their method identifies shifts in writing style across a user's posts, enabling forensic tracking of account takeovers by AI systems.
Further complicating the landscape, Gameiro et al.~\cite{gameiro2024llm} demonstrated that even state-of-the-art LLM detectors underperform in real-world settings, particularly when facing short news-like posts crafted by attackers who manipulate sampling parameters. Their findings stressed the urgent need for domain-specific benchmarks and cautioned against relying on existing detectors for operational deployment.
Tuck and Verma expanded this line of inquiry by evaluating how censorship removal (\eg, uncensoring models like LLaMA3 and Qwen2) affects detectability~\cite{tuck2024unmasking}. They showed that uncensored models produce more human-like outputs, significantly undermining standard detection methods, and called for detectors capable of adapting to new domains and generation strategies.
Finally, Ayoobi et al.~\cite{ayoobi2023looming} extended the concern to professional social networks like LinkedIn, where they introduced the SSTE (Section and Subsection Tag Embedding) method for detecting LLM-generated fake profiles. Their results demonstrated that even with limited training data, structured stylometric embeddings could robustly flag AI-crafted user profiles at registration time.

In conclusion, studies consistently show that detection accuracy drops significantly in real-world conditions~\cite{cui2023said, gameiro2024llm}, where generation parameters are unknown and text is often shorter and noisier. Moreover, the rapid adoption of increasingly human-like uncensored models further weakens traditional detection strategies~\cite{tuck2024unmasking}. However, novel approaches, such as stylometric change point analysis~\cite{kumarage2023stylometric}, multilingual platform specific tuning~\cite{macko2024multisocial}, and in-context LLM-based detection~\cite{lucas-etal-2023-fighting}, have shown promise in adapting to these challenges. Attribution and detection methods based on writing style rather than token probabilities offer improved generalization, while timeline and user level modeling introduce new avenues for forensic tracking.

\subsection{Image}\label{sec:Image_det}
While DMs are evolving rapidly and producing increasingly realistic images, they still often make certain mistakes and leave identifiable fingerprints. 

\begin{figure*}[t]
    \centering
    \includegraphics[width=1\textwidth]{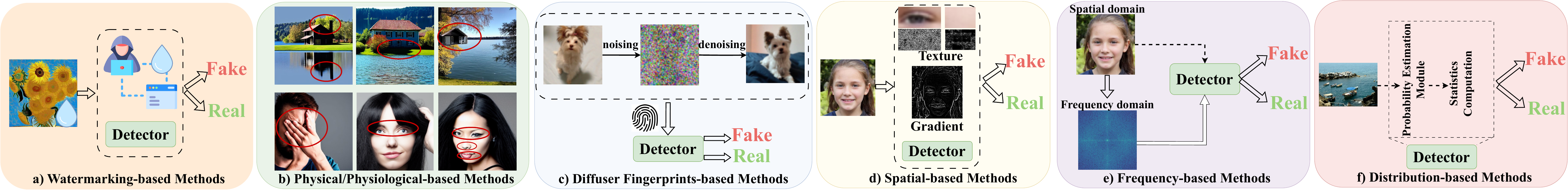}
    \vspace{-6mm}
    \caption{\small Illustrations of pure detection methodologies for LAIM-generated image. In a), The red circles highlight regions that violate physical or physiological principles, indicating artifacts commonly found in LAIM-generated images.}
    \label{fig:image_pure_detection_overview}
    \vspace{-4mm}
\end{figure*}

\subsubsection{Pure Detection}\label{sec:image_pure}
Research summarized here aims to identify DM-generated images by examining physical and physiological cues, as well as by focusing on enhancing detection accuracy. 

\smallskip
\noindent
$\clubsuit$ \textbf{Watermarking.} Recent watermarking methods enable the detection of LAIM-generated images by identifying embedded watermarks introduced during or after the generation process. see Fig.~\ref{fig:image_pure_detection_overview} a).

WaDiff~\cite{min2024watermark} embeds user-specific watermarks directly into the generation process using a conditioned input and custom loss functions, enabling detection and identification even in black-box settings. Stable Signature~\cite{fernandez2023stable} fine-tunes only the latent decoder to embed binary signatures but is limited to latent diffusion models and requires architectural access. In contrast, Tree-Ring Watermarking~\cite{wen2023tree} perturbs the initial noise in Fourier space without retraining, offering broad model compatibility and robustness to common image transformations.

Other approaches include watermark-embedded adversarial examples~\cite{zhu2024watermark}, which embed visible watermarks via adversarial perturbations to prevent diffusion models from mimicking copyrighted content—focusing more on protection than traceability. Finally, Gaussian Shading~\cite{yang2024gaussian} achieves performance-lossless watermarking by encoding signals into Gaussian latent codes without altering model parameters, offering strong robustness and seamless deployment across diffusion models.

\smallskip
\noindent
$\clubsuit$ \textbf{Physical/Physiological-based Methods.} 
Physical-based methods detect DM-generated images by examining inconsistencies with real-world physics, such as lighting and reflections.
Physiologically-based methods, on the other hand, investigate the semantic aspects of human faces~\cite{ciftci2020fakecatcher}, 
including cues such as symmetry, iris color, pupil shapes, skin, etc.,
see Fig.~\ref{fig:image_pure_detection_overview} b).
These methods have stronger interpretability than data-driven methods, which have been widely adapted to detect GAN-generated images~\cite{wang_etal_ecai23,kang2023scaling}. 

Borji~\cite{borji2023qualitative} outlines key cues for detecting DM-generated images that violate physical rules, as shown in Fig.~\ref{fig:image_pure_detection_overview} a) top. These cues include: i) \textit{Reflection}. Generated images can exhibit artificial reflections that are inconsistent with the natural lighting and environment, such as those in glasses, mirrors, or pupils. ii) \textit{Shadow}. Generated images might not include shadows, or have inconsistent shadows. iii) \textit{Objects without Support}. In generated images, an object or material appears to be floating in mid-air without any visible means of support, it gives the impression that the object is defying gravity or the laws of physics. 

The above cues provided by Borji are based on the observations of the failure cases of DM-generated images, while Farid analyzes the perspective~\cite{farid2022perspective} and lighting~\cite{farid2022lighting} consistency in images synthesized by DALL·E 2. Noting that DALL·E 2 sometimes fails to maintain geometric consistency, such as parallel lines cannot converge at a common vanishing point. Additionally,  while DALL·E 2 generally creates realistic lighting, there's a tendency for the lighting direction to be more frontal or rearward relative to the camera compared with natural photographs.

As for physiological-based forensics, Borji~\cite{borji2023qualitative} shows various physiological cues relative to eyes, teeth, ears, hair, skin, limbs, and fingers, etc, as shown in   Fig.~\ref{fig:image_pure_detection_overview} a) bottom. These artifacts suggest that generative models often fall short when accurately depicting the intricate details of human extremities.

\smallskip
\noindent
$\clubsuit$ \textbf{Diffuser Fingerprints-based Methods.} Every generation mechanism leaves its own unique fingerprints, which can be explored for detection, as shown in Fig.~\ref{fig:image_pure_detection_overview} c).
In \cite{sinitsa2023deep}, the authors elucidated that CNNs inherently manifest certain image artifacts. These artifacts are harnessed in their Deep Image Fingerprint (DIF) methodology to extract distinct fingerprints from images generated by these networks (GANs and DMs). This approach facilitates the identification of images originating from a specific model or its fine-tuned variants. Complementarily, the studies by Wang et al. \cite{wang2023dire} and Ma et al. \cite{ma2023exposing} delve into the realm of diffusion models. They have laid the groundwork for detecting DM-generated images by leveraging the deterministic reverse and denoising processes inherent diffusion models.

Wang et al. \cite{wang2023dire} propose a novel image representation called DIffusion Reconstruction Error (DIRE) based on their hypothesis that images produced by diffusion processes can be reconstructed more accurately by a pre-trained diffusion model compared to real images. DIRE measures the error between an input image and its reconstruction by a pre-trained diffusion model. The computation process of DIRE can be simply concluded as follows: the input image $\x_0$ is first gradually inverted into a noise image $\x_T$ by DDIM inversion~\cite{song2020denoising} and then is denoised step by step until getting a reconstruction $\x_0'$. DIRE is the residual image obtained from $\x_0$ and $\x_0'$, which can be used to differentiate real or generated images.

Though Wang et al.~\cite{wang2023dire} indeed leverages some deterministic aspects, their approach (\ie, DIRE) primarily targets the reconstruction at the initial time step $\x_0$. This method potentially overlooks the rich information present in the intermediate stages of the diffusion and reverse diffusion processes.
Ma et al.~\cite{ma2023exposing} exploits these intermediate steps later. They design Stepwise Error for Diffusion generated Image Detection (SeDID), particularly focusing on the errors between reverse and denoise samples at specific time steps during the generation process. 
Luo et al.~\cite{Luo_2024_CVPR} further improve efficiency and effectiveness with LaRE², a Latent REconstruction error-guided feature REfinement method. Instead of full inversion, LaRE² computes reconstruction error via single step denoising in the latent space, making it over 8× faster than multi-step approaches like DIRE \cite{wang2023dire}.

Rajan et al.~\cite{rajan2025aligned} focus on improving fingerprint-based detection by aligning real and fake images more precisely. Instead of generating fake images through the standard denoising pipeline, they reconstruct real images using only the LDM’s autoencoder, capturing artifacts introduced by the decoder while maintaining alignment in resolution, semantics, and format. This allows the detector to focus specifically on decoder induced fingerprints, avoiding spurious correlations. Brokman et al.~\cite{brokman2025manifold} present a theoretically grounded zero-shot detection method that leverages the score function of pre-trained diffusion models to quantify biases in generated images. Their method defines geometric criteria, curvature and gradient magnitude, on the implicit probability manifold learned by the model. These quantities, when computed over local perturbations of a given image, serve as indicators of whether the image is fake or real. 

\smallskip
\noindent
$\clubsuit$ \textbf{Spatial-based Methods.} This research collection focuses on mining spatial characteristics and features within images to detect DM-generated content. Each study utilizes different spatial aspects of images, such as texture, gradients, and local intrinsic dimensionality, for detection, see Fig.~\ref{fig:image_pure_detection_overview} d).
Motivated by the principle that pixels in rich texture regions exhibit more significant fluctuations than those in poor texture regions. Consequently, synthesizing realistic rich texture regions proves to be more challenging for existing generative models. Zhang et al. \cite{zhong2023rich} leverage such texture features and exploit the contrast in inter-pixel correlation between rich and poor texture regions within an image for DM-generated image forensics.
Nguyen et al. \cite{nguyen2023unmasking} use gradient-based features to differentiate DM-generated and human-made artwork. 
Lorenz et al. \cite {lorenz2023detecting} propose using the lightweight multi Local Intrinsic Dimensionality (multiLID) for effective detection. This approach stems from the principle that neural networks heavily rely on low-dimensional textures \cite{geirhos2018imagenet} and natural images can be represented as mixtures of textures residing on a low-dimensional manifold \cite{vacher2020texture}. Leveraging this concept, multiLID scores are calculated on the lower dimensional feature map representations derived from ResNet18. Then a classifier (random forest) is trained on these multiLID scores to distinguish between synthetic and real images. Their proposed multiLID approach exhibits superiority in diffusion detection. However, this solution does not offer good transferability.

\smallskip
\noindent
$\clubsuit$ \textbf{Frequency-based Methods.} Frequency-based methods analyze the frequency components of an image to extract information that is not readily apparent in the spatial domain, see Fig.~\ref{fig:image_pure_detection_overview} e).
The initial study by Wolter et al. \cite{wolter2022wavelet} highlights the limitations of traditional detection methods that primarily utilize spatial domain convolutional neural networks or Fourier transform techniques.
In response, they propose an innovative approach using multi-scale wavelet representation, which incorporates elements from both spatial and frequency domains. 
However, the study notes limited benefits from higher order wavelets in guided diffusion data, pinpointing this as a potential area for future research.
In a recent study, Xi et al. \cite{xi2023ai} develop a dual stream network that combines texture information and low frequency analysis. This approach identifies artificial images by focusing on low frequency forgeries and detailed texture information. It has shown to be efficient across various image resolutions and databases. 
\cite{poredi2023ausome} introduces AUSOME to detect DALL-E 2 images by performing a spectral comparison of Discrete Fourier Transform (DFT) and Discrete Cosine Transform (DCT).
Bammey~\cite{bammey2023synthbuster} uses the cross difference filter on the image to highlight the frequency artifacts from DM-generated images. It shows some generalization ability, as well as robustness to JPEG compression.
Lanzino et al. \cite{Lanzino_2024_CVPR} propose a real-time deepfake detector using Binary Neural Networks, incorporating FFT and LBP features to capture frequency and texture artifacts with minimal accuracy loss.

\smallskip
\noindent
$\clubsuit$ \textbf{Distribution-based Methods.} Instead of analyzing images in the spatial or frequency domain directly, distribution-based methods model the intrinsic statistical distribution of real images and detect synthetic images as distributional anomalies, see Fig.~\ref{fig:image_pure_detection_overview} f). Cozzolino et al.~\cite{cozzolino2024zero} propose a zero-shot detector (ZED) that leverages a lossless image compression model trained only on real images to predict the probability distribution of each pixel given its context.  By computing the gap between the actual negative log-likelihood (NLL) and the expected entropy across multiple resolution scales, ZED achieves strong detection performance without requiring any synthetic training data.

\subsubsection{Beyond Detection}\label{sec:image_beyond}
Research on detecting DM-generated images aims not only to enhance accuracy but also to add additional functionalities to detectors. It seeks to develop a more nuanced understanding and application of these detectors, unlocking new avenues and potential in the realm of DM-generated image analysis and detection. We summarize the existing work into the following categories:

\begin{figure*}[t]
    \centering
    \includegraphics[width=1\textwidth]{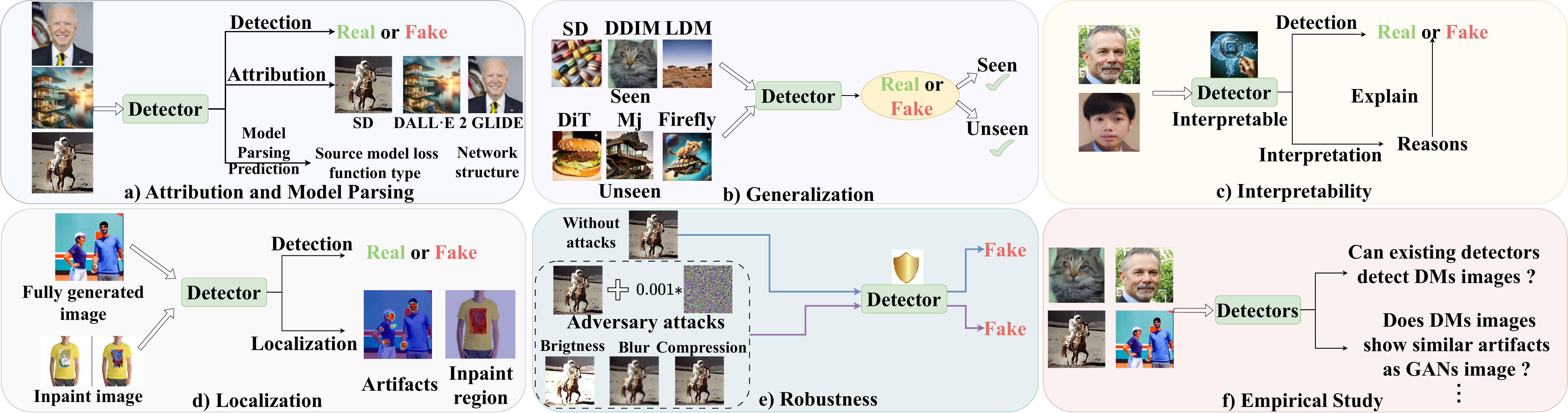}
    \vspace{-6mm}
    \caption{\small Illustrations of beyond detection methodologies for LAIM-generated image.}
    \label{fig:image_beyond_overview}
    \vspace{-4mm}
\end{figure*}

\smallskip
\noindent
$\clubsuit$ \textbf{Attribution and Model Parsing.} Attribution is to recognize the specific diffusion model that generates images. Model parsing refers to the process of analyzing and interpreting the structure and characteristics of generative models. The purpose of model parsing is to gain insight into how a particular model generates images, which can be crucial for tasks like identifying the source of a synthetic image, see Fig.~\ref{fig:image_beyond_overview} a).

\raisebox{-0.3ex}{\scalebox{1.3}{\ding{192}}} \textbf{Attribution}.
Guarnera et al. \cite{guarnera2024level} focus on attributing DM-generated images using a multi-level hierarchical approach. Each level solves a specific task: level 1 classifies images as real or AI-generated; level 2 defines whether the input images are created by GAN or DM technologies; level 3 solves the model attribution task. 

\raisebox{-0.3ex}{\scalebox{1.3}{\ding{193}}} \textbf{Model Parsing}. The model parsing task is first defined by Asnani et al. \cite{asnani2023reverse} as estimating Generative Model (GM) network architectures (\eg, convolutional layers, fully connected layers, and the number of layers) and their loss function types (\eg,  Cross-entropy, Kullback–Leibler divergence) by examining their generated images.
To tackle this problem, they compiled a dataset comprising images from 116 GMs, which includes detailed information about the network architecture types, structures, and loss functions utilized in each GM.
They introduce a framework with two components: a Fingerprint Estimation Network, which estimates a GM fingerprint from a generated image, and a Parsing Network, which predicts network architecture and loss functions from the estimated fingerprints. 

\smallskip
\noindent
$\clubsuit$ \textbf{Generalization.} 
A generalizable detector can successfully detect the unseen images generated by the newly released generators, as shown in  Fig.~\ref{fig:image_beyond_overview} b).

Epstein et al. \cite{epstein2023online} 
collect a dataset generated by 14 well-known diffusion models and simulate a real-world learning setting with incremental data from new DMs. They find that the classifiers generalize to unseen models, although performance drops substantially when there are major architectural changes. 
Different from \cite{epstein2023online}, \cite{ojha2023towards, cozzolino2023raising} utilize pre-trained models, ViT~\cite{dosovitskiy2020image} and CLIP~\cite{radford2021learning} respectively, to achieve high generalization and robustness in various scenarios.

In addition, \cite{yan2023transcending} uses data augmentation approaches to improve detection generalization. To avoid data dependency on particular generative models and improve generalizability, Jeong et al. \cite{jeong2022fingerprintnet} propose a framework that can generate various fingerprints and train the detector by adjusting the amplitude of these fingerprints or perturbations.
Their method outperforms the prior state-of-the-art detectors, even though only real images are used for training.
Reiss et al. \cite{reiss2023detecting} introduce FACTOR, particularly for fact-checking tasks. Although their method is training-free, it achieves promising performance. Dogoulis et al.~\cite{dogoulis2023improving} address the generalization of detectors in cross-concept scenario (\eg, when training a detector on human faces and testing on synthetic animal images), and propose a sampling strategy that considers image quality scoring for sampling training data. It shows better performance than existing methods that randomly sample training data. DNF~\cite{zhang2023diffusion} seeks to address generalization challenges through an ensemble representation that estimates the noise generated during the inverse diffusion process and achieves state-of-the-art effects of detection.

More recently, Baraldi et al. \cite{baraldi2024contrasting} proposed CoDE, a contrastive learning framework that enforces global-local similarities to improve generalization across different generated image types.
Yuan et al. \cite{yuan2025hrr} introduced HRR, a diffusion-based detection model that enhances generalization through multi-scale retrospection and feature refinement modules, mitigating dataset and generator biases. Yan et al.~\cite{yan2024sanity} propose AIDE, a hybrid detector that combines frequency-domain features and CLIP-based semantic features. AIDE achieves state-of-the-art results across both general (GenImage~\cite{zhu2023genimage}, AIGCD~\cite{zhong2023rich}) and challenging (Chameleon~\cite{yan2024sanity}) benchmarks. Cheng et al.~\cite{cheng2024stacking} address the catastrophic forgetting problem in incremental face forgery detection, which directly impacts generalization to new forgery types. Their method, SUR-LID, introduces Sparse Uniform Replay (SUR) to approximate the latent feature distribution of past tasks and Latent-space Incremental Detector (LID) to align new features “brick by brick.” The framework enables continual learning across datasets and forgery types while retaining high detection accuracy.  
To further enhance generalization, Guillaro et al.~\cite{guillaro2024bias} propose B-Free, a bias-free training paradigm that generates self-conditioned fake images from real ones to avoid semantic and compression biases. Combined with content augmentation, B-Free significantly improves generalization across 27 generative models. Meanwhile, Zheng et al.~\cite{zheng2024breaking} identify cross-scene generalization as a key challenge, where detectors struggle on images with unseen semantic content due to overfitting to “semantic artifacts.” They address this by introducing a patch-shuffling strategy and a patch-based classifier that disrupts global scene cues.  \cite{chen2025dual} introduces a dual (pixel + frequency) alignment strategy and new benchmark datasets (DDA‑COCO \& EvalGEN) to improve cross‑distribution generalization.

\smallskip
\noindent
$\clubsuit$ \textbf{Interpretability.} 
``\textit{This picture looks like someone I know, and if the AI algorithm tells it is fake or real, then what is the reasoning and should I trust?} " Detection with interpretability is working towards solving such question, see Fig.~\ref{fig:image_beyond_overview} c). 
In pursuit of creating interpretable detectors, Aghasanli et al. \cite{aghasanli2023interpretable} propose a method for DM-generated image detection, taking advantage of features from fine-tuned ViTs combined with existing classifiers such as Support Vector Machines.
This approach is notable for interpreting the model behavior through prototypes by analyzing SVM's closest support vectors.
using SVM's closest support vectors, enabling a unique form of interpretability-through prototypes. 
Color distributions offer a rich and interpretable cue for synthetic image detection. Uhlenbrock et al.~\cite{uhlenbrock2024did} show that perceptual loss functions in diffusion models emphasize luminance over chrominance, leading to unnatural relationships between color channels. By transforming images into multiple color spaces and computing residual-based co-occurrence statistics, their simple random forest detector achieves strong generalization across generators and post-processing variants.

\smallskip
\noindent
$\clubsuit$ \textbf{Localization.} A detector that can localize the artifacts in a DM-generated image or give a prediction with a localization map indicating which input regions have been manipulated, as shown in Fig.~\ref{fig:image_beyond_overview} d).

\raisebox{-0.3ex}{\scalebox{1.3}{\ding{192}}} \textbf{Fully-supervised}. Localization in a fully-supervised setting requires detectors to be  explicitly trained for localization as a segmentation problem with localization mask label.   
Guo et al. \cite{guo2023hierarchical} propose a hierarchical fine-grained Image Forgery Detection and Localization (IFDL) framework with three components: a multi-branch feature extractor, localization, and classification modules. The localization module segments pixel-level forgery regions, while the classification module detects image-level forgery. 
\cite{zhang2023perceptual} centers on detecting and segmenting artifact areas that are only noticeable to \textit{human perception}, not full manipulation region. 
In addition to providing a pixel-level localization map, the study by {Guillaro et al.} \cite{guillaro2023trufor} also offers an integrity score to assist users in their analysis.
Li et al.~\cite{Li_2024_CVPR} introduce UnionFormer, a unified-learning transformer framework for simultaneous detection and localization. It incorporates three complementary views: RGB, noise, and object consistency. The proposed BSFI-Net enhances boundary sensitivity and captures artifacts across scales, while a unified transformer integrates multi-view clues to yield both detection results and a pixel-level manipulation mask.

\raisebox{-0.3ex}{\scalebox{1.3}{\ding{193}}} \textbf{Weakly-supervised}. In contrast to the above work, which addresses localization in a fully-supervised setting. Tantaru et al. \cite{tantaru2023weakly} consider a weakly-supervised scenario
motivated by the fact that ground truth manipulation masks are usually unavailable, especially for newly developed local manipulation methods. They conclude that localization of manipulations for latent diffusion models~\cite{rombach2021highresolution} is very challenging in the weakly-supervised scenario.

\smallskip
\noindent
$\clubsuit$ \textbf{Robustness.} A robust detector is strategically developed to counteract different attacks consisting of intentionally designed perturbations or noise, as shown in Fig.~\ref{fig:image_beyond_overview} e).  It is also designed to maintain consistent detection performance, for example, if the input image is fake, a robust detector should consistently predict it as fake, regardless of whether the input has been subjected to attacks such as compression, blurring, or other real-world degradations.


\raisebox{-0.3ex}{\scalebox{1.3}{\ding{192}}} \textbf{Adversarial Attack Robustness}. De Rosa et al.\cite{de2024exploring} study the adversarial robustness of CLIP-based forensic detectors compared to CNN-based ones. They show that while both architectures are vulnerable to white-box attacks, adversarial perturbations exhibit different frequency-domain patterns between CNNs and CLIP-based models, making attacks less transferable across architectures. Hooda et al.\cite{Hooda2024D4} propose D4, an ensemble-based detector partitioning frequency components across models. This disjoint design reduces the adversarial subspace dimensionality and achieves strong robustness against black-box adversarial attacks, outperforming conventional ensemble defenses

\raisebox{-0.3ex}{\scalebox{1.3}{\ding{193}}} \textbf{Post-Processing Robustness}. Ju et al.~\cite{ju2023glff} address robustness against post-processing such as JPEG compression, Gaussian blurring. They propose GLFF, a Global and Local Feature Fusion framework that combines multi-scale global and subtle local features, achieving improved performance on the challenging DF$^3$ dataset with diverse post-processing operations.  Xu et al. \cite{xu2023exposing} also leverages multi-level feature representation for enhancing robustness. Different from the above works, \cite{jer2023local} introduces a non-deep learning method based on Bayer pattern statistics and local variance features to detect diffusion-generated images. Their method remains effective under image resizing and JPEG compression, showing resilience to perturbations. \cite{yang2025all} proposes a panoptic patch learning framework that leverages multi‑scale, context‑aware patch selection to boost detection robustness and explainability.

\smallskip
\noindent
$\clubsuit$ \textbf{Empirical Study.} The empirical study serves as a crucial foundation for devising methods to detect images generated by rapidly advancing DMs. Despite DMs' rapid progression, detection methods have not evolved at the same pace. Therefore, thorough experiments and insights from these empirical studies are vital for advancing detection technologies, as illustrated in Fig.~\ref{fig:image_beyond_overview} f).

Recent studies by Corvi et al.~\cite{corvi2023detection} and Ricker et al.~\cite{ricker2022towards} investigate the challenge of differentiating photorealistic synthetic images from real ones. They find that detectors trained only on GAN images work poorly on DM-generated images. Additionally, DMs exhibit much subtler fingerprints in the frequency domain than GANs.
 
Realizing it is challenging for existing methods to detect DM-generated images, Corvi et al. \cite{corvi2023intriguing} extend their previous work \cite{corvi2023detection} to gain insight into which image features better discriminate fake images from real ones. Their experimental results shed light on some intriguing properties of synthetic images:
\begin{itemize}[leftmargin=*]
    \item To date, no generator appears to be artifact-free. Unnatural regular patterns are still observed in the autocorrelation of noise residuals;
    \item When a model is trained on a limited-variety dataset, its biases can transfer to the generated images;
    \item Synthetic and real images differ notably in their mid-high frequency signal content.
\end{itemize}
These traces exploited in this empirical study can be instrumental in developing forensic detectors.

Cocchi et al. \cite{cocchi2023unveiling} investigate the robustness of different pre-trained model backbones against image augmentations and transformations. Their findings reveal that these transformations significantly affect classifier performance. 

Other works \cite{porcile2023finding, papa2023use} empirically study human face detection. \cite{porcile2023finding} focuses on detecting profile photos of online accounts, finding that a detector (EfficientNet-B1~\cite{tan2019efficientnet}) only trained on face images completely fails when detecting non-face images, and the detector can learn the semantic level artifacts because it remains highly accurate even when the image is compressed to a low resolution.
\cite{papa2023use} investigates the difficulty of non-expert humans in recognizing fake faces and further concludes that trained models outperform non-expert human users, which brings out the need for solutions to contrast the spread of disinformation.  

Furthermore, \cite{mandelli2022forensic} centers on the biological images and narrows on detecting synthetic western blot images. They explore a range of detection strategies, such as binary classification and one-class detectors, and notably, they do this without utilizing synthetic images in training. Their findings demonstrate the successful identification of synthetic biological images. \cite{carriere2023beyond} is the first work to detect authentic looking handwriting generated by DMs. Their experiments reveal that the strongest discriminative features come from real-world artifacts in genuine handwriting that are not reproduced by generative methods.

\subsubsection{Analysis} 
\noindent
$\clubsuit$ \textbf{Insights}.
Across the surveyed detection methods, several shared insights emerge regarding the vulnerabilities of DM-generated images and the design principles of effective detectors. Despite the rapid improvement of diffusion models, their outputs still exhibit subtle artifacts that remain detectable across multiple domains. These include geometric inconsistencies, such as perspective distortions and unnatural reflections~\cite{borji2023qualitative, farid2022perspective}, as well as low-level generation artifacts left by decoder structures or denoising steps~\cite{rajan2025aligned, wang2023dire}. Many successful detection frameworks exploit the deterministic behavior of diffusion processes, using inversion and reconstruction-based strategies to quantify deviations between real and generated images~\cite{ma2023exposing, brokman2025manifold}. In parallel, spatial and frequency domain features continue to serve as strong indicators, with studies showing that texture irregularities~\cite{zhong2023rich} and abnormal frequency patterns~\cite{xi2023ai, wolter2022wavelet} often co-occur in synthetic images. However, generalization remains a persistent challenge: detectors trained on specific generators or domains often fail when applied to unseen models or image categories, especially when semantic content shifts significantly~\cite{zheng2024breaking, corvi2023detection}. Moreover, robustness to post-processing and adversarial perturbations is limited, with even strong detectors showing vulnerability to common transformations like JPEG compression, resizing, or blurring~\cite{de2024exploring, ju2023glff}.

\smallskip
\noindent
$\clubsuit$ \textbf{Recommendations.} To address these challenges, several design principles can be recommended. First, hybrid detectors that combine spatial, frequency, and latent cues tend to exhibit improved robustness and generalizability~\cite{Li_2024_CVPR, ju2023glff}. Second, inversion-based approaches that leverage model-induced consistency, such as reconstruction error or decoder fingerprinting, offer interpretable and transferable signals~\cite{ma2023exposing, rajan2025aligned}. Third, promoting open-world generalization requires methods to avoid overfitting to dataset-specific or generator-specific patterns. Strategies such as patch-based classification~\cite{zheng2024breaking}, fingerprint regularization~\cite{jeong2022fingerprintnet}, or contrastive learning~\cite{baraldi2024contrasting} show promise in this direction. Fourth, interpretability remains crucial for user trust and adoption, particularly in forensic applications; color-space analysis and prototype-based reasoning have emerged as promising paths for interpretable detection~\cite{aghasanli2023interpretable, uhlenbrock2024did}. Lastly, to ensure real-world applicability, detectors must be evaluated under diverse image transformations and compression schemes, reflecting practical deployment environments like social media or messaging platforms~\cite{corvi2023detection, jer2023local}.

\smallskip
\noindent
$\clubsuit$ \textbf{Social Media.}
The detection of LAIM-generated images on social media has emerged as a critical research challenge, unlike controlled lab settings, social media platforms introduce compression, resizing, adversarial editing, and a lack of reliable metadata, causing lab-trained detectors to underperform in real-world environments. 
ODDN~\cite{tao2025oddn} tackles the unpaired data problem by training directly on compressed fake images without requiring pristine counterparts, demonstrating the need for explicitly training detectors for robustness against compression artifacts.

To enable real-world deployment, Cavia et al.\cite{cavia2024real} propose LaDeDa, a lightweight detector suited for real-time use. While it performs well on lab datasets, tests on the WildRF benchmark, comprising real social media images, reveal notable performance drops.
A critical reassessment by Yan et al.\cite{yan2024sanity} further shows that even top detectors often misclassify highly realistic LAIM images. 
Advancing the field, Huang et al.\cite{huang2024sida} present SIDA,  the first framework for both detection and localization of LAIM images on social media. Leveraging large vision-language models and the newly curated SID-Set dataset, SIDA enables not only accurate detection but also pixel-level localization and interpretable explanations.

Key findings across these studies underline that detecting LAIM-generated images on social media is significantly more difficult than in controlled lab environments. Real-world images often undergo aggressive transformations such as JPEG compression, resizing, filtering, and adversarial post-processing, all of which obscure generative artifacts that detectors typically exploit. Moreover, the dominant sources of AI images on social platforms have shifted from earlier GAN models to advanced diffusion models, which produce much higher quality and contextually realistic outputs. Lab-generated datasets, in contrast, are curated, balanced, and captured under idealized conditions, leading to overestimation of detector performance.

\subsection{Video}\label{sec:Video_det}
Recently, there has been limited research focused on detecting videos generated by LAIMs. This is primarily because video generation techniques are more complex than those for images. Additionally, as outlined in Section \ref{sec:LAIMs_generated_Multimedia}, video generation technology is still in its nascent stages compared with image generation.

\begin{figure*}[t]
  \centering
  \includegraphics[width=1\linewidth]{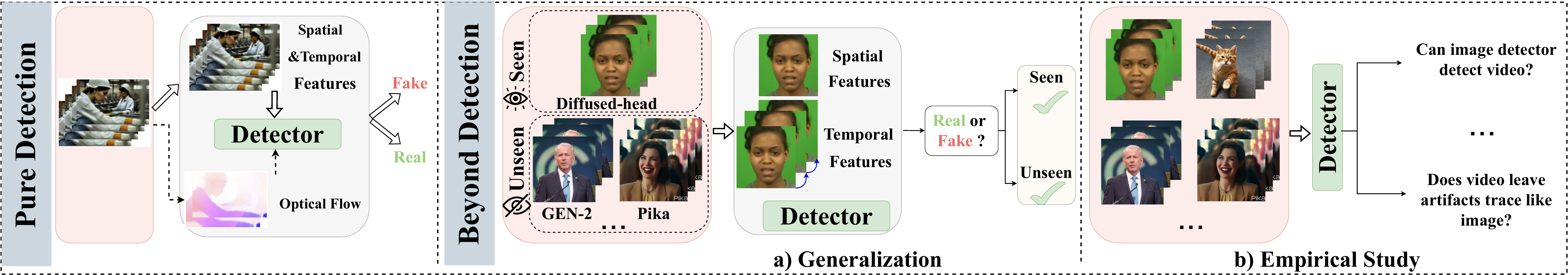}
  \vspace{-6mm}
  \caption{\small Illustration of detection methodology for LAIM-generated video.}
  \vspace{-4mm}
  \label{fig:video_detection}
\end{figure*}

\subsubsection{Pure Detection}\label{sec:video_pure} 
$\clubsuit$ \textbf{Spatial\&Temporal-based Methods}.
Ji et al.~\cite{ji2024distinguish} propose DuB3D, a Dual-Branch 3D Transformer that leverages both spatiotemporal and motion cues for LAIM-generated video detection. DuB3D integrates appearance and optical flow branches through a late fusion mechanism. Extensive experiments show its superior performance. He et al.~\cite{he2024exposing} propose a detection method that jointly models local motion inconsistencies and global appearance variations leveraging a frame prediction error module and a temporal transformer over BEiT~\cite{peng2022beit} features. These are fused via a channel attention module to adaptively highlight defects.

\subsubsection{Beyond Detection}\label{sec:video_beyond} 

\smallskip
\noindent
$\clubsuit$ \textbf{Generalization}. Many works address generalization in deepfake video detection. However, they may overlook one challenge:
the standard out-of-domain evaluation datasets are very similar in form to the training data,
failing to keep up with the advancements in DM video generation. 
\cite{kamat2023revisiting} addresses such an issue by 
introducing the Simulated Generalizability Evaluation (SGE) method, which involves simulating spatial and temporal deepfake artifacts in videos of human faces using a Markov process. SGE aims to improve the generalizability of video detection by creating artifacts that reflect implausibilities in facial structures, which could accompany unseen manipulation types. 

\smallskip
\noindent
$\clubsuit$ \textbf{Empirical Study}. Vahdati et al.~\cite{vahdati2024beyond} conduct a comprehensive study revealing that detectors trained on synthetic images fail to generalize to AI-generated videos. Their experiments show that this failure is not primarily due to compression artifacts (\eg, H.264), but rather stems from fundamentally different forensic traces left by video generators compared to image generators. To address this gap, they demonstrate that synthetic video traces can be learned directly, enabling reliable detection and generator source attribution even after compression. Furthermore, their study highlights that while zero-shot detection of new video generators remains challenging, few-shot learning can effectively adapt detectors to unseen generators with minimal data. 

\subsubsection{Analysis.}
\noindent
$\clubsuit$ \textbf{Insights}.
Although video generation is still evolving, early detection methods already reveal several shared insights about LAIM-generated video artifacts. Most notably, LAIM-generated videos introduce temporal inconsistencies, such as unnatural motion patterns or frame prediction anomalies, which differ from spatial inconsistencies observed in synthetic images~\cite{ji2024distinguish, he2024exposing}. These motion related artifacts require models to incorporate both spatial and temporal reasoning, often through dual-branch or transformer based architectures. Additionally, empirical studies highlight that forensic traces in videos are fundamentally different from those in images, and detectors trained on image data perform poorly on synthetic videos~\cite{vahdati2024beyond}. Generalization remains a central challenge: video detectors often fail on unseen generators due to overfitting to known artifacts or reliance on overly constrained datasets~\cite{kamat2023revisiting}. 

\smallskip
\noindent
$\clubsuit$ \textbf{Recommendations.} To address current limitations, detection systems should exploit temporal dynamics and motion inconsistencies unique to synthetic videos. Dual-branch or transformer based models that integrate spatial and motion cues, enhanced by optical flow and temporal attention, have shown promise~\cite{ji2024distinguish, he2024exposing}. Since detectors trained on images often fail on video~\cite{vahdati2024beyond}, training on synthetic video data is crucial to capture video-specific forensic signals. Domain adaptation and few-shot fine-tuning can help generalize across generation methods. A good benchmark should cover diverse sources, compression levels, and durations to reflect real-world conditions. Incorporating frame-level localization and temporal consistency analysis will also improve interpretability and traceability.

\smallskip
\noindent
$\clubsuit$ \textbf{Social Media.}
Efforts such as PDID~\cite{walker2024merging} compile real-world political deepfake incidents, emphasizing that the deployment of generative AI in sociopolitical contexts requires not only detection but also content tracking and contextual analysis. In parallel, benchmarks like WildDeepfake~\cite{zi2020wilddeepfake} reveal that conventional artifact-based detection methods, effective in controlled settings, often fail against social media deepfakes characterized by compression, noise, and diverse manipulation techniques. Yu et al.\cite{yu2023augmented} propose AMSIM, a model that magnifies subtle spatiotemporal inconsistencies through global and multi-timescale local observations, and introduce adversarial data augmentation to simulate unseen social media degradations, improving robustness to wild conditions. Wu et al.\cite{wu2023interactive} develop ITSNet, an interactive two-stream network that enhances deepfake detection by jointly modeling spatiotemporal inconsistencies across RGB and frequency domains, achieving strong performance even under social media distortions and compression artifacts.

\subsection{Audio}\label{sec:Audio_det}
The research on the detection of LAIM-generated audio is also very limited. The reason is that visual and text content are more dominant and widespread in media and online platforms. This prevalence makes fake text/images more common and thus a higher priority for detection. However, as voice synthesis and manipulation technologies improve and become more accessible, the importance and focus on detecting fake audio are likely to increase.
\begin{figure}[t]
  \centering
  \includegraphics[width=1\linewidth]{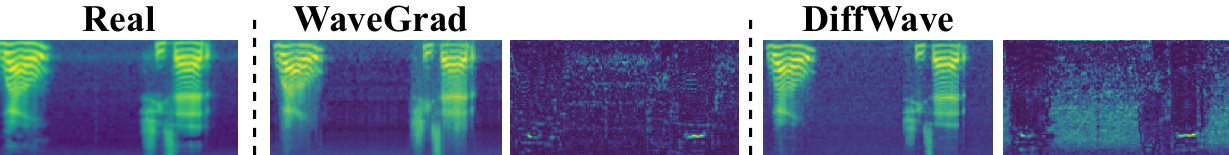}
  \vspace{-6mm}
  \caption{\small The artifacts introduced by DM-based neural vocoders (WaveGrad~\cite{chen2020wavegrad} and DiffWave~\cite{kong2020diffwave}) to a voice signal \cite{sun2023ai}. The differences in mel-spectrograms between real and generated ones are illustrated in the third and fifth columns.}
  \vspace{-4mm}
  \label{fig:audio_mel_spectrogram}
\end{figure}


\subsubsection{Pure Detection}\label{sec:audio_pure}

\smallskip
\noindent
$\clubsuit$ \textbf{Vocoder-based.} Sun et al. \cite{sun2023ai}  
focus on detecting unique artifacts generated by neural vocoders in audio signals. Visible artifacts introduced by different neural vocoder models can be observed in Fig.~\ref{fig:audio_mel_spectrogram}. The study employs a multi-task learning framework incorporating a binary-class RawNet2 \cite{tak2021end} model. This model uniquely shares a feature extractor with a vocoder identification module. By employing this shared structure, the feature extractor is effectively guided to concentrate on vocoder-specific artifacts. However, in their method, the detector may become overly specialized in detecting specific forgery technologies, hindering generalization to unseen vocoder artifacts.

\subsubsection{Beyond Detection}\label{sec:audio_beyond}
\smallskip
\noindent
$\clubsuit$ \textbf{Generalization.} Ren et al.\cite{ren2025improving} propose a disentanglement-based framework that targets domain-agnostic vocoder artifacts for generalization. Their model leverages a dual encoder architecture, separating content and artifact features, paired with multi-task and contrastive learning to classify domain-specific and domain-agnostic features. Their method improves detection across both seen and unseen vocoders. 

\subsubsection{Analysis} 
\noindent
$\clubsuit$ \textbf{Insights}.
Though research on detecting LAIM-generated audio is still emerging, it reveals key patterns distinguishing synthetic from human speech. Neural vocoders often leave detectable artifacts, such as unnatural harmonics or frequency discontinuities, in spectrotemporal features~\cite{sun2023ai}. However, detectors trained on specific vocoders often overfit, limiting generalization to unseen methods. To address this, recent studies propose disentangling content from vocoder-specific artifacts to capture domain-agnostic cues~\cite{ren2025improving}. Incorporating vocoder classification into multitask learning further encourages models to prioritize forensic over linguistic or speaker features.

\smallskip
\noindent
$\clubsuit$ \textbf{Recommendations.} Disentangling content from vocoder-specific artifacts, using dual-encoder architectures and contrastive learning, can improve generalization across unseen synthesis models~\cite{ren2025improving}. Multitask learning that jointly performs vocoder classification and authenticity detection has shown promise, as it encourages feature extractors to focus on generation specific anomalies while remaining robust to linguistic variation~\cite{sun2023ai}. Additionally, leveraging spectrogram-based representations enriched with perceptually motivated features (\eg, phase, pitch, or formant structure) can help detect subtle inconsistencies introduced by synthesis pipelines. Robust evaluation should further consider compression artifacts and language diversity, as both factors significantly affect detection reliability. Building comprehensive, multilingual, and vocoder-diverse benchmarks will be essential to support the development and evaluation of generalizable audio detection systems.

\smallskip
\noindent
$\clubsuit$ \textbf{Social Media.}
A benchmark conducted by Müller et al.~\cite{muller2022does} shows that detectors that perform well on curated datasets like ASVspoof fail dramatically when faced with real-world, social media processed audio, experiencing up to a tenfold increase in error rates. Social media introduces platform specific degradations such as compression, resampling, and added noise, which obscure the synthetic artifacts that most current detectors rely on. Models using fixed-length inputs and limited feature representations, such as melspectrograms, were particularly fragile, while raw waveform-based models showed relatively better, though still insufficient, resilience. 

\subsection{Multimodal}\label{sec:Multimodal_det}
Here, we categorize literature utilizing multimodal learning, which takes multiple modalities for detecting forgeries in single or multiple modalities. In particular, multimodal learning \cite{khalid2021evaluation} refers to an embodied learning situation that includes learning multiple data modalities such as text, image, video, and audio. 

\begin{figure}[t]
  \centering
  \includegraphics[width=1\linewidth]{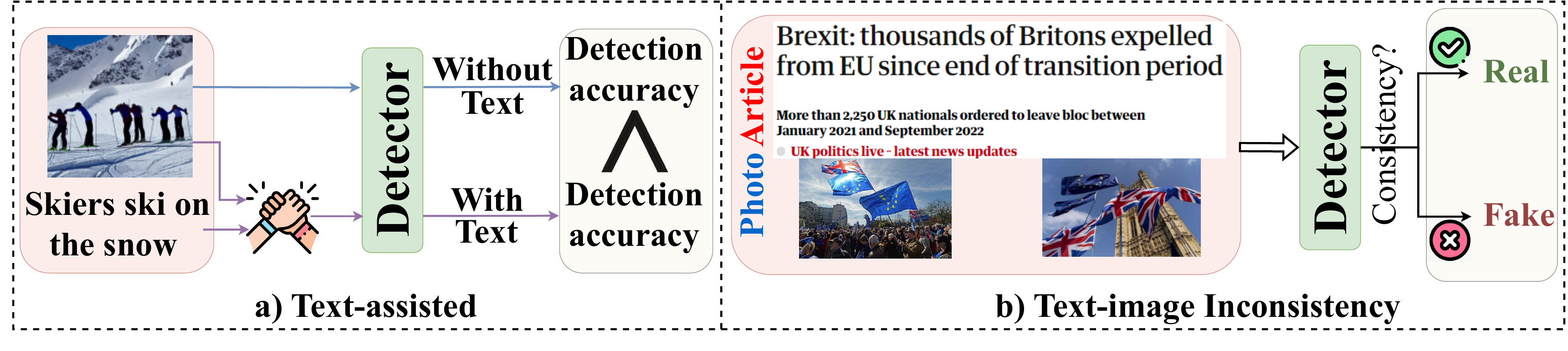}
  \vspace{-6mm}
  \caption{\small Illustrations of pure detection methodologies for LAIM-generated multimodal media.}
  \vspace{-4mm}
  \label{fig:multimodal_pure_overview}
\end{figure}

\subsubsection{Pure Detection}\label{sec:multimodal_pure}

\smallskip
\noindent
$\clubsuit$ \textbf{Prompt-guided.} 
This category includes methods that leverage prompts, instructions, or text-image pairs to guide the learning process of image or video forgery detection models. These approaches typically extract both visual and semantic features, using the language modality to improve performance in DM-generated media detection, as shown in Fig.~\ref{fig:multimodal_pure_overview} a).
Amoroso et al. \cite{amoroso2024parents} leverage the semantic content of textual descriptions alongside visual data. They introduce a contrastive-based disentangling strategy to analyze the role of the semantics of textual descriptions and low-level perceptual cues. The experiments are conducted on their proposed COCOFake dataset, 
where CLIP~\cite{radford2021learning} is employed as the backbone for feature extraction. These extracted features are then used to train a logistic regression model.
They find that the visual features extracted from the generated image still retain the semantic information of the original caption used to create it, which allows them to distinguish between natural images and the generated ones using only semantic cues while neglecting the perceptual ones.
Song et al.~\cite{song2024learning} propose MM-Det, a diffusion-generated video detector that learns a Multi-Modal Forgery Representation from large vision-language models via instruction tuning. Alongside the MMFR, the authors introduce an In-and-Across Frame Attention mechanism that captures spatial artifacts and temporal inconsistencies across frames. A dynamic fusion strategy adaptively weights both streams for final prediction.
Bohacek et al.~\cite{bohacek2024human} propose Human Action CLIPS, a detection method targeting AI-generated human motion videos. Instead of relying on low-level artifacts, their approach uses multi-modal semantic embeddings (\eg, CLIP, SigLIP) combined with lightweight classifiers. Evaluated on their proposed DeepAction dataset, which includes videos from seven AI models and real footage, their method achieves robust detection even after compression.

\smallskip
\noindent
$\clubsuit$ \textbf{Text-image Inconsistency.} This refers to a type of misinformation where the textual content does not align with or accurately represent the original meaning or intent of the associated image. This type of misinformation is commonly presented in the guise of news~\cite{luo2021newsclippings} to mislead the public, see Fig.~\ref{fig:multimodal_pure_overview} b). 
Tan et al.~\cite{tan2020detecting} create a {NeuralNews} dataset which contains articles with LLM-manipulated text and authentic images tailored for such task.
They further propose DIDAN (\textbf{D}etecting Cross-Modal \textbf{I}nconsistency to \textbf{D}efend \textbf{A}gainst \textbf{N}eural Fake News) framework that effectively leverages visual-semantic inconsistencies between article text, images, and captions, offering a novel approach to counteract neural fake news. 
Huang et al.~\cite{huang2023exposing} introduces D-TIIL (Diffusion-based Text-Image
Inconsistency Localization), which employs text-to-image diffusion models to localize semantic inconsistencies in text and image pairs.

\begin{figure*}[t]
    \centering
    \includegraphics[width=1\textwidth]{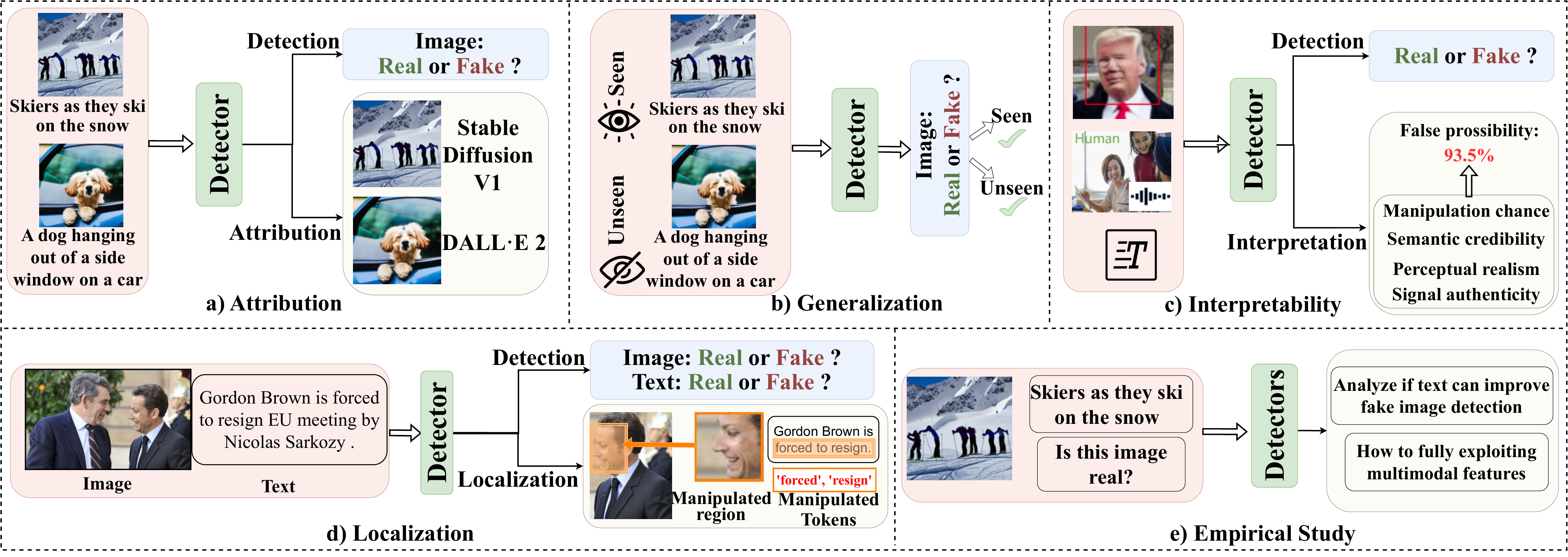}
    \vspace{-6mm}
    \caption{\small Illustrations of beyond detection methodologies for LAIM-generated multimodal media.}
    \label{fig:multimodal_beyond_overview}
    \vspace{-4mm}
\end{figure*}

\subsubsection{Beyond Detection}\label{sec:multimodal_beyond}

\smallskip
\noindent
$\clubsuit$ \textbf{Attribution}. This refers to research works that utilize a multimodal learning approach, combining prompt information with image features, to enhance the accuracy of attributing DM-generated images to their source model, shown in Fig.~\ref{fig:multimodal_beyond_overview} a). Sha et al. \cite{sha2023fake} build an image-only detector (ResNet18) and a hybrid detector (take advantage of CLIP’s \cite{radford2021learning} image and text encoders for feature extraction). Empirical results show that learning images with prompts achieves better performance than image-only attribution regardless of the dataset. Building upon this, Keita et al. \cite{keita2025fidavl} introduced FIDAVL, a multitask framework leveraging vision-language models and soft prompt-tuning to jointly detect and attribute synthetic images to their originating generative models. 

\smallskip
\noindent
$\clubsuit$ \textbf{Generalization}. This area of research focuses on developing a fake image detector that is guided by language, enhancing its ability to detect a wider range of new, unseen DM-generated images, shown in Fig.~\ref{fig:multimodal_beyond_overview} b). 


\raisebox{-0.3ex}{\scalebox{1.3}{\ding{192}}} \textbf{Prompt Tuning}. In this vein, Chang et al. \cite{chang2023antifakeprompt} employ soft prompt tuning on a language-image model to treat image detection as a visual question-answering problem. 
These prompt questions (\eg, ``Is this photo real?'') are then converted into vector representations, later are fed into Q-former and LLM with the image features extracted by the image encoder, therefore the DM-generated image detection problem is formulated as a visual question-answering problem, the LLM gives the detection output ``Yes'' for real images and ``No'' for fake images.

\raisebox{-0.3ex}{\scalebox{1.3}{\ding{193}}} \textbf{Contrastive Learning}. \cite{wu2023generalizable} adopt a language-guided contrastive learning approach, augmenting training images with textual labels (\eg, ``Real/Synthetic Photo'' and ``Real/Synthetic Painting'') for forensic feature extraction. 

\smallskip
\noindent
$\clubsuit$ \textbf{Interpretability}. It provides evidence with detection results by leveraging multimodal media, shown in Fig.~\ref{fig:multimodal_beyond_overview} c). 
To combat misinformation and emphasize the significance of scalability and explainability in misinformation detection, Xu et al.~\cite{xu2023combating} propose a conceptual multi-modal framework composed of 4 levels:
signal, perceptual, semantic, and human, for explainable multimodal misinformation detection.
Huang et al.~\cite{huang2024ffaa} introduce FFAA, a face forgery analysis assistant that redefines detection as an Open-World Face Forgery VQA task. It fine-tunes MLLMs (Multimodal Large Language Models) using hypothetical prompts and introduces a Multi-answer Intelligent Decision System to disambiguate difficult cases by reasoning over multiple candidate answers. FFAA not only provides fine-grained reasoning (\eg, forgery type, difficulty) but also outperforms prior detectors across six benchmark datasets. Chen et al.~\cite{chen2024textit} propose X²-DFD, a framework that identifies strong and weak forgery-related features—such as facial structure and blending artifacts—through a feature assessment module. It enhances detection by combining prompt-based fine-tuning with external detectors, and goes beyond image-level predictions by providing feature-grounded textual explanations, thereby improving the transparency of detection results.

\smallskip
\noindent
$\clubsuit$ \textbf{Localization}. Detection methods with localization aim to not only detect the authenticity of media content but also locate the manipulated content (\ie, image bounding boxes and text tokens). Unlike leveraging multimodal learning for enhanced single-modal forgery detection tasks, the methods we summarize here can detect and locate multi-modal media, shown in Fig.~\ref{fig:multimodal_beyond_overview} d). 

\raisebox{-0.3ex}{\scalebox{1.3}{\ding{192}}} \textbf{Spatial-based}. Shao et al,~\cite{shao2023detecting} initiate a significant development in the field of multimodal media manipulation detection with their dataset ``Detecting and Grounding MultiModal Media Manipulation" ($DGM^4$), where image-text pairs are manipulated by various approaches, to cope with the threat that enormous text fake news is generated or manipulated by LLMs. They further use contrastive learning to align the semantic content across different modalities (image and text) for detection and localization.
Subsequent research \cite{shao2023detectingbeyond, liu2023unified, wang2023exploiting} in this area has built upon this dataset.
Specifically, \cite{shao2023detectingbeyond} is an extension of \cite{shao2023detecting} by integrating Manipulation-Aware Contrastive Loss with Local View and constructing a more advanced model HAMMER++~\cite{shao2023detectingbeyond}, which further improves detection and localization performance.
Wang et al. \cite{wang2023exploiting} construct a simple and novel transformer-based framework for the same task. Their designed dual-branch cross-attention and decoupled fine-grained classifiers can effectively model cross-modal correlations and exploit modality-specific features, demonstrating superior performance compared to HAMMER~\cite{shao2023detecting} and UFAFormer\cite{liu2023unified} on $DGM^4$.

\raisebox{-0.3ex}{\scalebox{1.3}{\ding{193}}} \textbf{Frequency-based}. UFAFormer (a Unified Frequency-Assisted transFormer framework) \cite{liu2023unified} incorporates frequency domain to address the detection and localization problem. It simultaneously integrates and processes the detection and localization processes for manipulated faces, text, as well as image-text pair, simplifying the architecture in \cite{shao2023detecting} and facilitating the optimization process. 

\raisebox{-0.3ex}{\scalebox{1.3}{\ding{194}}} \textbf{MLLM-based}. While existing multimodal localization methods often rely on spatial or frequency alignment across image-text pairs,  recent advances leverage MLLMs (Multimodal Large Language Models) to enable localization with accompanying textual explanations. Xu et al.\cite{xu2024fakeshield} introduce FakeShield, a pioneering framework that combines MLLMs with pixel-level segmentation and language-based justification. It produces both textual descriptions and tampering masks using a Domain Tag-guided Explainable Forgery Detection Module and a SAM-based localization module. Liu et al.\cite{liu2024forgerygpt} present ForgeryGPT, which supports multi-turn dialogue for forgery analysis by integrating a Mask-Aware Forgery Extractor into a customized LLM. It employs an Object-agnostic Forgery Prompt and a Vocabulary-enhanced Vision Encoder to capture multi-scale forgery patterns and semantic-level inconsistencies, enabling precise localization and rich textual explanation.

\smallskip
\noindent
$\clubsuit$ \textbf{Empirical Study}. It provides insights into the feasibility of detecting generated images by leveraging multimodal media shown in Fig.~\ref{fig:multimodal_beyond_overview} e). 
\cite{coccomini2024detecting} shows that it is possible to detect the generated images using simple Multi-Layer Perceptrons (MLPs), starting from
features extracted by CLIP~\cite{radford2021learning} or traditional CNNs. They found that incorporating the associated textual information with the images rarely leads to significant improvement in detection results but that the type of subject depicted in the image can significantly impact performance. 
Similarly, ~\cite{khan2024clipping} explores the effectiveness of CLIP with four distinct transfer learning strategies, including fine-tuning, linear probing, prompt tuning, and training an adapter network. They reveal that prompt tuning strategy which leverages both the image and text components of CLIP consistently achieves better performance compared with baselines and SOTA methods (\eg, Corvi et al.~\cite{corvi2023detection}, Ojha et al.~\cite{ojha2023towards}
Papadopoulos et al.~\cite{papadopoulos2024verite} highlight that specific patterns and biases in multimodal misinformation detection benchmarks can result in biased or unimodal models outperforming their multimodal counterparts on a multimodal task. Therefore they create a ''VERification of Image-TExt pairs'' (VERITE) benchmark to address the unimodal bias in multimodal misinformation detection. 
Jia et al.~\cite{Jia_2024_CVPR} investigate the capabilities of multimodal large language models (LLMs) in DeepFake detection. They input fake and real images with different prompts that embody the instruction and request to the LLM (\eg, GPT4) to detect DeepFake faces. They find that LLMs can achieve a notable AUC score of 75\%. Despite their promise, these models face challenges in accurately recognizing authentic images and require sophisticated prompting techniques for optimal performance, all without depending on signal cues or traditional data-driven detection methods.

\subsubsection{Analysis}
\noindent
$\clubsuit$ \textbf{Insights}.
Multimodal detection methods capitalize on the interplay between visual, linguistic, and occasionally auditory signals to identify inconsistencies or artifacts that may not be evident within a single modality. A key insight is that semantic misalignment—such as mismatched captions or fabricated prompts—serves as a strong cross-modal signal for detection~\cite{tan2020detecting, amoroso2024parents}. Prompt-guided methods, contrastive frameworks, and visual question answering (VQA)-based systems leverage this to enhance both detection and attribution~\cite{chang2023antifakeprompt, song2024learning, sha2023fake}. However, empirical studies caution that naive use of language features often yields marginal gains unless modality interaction is explicitly modeled~\cite{coccomini2024detecting, papadopoulos2024verite}. Moreover, advances in MLLMs introduce new opportunities for interpretability and localization, enabling models to generate rationales and segment manipulated regions simultaneously~\cite{huang2024ffaa, xu2024fakeshield}. 

These developments underscore the growing feasibility of multimodal detection systems in practical settings. Recent works such as FakeShield~\cite{xu2024fakeshield}, ForgeryGPT~\cite{liu2024forgerygpt}, and FFAA~\cite{huang2024ffaa} demonstrate that MLLMs can be effectively fine-tuned or guided to perform both detection and explanation tasks. These systems support multi-turn reasoning, open-ended prompts, and produce interpretable outputs like textual justifications and pixel-level mask, making them adaptable for detecting complex, real-world manipulations. As such, the feasibility of deploying MLLM-based detectors is supported by strong empirical evidence, especially in applications where scalable, explainable, and cross-modal understanding is essential.
Despite this progress, challenges remain: multimodal detectors are susceptible to unimodal biases, often over-relying on dominant modalities, and performance varies significantly with image subject type or text domain~\cite{papadopoulos2024verite, khalid2021evaluation}. 

\smallskip
\noindent
$\clubsuit$ \textbf{Recommendations.} To improve the robustness and effectiveness of multimodal detection, several directions are promising. First, enhancing cross-modal alignment through fine-grained contrastive learning can better capture subtle inconsistencies across modalities. To mitigate unimodal bias, training with balanced and semantically diverse data, alongside attention regularization, ensures the model does not over-rely on a dominant modality. Soft prompt tuning and instruction-guided learning can further support generalization to unseen domains by leveraging linguistic context as a guiding signal. Integrating detection and localization within unified transformer-based frameworks improves interpretability and supports pixel-level forensics. Additionally, incorporating MLLMs enables explainable detection through natural language rationales, though care must be taken to avoid hallucinated explanations. Finally, hybrid architectures combining spatial, frequency, and semantic features, along with well-constructed benchmarks such as DGM\textsuperscript{4}~\cite{shao2023detecting} and MMFakeBench~\cite{liu2024mmfakebench}, are critical to evaluating detectors under realistic, cross-modal manipulation scenarios.

\smallskip
\noindent
$\clubsuit$ \textbf{Social Media.}
In addressing the challenge of detecting social-media-processed AI-generated multimodal content, MMFakeBench~\cite{liu2024mmfakebench} introduces MMD-Agent, a specialized detection model designed to tackle mixed-source misinformation. MMD-Agent decomposes the detection task into three complementary branches—textual, visual, and cross-modal inconsistency reasoning—leveraging large vision-language models (LVLMs) to handle the noisy, stylized, and semantically complex content commonly found on social platforms. By explicitly modeling reasoning across modalities and training on mixed-source forgeries, MMD-Agent achieves more robust performance under real-world distortions compared to traditional artifact-based models. The ILLUSION benchmark~\cite{thakralillusion} reveals significant performance degradation when detectors are exposed to social-media-sourced deepfakes through extensive benchmarking of state-of-the-art unimodal and multimodal models, highlighting critical gaps in cross-modal generalization, robustness to noise, and zero-shot generalizability. 

\section{Tools and Evaluation Metrics}\label{sec:ToolsCompetitionsMetrics}
This section reviews practical tools and standard evaluation metrics commonly used in detecting LAIM-generated content. 

\subsection{Tools}\label{sec:tools}
In this section, we present a selection of popular, user-friendly detection tools we have evaluated, see Table \ref{tab:my-table}.  Several tools discussed in the multimodal section offer support for image, video, and audio detection. In the modality-specific subsections (image, video, and audio), we focus on tools that are specifically designed for a single modality. 

\smallskip
\noindent
$\clubsuit$ \textbf{Text}.
In the domain of LLM-generated text detection tools, diverse instances illuminate specific strengths and drawbacks. 
Copyleaks~\cite{copyleaks} offers high accuracy across 30 languages but faces performance delays and lacks text exclusion options. ZeroGPT~\cite{zerogpt} is user-friendly but doesn't support file analysis. Though Winston AI\cite{winstonai} presents a free OCR (Optical Character Recognition)-powered alternative with an accessible interface, it needs enhancements in reporting and language support.
Simplicity and speed take center stage with Crossplag's\cite{crossplag},
but it is confined to English and lacks graphical elements. GLTR\cite{gltr} delivers in-depth, color-coded text analysis but is less intuitive for beginners and struggles with newer technologies. Brandwell~\cite{contentatscale_text} yields easy-to-interpret results but may exhibit inconsistencies in AI rewrites. Undetectable.ai \cite{undetectableai} supports various models but may discourage users due to pricing concerns. Illuminarty\cite{illuminarty_text} excels in detecting AI-generated texts but introduces complexity and slower processing for basic needs. Finally, Is it AI\cite{isitai_text} matches other tools in function but faces reliability issues, especially with short or non-English texts.

\begin{table*}[]
\centering
\caption{\small \textit{{Summary of existing popular detection tools for exposing LAIM-generated multimedia. The `Multi-platform' label in the Type column indicates that these tools are not limited to web applications but can also be used or deployed on other platforms, such as mobile apps or Chrome extensions.}} }
\vspace{-2mm}
\resizebox{\linewidth}{!}{%
\begin{tabular}{|c|c|c|c|c|c|c|c|}
\hline
\textbf{Modality} &
  \textbf{Tool} &
  \textbf{Company} &
  \textbf{Link} &
  \multicolumn{1}{l|}{\textbf{Reference}} &
  \textbf{Type} & \textbf{Open Source} & \textbf{Cost} \\ \hline

\rowcolor[HTML]{C8E4B2} 
  {\cellcolor[HTML]{FFFFFF}} &
  AI Content Detector &
  Copyleaks & 
  {\color[HTML]{0000EE}\href{https://copyleaks.com/ai-content-detector}{\color[HTML]{0000EE} { Link}}} &
  \cite{copyleaks} & Webapp \& API & \xmark & Limited free usage \\ \cline{2-8}
  
\rowcolor[HTML]{C8E4B2} 
  {\cellcolor[HTML]{FFFFFF}} &
  AI Content Detector, ChatGPT detector &
  ZeroGPT   &
  {\color[HTML]{0000EE}\href{  https://zerogpt.net/zerogpt-results}{\color[HTML]{0000EE} { Link}}} &
  \cite{zerogpt} & Webapp \& API & \xmark & Free usage\\ \cline{2-8}

\rowcolor[HTML]{C8E4B2} 
  {\cellcolor[HTML]{FFFFFF}} &
  AI Detector &
  GPTZero   &
  {\color[HTML]{0000EE}\href{  https://gptzero.me}{\color[HTML]{0000EE} { Link}}} &
  \cite{GPTzero2023} & Multi-platform & \xmark & Limited free usage\\ \cline{2-8}

\rowcolor[HTML]{C8E4B2} 
  {\cellcolor[HTML]{FFFFFF}} &
  AI Content Detector &
  Winston AI  &
  {\color[HTML]{0000EE}\href{  https://gowinston.ai/}{\color[HTML]{0000EE} { Link}}} &
  \cite{winstonai} & Webapp \& API & \xmark & Limited free usage\\ \cline{2-8}
\rowcolor[HTML]{C8E4B2} 
 {\cellcolor[HTML]{FFFFFF}}  &
  AI Content Detector &
  Crossplag  &
  {\color[HTML]{0000EE}\href{  https://crossplag.com/ai-content-detector/}{\color[HTML]{0000EE} { Link}}} &
  \cite{crossplag} & Webapp & \xmark & Limited free usage \\ \cline{2-8}
\rowcolor[HTML]{C8E4B2} 
 {\cellcolor[HTML]{FFFFFF}}  &
  Giant Language model Test Room &
  GLTR  &
  {\color[HTML]{0000EE}\href{  http://gltr.io/}{\color[HTML]{0000EE} { Link}}} &
  \cite{gltr} & Webapp & \cmark & Free usage\\ \cline{2-8}

\rowcolor[HTML]{C8E4B2} 
 {\cellcolor[HTML]{FFFFFF}}  &
  The AI Detector &
  Brandwell &
  {\color[HTML]{0000EE}\href{  https://contentatscale.ai/ai-content-detector/}{\color[HTML]{0000EE} { Link}}} &
  \cite{contentatscale_text} & Webapp & \xmark & Limited free usage  \\ \cline{2-8}

\rowcolor[HTML]{C8E4B2} 
 {\cellcolor[HTML]{FFFFFF}}  &
  AI Checker &
  Originality ai  &
  {\color[HTML]{0000EE}\href{  https://originality.ai/}{\color[HTML]{0000EE} { Link}}} &
  \cite{originalityai} & Webapp \& API  & \xmark & Limited free usage\\ \cline{2-8}
  
\rowcolor[HTML]{C8E4B2} 
 {\cellcolor[HTML]{FFFFFF}}  &
  Advanced AI Detector and Humanizer &
  Undetectable ai &
  {\color[HTML]{0000EE}\href{  https://undetectable.ai/}{\color[HTML]{0000EE} { Link}}} &
  \cite{undetectableai} & Webapp \& API & \xmark & Limited free usage\\ \cline{2-8}

\rowcolor[HTML]{C8E4B2} 
 {\cellcolor[HTML]{FFFFFF}}  &
  AI Content Detector &
  Writer &
  {\color[HTML]{0000EE}\href{  https://writer.com/ai-content-detector/}{\color[HTML]{0000EE} { Link}}} &
  \cite{writer} & Webapp \& API & \xmark & Limited free usage\\ \cline{2-8}

\rowcolor[HTML]{C8E4B2} 
 {\cellcolor[HTML]{FFFFFF}}  &
  AI Content Detector &
  Conch &
  {\color[HTML]{0000EE} \href{  https://www.getconch.ai/}{\color[HTML]{0000EE} { Link}}} &
  \cite{conch} & Webapp & \xmark & Limited free usage\\ \cline{2-8}
  
\rowcolor[HTML]{C8E4B2} 
 {\cellcolor[HTML]{FFFFFF}}  &
  Illuminarty Text &
  Illuminarty  &
  {\color[HTML]{0000EE}\href{ https://app.illuminarty.ai/\#/text}{\color[HTML]{0000EE} { Link}}} &
  \cite{illuminarty_text} & Webapp \& API  & \xmark & Limited free usage\\ \cline{2-8}
\rowcolor[HTML]{C8E4B2} 
\multirow{-12}{*}{\cellcolor[HTML]{FFFFFF}\textbf{Text}} &
  AI-Generated Text Detector &
  Is it AI  &
  {\color[HTML]{0000EE}\href{ https://isitai.com/ai-text-detector/}{\color[HTML]{0000EE} { Link}}} &
  \cite{isitai_text} & Webapp \& API & \xmark & Limited free usage\\ \hline

\rowcolor[HTML]{ffcfd2} 
 {\cellcolor[HTML]{FFFFFF}}  &
  Liveness Detection, Facial Recognition &
  Incode &
  {\color[HTML]{0000EE}\href{https://incode.com/use-cases/id-verification/}{\color[HTML]{0000EE} { Link}}} &
  \cite{incode} & Multi-platform & \xmark & Paid\\ \cline{2-8}

\rowcolor[HTML]{ffcfd2} 
 {\cellcolor[HTML]{FFFFFF}}  &
  AI or Not image &
  AI or Not &
  {\color[HTML]{0000EE}\href{  https://www.aiornot.com/}{\color[HTML]{0000EE} { Link}}} &
  \cite{aiornot} & Webapp \& API & \xmark & Limited free usage\\ \cline{2-8}

\rowcolor[HTML]{ffcfd2} 
 {\cellcolor[HTML]{FFFFFF}}  &
  AI-Generated Image Detector &
  Is it AI  &
  {\color[HTML]{0000EE}\href{  https://isitai.com/ai-image-detector/}{\color[HTML]{0000EE} { Link}}} &
  \cite{isitai_image} & Webapp \& API  & \xmark & Limited free usage\\ \cline{2-8}
\rowcolor[HTML]{ffcfd2} 
 {\cellcolor[HTML]{FFFFFF}}  &
  Illuminarty Image &
  Illuminarty  &
  {\color[HTML]{0000EE}\href{  https://app.illuminarty.ai/\#/image}{\color[HTML]{0000EE} { Link}}} &
  \cite{illuminarty_image} & Webapp \& API  & \xmark & Limited free usage\\ \cline{2-8}

\rowcolor[HTML]{ffcfd2} 
 {\cellcolor[HTML]{FFFFFF}}  &
  AI Image Detector &
  Undetectable ai  &
  {\color[HTML]{0000EE}\href{https://undetectable.ai/ai-image-detector}{\color[HTML]{0000EE} { Link}}} &
  \cite{undetectableai} & Webapp \& API & \xmark & Limited free usage \\ \cline{2-8}

\rowcolor[HTML]{ffcfd2} 
 {\cellcolor[HTML]{FFFFFF}}  &
  SynthID &
  Google  &
  {\color[HTML]{0000EE}\href{  https://deepmind.google/discover/blog/identifying-ai-generated-images-with-synthid/}{\color[HTML]{0000EE} {\color[HTML]{0000EE} { Link}}}} &
  \cite{synthid} & Webapp & \xmark & Limited free usage\\ \cline{2-8}

\rowcolor[HTML]{ffcfd2} 
 {\cellcolor[HTML]{FFFFFF}}  &
  The AI image detector &
  Winston &
  {\color[HTML]{0000EE}\href{https://gowinston.ai/ai-image-detector/}{\color[HTML]{0000EE} { Link}}} &
  \cite{winstonai} & Webapp \& API & \xmark &
 Limited free usage\\ \cline{2-8}

\rowcolor[HTML]{ffcfd2} 
 \multirow{-7}{*}{\cellcolor[HTML]{FFFFFF}\textbf{Image}}  &
  Advanced AI Image Detector &
  Brandwell  &
  {\color[HTML]{0000EE}\href{  https://contentatscale.ai/ai-image-detector/}{\color[HTML]{0000EE} { Link}}} &
  \cite{contentatscale_image} & Webapp & \xmark &
 Limited free usage\\ \hline
\rowcolor[HTML]{fec89a} 
 {\cellcolor[HTML]{FFFFFF}}  &
  Deepware Scanner &
  Deepware  &
  {\cellcolor[HTML]{fec89a}\href{https://scanner.deepware.ai/}{\color[HTML]{0000EE} { Link}}} &
  \cite{deepware} & Webapp \& API & \cmark &
 Free usage\\ \cline{2-8}

\rowcolor[HTML]{fec89a} 
 \multirow{-2}{*}{\cellcolor[HTML]{FFFFFF}\textbf{Video}}  &
  Attestiv Deepfake Video Detection &
  Attestiv  &
  {\color[HTML]{0000EE}\href{https://attestiv.com/deepfake-video-detection-software/}{\color[HTML]{0000EE} { Link}}} &
  \cite{attestiv} & Webapp \& API  & \xmark &
 Limited free usage\\ \hline

\rowcolor[HTML]{C3ACD0} 
 {\cellcolor[HTML]{FFFFFF}}  &
  Pulse Inspect &
  Pindrop  &
  {\color[HTML]{0000EE}\href{https://www.pindrop.com}{\color[HTML]{0000EE} { Link}}} &
  \cite{pindrop} & Multi-platform & \xmark &
 Paid\\ \cline{2-8}

\rowcolor[HTML]{C3ACD0} 
  {\cellcolor[HTML]{FFFFFF}}  &
  AI Voice Detector
 &
  AI Voice Detector  &
  {\color[HTML]{0000EE}\href{https://aivoicedetector.com/}{\color[HTML]{0000EE} { Link}}} &
  \cite{aivoice} & Webapp \& API & \xmark &
 Limited free usage\\ \cline{2-8}

 \rowcolor[HTML]{C3ACD0} 
  {\cellcolor[HTML]{FFFFFF}}  &
  AI Speech Classifier
 &
  ElevenLabs  &
  {\color[HTML]{0000EE}\href{https://elevenlabs.io/ai-speech-classifier}{\color[HTML]{0000EE} { Link}}} &
  \cite{elevenlabs} & Webapp \& API & \xmark &
 Limited free usage\\ \cline{2-8}

 \rowcolor[HTML]{C3ACD0} 
 \multirow{-3}{*}{\cellcolor[HTML]{FFFFFF}\textbf{Audio}}   &
  AI or Not audio
 &
  AI or Not  &
  {\color[HTML]{0000EE}\href{https://www.aiornot.com/}{\color[HTML]{0000EE} { Link}}} &
  \cite{aiornot} & Multi-platform  & \xmark &
 Limited free usage\\ \hline

\rowcolor[HTML]{f3d5b5} 
 {\cellcolor[HTML]{FFFFFF}}  &
  Video, Image, and Audio Detector &
  Deep Media  &
  {\color[HTML]{0000EE}\href{https://deepmedia.ai/product}{\color[HTML]{0000EE} { Link}}} &
  \cite{deepmedia} & Multi-platform & \xmark &
 Limited free usage\\ \cline{2-8}

\rowcolor[HTML]{f3d5b5} 
 {\cellcolor[HTML]{FFFFFF}}  &
  Deepfake Detection &
  Sensity AI  &
  {\color[HTML]{0000EE}\href{https://sensity.ai/}{\color[HTML]{0000EE} { Link}}} &
  \cite{sensityai} & Multi-platform & \xmark &
 Paid\\ \cline{2-8}


\rowcolor[HTML]{f3d5b5} 
 {\cellcolor[HTML]{FFFFFF}}  &
 Hive AI's Deepfake Detection API &
 Hive AI &
 {\color[HTML]{0000EE}\href{https://thehive.ai/blog/spot-deepfakes-with-hives-new-deepfake-detection-api}{\color[HTML]{0000EE}{Link}}} &
 \cite{hiveapi} &
 API &
 \xmark &
 Limited free usage \\ \cline{2-8}

\rowcolor[HTML]{f3d5b5} 
 {\cellcolor[HTML]{FFFFFF}}  &
 Resemble Detect &
 Resemble AI &
 {\color[HTML]{0000EE}\href{https://www.resemble.ai/detect/}{\color[HTML]{0000EE}{Link}}} &
 \cite{resembledetect} &
 Webapp \& API &
 \xmark &
 Limited free usage \\ \cline{2-8}

\rowcolor[HTML]{f3d5b5} 
 {\cellcolor[HTML]{FFFFFF}}  &
 DuckDuckGoose AI (Phocus) &
 DuckDuckGoose AI &
 {\color[HTML]{0000EE}\href{https://www.duckduckgoose.ai/phocus}{\color[HTML]{0000EE}{Link}}} &
 \cite{duckduckgoose} &
 Webapp &
 \xmark &
 Paid \\ \cline{2-8}

\rowcolor[HTML]{f3d5b5} 
 {\cellcolor[HTML]{FFFFFF}}  &
 Sentinel &
 Sentinel &
 {\color[HTML]{0000EE}\href{https://thesentinel.ai/}{\color[HTML]{0000EE}{Link}}} &
 \cite{sentinel} &
 Webapp \& API &
 \xmark &
 Paid \\ \cline{2-8}


\rowcolor[HTML]{f3d5b5} 
 {\cellcolor[HTML]{FFFFFF}}  &
 Deepfake Detector &
 Deepfake Detector &
 {\color[HTML]{0000EE}\href{https://deepfakedetector.ai/}{\color[HTML]{0000EE}Link}} &
 \cite{aivoicedetector} &
 Multi-platform &
 \xmark &
 Limited free usage \\ \cline{2-8}

 \rowcolor[HTML]{f3d5b5} 
 {\cellcolor[HTML]{FFFFFF}}  &
 DeepFake-o-meter &
 U of Buffalo & 
 {\color[HTML]{0000EE}\href{https://zinc.cse.buffalo.edu/ubmdfl/deep-o-meter/landing_page}{\color[HTML]{0000EE}Link}} &
 \cite{deepfakeometerv2} &
 Webapp &
 \cmark &
 Free usage \\ \cline{2-8}

\rowcolor[HTML]{f3d5b5} 
 {\cellcolor[HTML]{FFFFFF}}  &
 BioID &
 BioID &
 {\color[HTML]{0000EE}\href{https://www.bioid.com}{\color[HTML]{0000EE}Link}} &
 \cite{bioid} &
 Webapp \& API &
 \xmark &
 Limited free usage \\ \cline{2-8}

\rowcolor[HTML]{f3d5b5} 
 {\cellcolor[HTML]{FFFFFF}}  &
  Get Real Protect &
  Get Real  &
  {\color[HTML]{0000EE}\href{https://www.getrealsecurity.com/solutions}{\color[HTML]{0000EE} { Link}}} &
  \cite{getrealsecurity} & Multi-platform & \xmark &
 Paid \\ \cline{2-8}

\rowcolor[HTML]{f3d5b5}
 \multirow{-14}{*}{\cellcolor[HTML]{FFFFFF}\textbf{Multi-modal}}  &
  Reality Defender &
  Reality Defender  &
  {\color[HTML]{0000EE}\href{https://www.realitydefender.com/technology}{\color[HTML]{0000EE} { Link}}} &
  \cite{realityDefender} & Multi-platform & \xmark &
 Paid\\ \hline

\end{tabular}%
}

\label{tab:my-table}
\end{table*}

\begin{table*}[t!]
\centering
\caption{\small \textit{{Summary of evaluation metrics used for detecting LAIM-generated multimedia.}} }
\vspace{-2mm}
\scalebox{0.72}{
\begin{tabular}{|c|c|c|l|}
\hline
Modality                    & Task                                   & Metric                    & Description                                                                                                                                                                                                                         \\ \hline
\multirow{7}{*}{Common}     & \multirow{7}{*}{Binary Classification} & \cellcolor[HTML]{e0e1dd}AUC~\cite{macko2023multitude}                       & \cellcolor[HTML]{e0e1dd}Area under the ROC curve, measures discriminative ability.                                                                                                                                                                          \\ \cline{3-4} 
                            &                                        & \cellcolor[HTML]{e0e1dd}F1-score~\cite{koike2023outfox}                        & \cellcolor[HTML]{e0e1dd}Harmonic mean of precision and recall.                                                                                                                                                                                              \\ \cline{3-4} 
                            &                                        & \cellcolor[HTML]{e0e1dd}FPR~\cite{xu2023generalization}                       & \cellcolor[HTML]{e0e1dd}Proportion of real instance incorrectly classified as generated.                                                                                                                                                                    \\ \cline{3-4} 
                            &                                        & \cellcolor[HTML]{e0e1dd}Accuracy~\cite{liu2024detectability}                  & \cellcolor[HTML]{e0e1dd}Proportion of correctly classified instances.                                                                                                                                                                                       \\ \cline{3-4} 
                            &                                        & \cellcolor[HTML]{e0e1dd}Precision~\cite{shao2023detecting}                 & \cellcolor[HTML]{e0e1dd}Proportion of correctly identified generated instances among all identified generated instances.                                                                                                                                    \\ \cline{3-4} 
                            &                                        & \cellcolor[HTML]{e0e1dd}Recall~\cite{fagni2021tweepfake}                    & \cellcolor[HTML]{e0e1dd}Proportion of correctly identified generated instances among all actual generated instances.                                                                                                                                        \\ \cline{3-4} 
                            &                                        & \cellcolor[HTML]{e0e1dd}Average Precision~\cite{tantaru2023weakly}         & \cellcolor[HTML]{e0e1dd}The area under the precision-recall curve, summarizing precision over all recall levels as the threshold varies                                                                                                                     \\ \hline
 Text                        & Binary Classification                  & \cellcolor[HTML]{C8E4B2}AvgRec~\cite{li2023deepfake}                    & \cellcolor[HTML]{C8E4B2}Average of recall on human-written and LLM-generated texts.                                                                                                                                                                         \\ \hline
\multirow{9}{*}{Image}     & \multirow{4}{*}{Localization}          &   \cellcolor[HTML]{ffcfd2}IoU (mIoU)~\cite{tantaru2023weakly}                & \cellcolor[HTML]{ffcfd2}Intersection over Union (mean IoU).                                                                                                                                                                                                 \\ \cline{3-4} 
                            &                                        & \cellcolor[HTML]{ffcfd2}Pixel-level AUC~\cite{guo2023hierarchical}           & \cellcolor[HTML]{ffcfd2}Area Under the ROC Curve computed at the pixel level.                                                                                                                                                                               \\ \cline{3-4} 
                            &                                        & \cellcolor[HTML]{ffcfd2}Pixel-level F1~\cite{guo2023hierarchical}            & \cellcolor[HTML]{ffcfd2}Harmonic mean of pixel-wise precision and recall.                                                                                                                                                                                   \\ \cline{3-4} 
                            &                                        & \cellcolor[HTML]{ffcfd2}PBCA~\cite{tantaru2023weakly}                      & \cellcolor[HTML]{ffcfd2}Pixel-wise Binary Classification Accuracy                                                                                                                                                                                           \\ \cline{2-4} 
                            & \multirow{5}{*}{Fairness}              & \cellcolor[HTML]{ffcfd2}Demographic Parity $F_{DP}$~\cite{lin2024ai}        & \cellcolor[HTML]{ffcfd2}Measures the maximum difference in prediction rates across all demographic groups.                                                                                                                                                  \\ \cline{3-4} 
                            &                                        & \cellcolor[HTML]{ffcfd2}Max Equalized Odds $F_{MEO}$~\cite{lin2024ai}       & \cellcolor[HTML]{ffcfd2}Captures the largest disparity in prediction outcomes (TPR or FPR) between any two demographic groups.                                                                                                                              \\ \cline{3-4} 
                            &                                        & \cellcolor[HTML]{ffcfd2}Equal Odds $F_{EO}$~\cite{lin2024ai}               &\cellcolor[HTML]{ffcfd2}Measures the disparity in TPR and FPR between each subgroup and the overall population.                                                                                                                                             \\ \cline{3-4} 
                            &                                        & \cellcolor[HTML]{ffcfd2}Overall Accuracy Equality $F_{OAE}$~\cite{lin2024ai} & \cellcolor[HTML]{ffcfd2}Measures the maximum accuracy gap across all demographic groups.                                                                                                                                                                    \\ \cline{3-4} 
                            &                                        & \cellcolor[HTML]{ffcfd2}Individual Fairness $F_{IND}$~\cite{lin2024ai}      & \cellcolor[HTML]{ffcfd2}Measures if similar individuals receive similar model outputs.                                                                                                                        \\ \hline
Video                       & Binary Classification                  & \cellcolor[HTML]{fec89a}Video-level AUC~\cite{song2024learning}           & \cellcolor[HTML]{fec89a}Area Under the ROC Curve computed at the video level.                                                                                                                                                                               \\ \hline
\multirow{2}{*}{Audio}      & \multirow{2}{*}{Binary Classification} &  \cellcolor[HTML]{C3ACD0}EER~\cite{sun2023ai}                       &  \cellcolor[HTML]{C3ACD0}Equal Error Rate. The point on the ROC curve where the false positive rate (FPR) equals the false negative rate (FNR).                                                                                                              \\ \cline{3-4} 
                            &                                        & \cellcolor[HTML]{C3ACD0}t-DCF~\cite{ren2025improving}                     & \cellcolor[HTML]{C3ACD0}Tandem Detection Cost Function. Evaluate the trade-off between detection errors and the cost of those errors.                                                                                                                       \\ \hline
\multirow{6}{*}{Multimodal} & Image Grounding                        & \cellcolor[HTML]{f3d5b5}mIoU~\cite{guo2023hierarchical}                      & \cellcolor[HTML]{f3d5b5}Measures average overlap between predicted and true bounding boxes.                                                                                                                                                                 \\ \cline{2-4} 
                            & \multirow{3}{*}{Text Grounding}        & \cellcolor[HTML]{f3d5b5}Token-level Precision~\cite{shao2023detecting}     & \cellcolor[HTML]{f3d5b5}Percentage of correctly predicted manipulated tokens among those predicted as manipulated.                                                                                                                                          \\ \cline{3-4} 
                            &                                        & \cellcolor[HTML]{f3d5b5}Token-level Recall~\cite{shao2023detecting}        & \cellcolor[HTML]{f3d5b5}Percentage of correctly predicted manipulated tokens among all ground truth manipulated tokens.                                                                                                                                     \\ \cline{3-4} 
                            &                                        & \cellcolor[HTML]{f3d5b5}Token-level F1~\cite{shao2023detecting}            & \cellcolor[HTML]{f3d5b5}Harmonic mean of precision and recall for manipulated token detection.                                                                                                                                                              \\ \cline{2-4} 
                            & \multirow{2}{*}{Text Explanation}      & \cellcolor[HTML]{f3d5b5}CSS~\cite{xu2024fakeshield}                       & \cellcolor[HTML]{f3d5b5}\begin{tabular}[c]{@{}l@{}}Cosine Semantic Similarity. Assess the similarity between the predicted text and ground truth text\\  by calculating the cosine similarity between their high-dimensional semantic vectors.\end{tabular} \\ \cline{3-4} 
                            &                                        & \cellcolor[HTML]{f3d5b5}ROUGE~\cite{liu2024forgerygpt}                     & \cellcolor[HTML]{f3d5b5}Recall-Oriented Understudy for Gisting Evaluation. Assess the overall accuracy of the generated text explanations.                                                                                                                  \\ \hline
\end{tabular}
}
\label{tab:metric}
\end{table*}

\smallskip
\noindent
$\clubsuit$ \textbf{Image}.
In evaluating tools for detecting images generated by DMs, 
AI or Not \cite{aiornot} demonstrated fast execution while identifying fake images and supported diverse formats. However, it cannot provide stable results.
In addition, Is it AI\cite{isitai_image} and Brandwell~\cite{contentatscale_image} offer a user-friendly interface with limitations in detecting certain content types. 
Illuminarty\cite{illuminarty_image} provides in-depth analysis but faces usability complexities, and Huggingface\cite{huggingface_image} excelled in accessibility with limitations in advanced analysis.
 Google has developed SynthID\cite{synthid}, specifically designed for images generated by Imagen\cite{imagen}.
\smallskip
\noindent
$\clubsuit$ \textbf{Video}.
Deepware Scanner~\cite{deepware} is a web-based platform designed for quick and accessible deepfake analysis. It allows users to link videos, particularly from platforms like YouTube, Twitter, and Facebook, and provides a probability score indicating the likelihood of manipulation. Aimed at general users such as journalists or content moderators, it prioritizes ease of use over technical depth. 
Attestiv~\cite{attestiv} targets enterprise use cases with an AI-driven framework offering forensic media analysis, manipulation localization, and verification via digital fingerprinting. It provides a detailed Suspicion Score and supports API integration, making it suitable for industries like insurance, finance, and cybersecurity. While Deepware emphasizes usability, Attestiv offers greater analytical depth and system-level scalability.

\smallskip
\noindent
$\clubsuit$ \textbf{Audio}. Commercial tools for detecting AI-generated audio are newer and often packaged as enterprise fraud solutions. Pindrop’s Pulse Inspect is one prominent example: it is a web-based tool for detecting AI-generated speech in any digital audio or video file with what it claims is a significantly high degree of accuracy: 99\%~\cite{pindrop_deepfake_detection_2025}. AI Voice Detector~\cite{aivoice} lets users upload audio files and analyzes them to ascertain the likelihood of AI generation. Key features include background noise reduction to enhance detection accuracy and a browser extension for real-time analysis of audio content on platforms like YouTube and Twitter. The tool is optimized for short audio clips and may struggle with longer recordings. AI or Not audio detector~\cite{aiornot} allows users to upload content or provide URLs, analyzing the input to determine the likelihood of AI generation.  ElevenLabs AI Speech Classifier~\cite{elevenlabs} provides a mechanism to verify the origin of synthetic speech, but it can only detect its own generated voices (audio clip created using ElevenLabs technology).

\smallskip
\noindent
$\clubsuit$ \textbf{Multi-modal}. As multimodal deepfake threats escalate across industries, many organizations turn first to enterprise-grade, subscription-based platforms. Solutions like Sensity AI \cite{sensityai} and Reality Defender\cite{realityDefender} build on large, multimodal neural ensembles to scan video, image, audio, and even metadata at scale, trading transparency for robust protection under custom-priced plans. DuckDuckGoose AI (Phocus)\cite{duckduckgoose} and Sentinel\cite{sentinel} extend similar capabilities via paid web interfaces and per-device licensing, while Get Real Protect \cite{getrealsecurity} offers a cross-platform defense suite—each promising high-throughput detection but requiring significant investment and relying on proprietary models that may lag behind the latest adversarial attacks. 

At the other end of the spectrum, a wave of freemium and API-first services democratizes access to deepfake detection. Hive AI’s \cite{hiveapi} Deepfake Detection API and Resemble Detect \cite{resembledetect} let developers plug lightweight classification and voice-benchmarking endpoints into their own workflows, supplemented by free browser extensions or usage quotas. Attestiv \cite{attestiv} and Deepfake Detector \cite{aivoicedetector} flag suspicious video and audio using generative-AI context checks, while BioID \cite{bioid} zeroes in on facial authenticity via simple web and API calls. These offerings balance ease of integration and cost, inviting smaller teams and researchers to experiment without large upfront commitments—albeit with limited coverage and a few caveats.

Meanwhile, the open-source and academic communities offer transparency and reproducibility in detection research. Tools like Deepware \cite{deepware} Scanner provide a free web UI and SDK for spotting visual and auditory anomalies, and the University at Buffalo’s DeepFake-o-meter\cite{deepfakeometerv2} experiments evolve public web-based detectors with no licensing fees.

\subsection{Metrics}
Evaluation metrics can help quantify the performance and compare the effectiveness of different models.
In LAIM-generated media detection tasks, the metrics are mainly from classification scenarios, as we summarize in Table~\ref{tab:metric}.

\subsubsection{Common Metrics Across Modalities}
The \textbf{most commonly} used measures among different modalities include \textit{AUC} (Area Under the ROC Curve)~\cite{macko2023multitude,shao2023detecting,corvi2023detection,mandelli2022forensic,may2023comprehensive, yan2024df40, wu2024detectrl, dugan-etal-2024-raid}, \textit{F1-score} \cite{koike2023outfox,he2024mgtbench,guo2023close,macko2023multitude,uchendu2021turingbench}, \textit{FPR (False Positive Rate)}  \cite{chen2023gpt,macko2023multitude,xu2023generalization,liu2023unforgeable,jia2023autosplice}, \textit{Accuracy} \cite{uchendu2021turingbench,liu2024detectability,wang2023m4,verma2023ghostbuster,macko2023multitude, guillaro2023trufor}, \textit{Precision} \cite{wang2023m4,fagni2021tweepfake,rosati-2022-synscipass,bird2023cifake,shao2023detecting, wu2023llmdet}, \textit{Recall} \cite{bird2023cifake,shao2023detecting,rosati-2022-synscipass,fagni2021tweepfake,macko2023multitude}, \textit{Average Precision}~\cite{tantaru2023weakly, lin2024ai, cozzolino2023raising, macko2023multitude, zhang-etal-2024-machine}.

\subsubsection{Modality-Specific Metrics}
\noindent
$\clubsuit$ \textbf{Text-Specific Metrics}.
In the text modality, except for above common metrics, \cite{li2023deepfake, koike2023outfox} used \textit{AvgRec} (average recall) for binary classification, which is calculated by averaging the recall scores on human-written texts (HumanRec) and LAIM-generated texts (MachineRec).

\smallskip
\noindent
$\clubsuit$ \textbf{Image-Specific Metrics}. In the context of LAIM-generated image detection, specific evaluation metrics are employed for tasks such as localization and fairness. For forgery localization~\cite{guo2023hierarchical, tantaru2023weakly, zhang2023perceptual}, \textit{IoU} (Intersection over Union), or its averaged form (\textit{mIoU}), is the most widely used metric. It quantifies the degree of overlap between the predicted and ground truth masks, providing a spatially sensitive measure of localization accuracy. In addition to IoU, some studies~\cite{Li_2024_CVPR, guo2023hierarchical, guillaro2023trufor} have adopted pixel-level evaluation metrics such as \textit{pixel-level AUC} and \textit{pixel-level F1 score}, which assess discriminative performance and the balance between precision and recall at the pixel level. \textit{PBCA} (Pixel-wise Binary Classification Accuracy), which treats each pixel as an independent sample to measure classification accuracy, is also used in some works~\cite{guo2023hierarchical, tantaru2023weakly}. Fairness evaluation in LAIM-generated image detection involves assessing disparities in performance across demographic groups~\cite{lin2024ai}. Demographic Parity $F_{DP}$ measures group-level prediction rate differences. Max Equalized Odds $F_{MEO}$ captures the largest disparity in true or false positive rates between any two groups, while Equal Odds $F_{EO}$ quantifies disparities between subgroup and overall performance. Overall Accuracy Equality $F_{OAE}$ reports the maximum gap in accuracy among all demographic groups. Individual Fairness $F_{IND}$ evaluates whether similar individuals receive similar model outputs.

\smallskip
\noindent
$\clubsuit$ \textbf{Video-Specific Metrics}.
For video detection tasks, \textit{Video-level AUC} is used to evaluate model performance in works~\cite{cheng2024stacking, song2024learning}, where each video is treated as a single instance by averaging frame-level predictions to compute a single score for AUC calculation.

\smallskip
\noindent
$\clubsuit$ \textbf{Audio-Specific Metrics}.
In audio detection, \textit{EER} (Equal Error Rate) remains a common choice~\cite{sun2023ai, ren2025improving}. It is the location on a ROC curve where the false positive rate and false negative rate are equal. Generally, biometric systems have higher accuracy when the EER is lower.
Additionally, the \textit{t-DCF} (Tandem Detection Cost Function) was created for the ASVspoof challenge~\cite{nautsch2021asvspoof} to evaluate spoofing countermeasures (CM) in conjunction with Automatic Speaker Verification (ASV).
It is used in ~\cite{ren2025improving} to quantify the trade-off between different types of detection errors and the cost associated with them.

\smallskip
\noindent
$\clubsuit$ \textbf{Multimodal-Specific Metrics}.
In the multimodal setting, there are tasks associated with specific evaluation metrics tailored to their objectives. For instance, in~\cite{shao2023detecting, wang2023exploiting, liu2023unified}, the framework involves both image and text grounding, where the detector is required to localize manipulated regions by predicting bounding boxes on the image and identifying corresponding forged segments in the text. For image grounding, the \textit{mIoU} is used to measure the average overlap between predicted and true bounding boxes. In text grounding, considering the class imbalance scenario that manipulated tokens are much fewer than original tokens, \textit{Precision}, \textit{Recall}, and \textit{F1 Score} are used as evaluation metrics for the manipulated text token grounding.

More recently, researchers have begun using MLLMs (Multimodal Large Language Models), such as GPT-4o and LLaMA2, for LAIM-generated image detection and explanation tasks~\cite{xu2024fakeshield, liu2024forgerygpt}. These models go beyond binary classification by providing natural language explanations that justify whether an image is real or fake. For evaluating such explanation tasks, metrics like \textit{CSS} (Cosine Semantic Similarity) and \textit{ROUGE} (Recall-Oriented Understudy for Gisting Evaluation) are utilized in ~\cite{xu2024fakeshield} and ~\cite{liu2024forgerygpt}, respectively. CSS measures the semantic alignment between the generated and reference explanations using high-dimensional embeddings, while ROUGE~\cite{lin2004rouge} quantifies the overall accuracy of the generated explanations.

\section{Discussion}\label{sec:Discussion}
In this section, we summarize shared detection strategies and key insights across modalities. We further discuss both common and modality-specific challenges, with particular emphasis on real-world environments such as social media. Guided by our analysis, we outline promising research directions in a modality-specific manner that correspond to addressing these challenges.

\subsection{Shared Techniques and Cross-Modal Insights}
Despite differences in data type, detection of AI-generated content often relies on a similar toolkit of strategies.

\smallskip
\noindent
$\clubsuit$ \textbf{Statistical Anomaly Detection}. AI-generated content often exhibits characteristics that deviate from those of human-produced data. For instance, LAIM-generated text tends to be more objective and formal, less prone to biased or harmful language, and generally longer and more detailed than human-authored text~\cite{chen2023can, Gehrmann2019GLTRSD}, and artifacts appear in co-occurrence patterns~\cite{pu2022unraveling}. Similarly, LAIM-generated images frequently contain frequency domain artifacts (\eg, periodic noise from upsampling) that diverge from the statistical regularities of natural photographs~\cite{corvi2023detection, ricker2022towards}. In the audio domain, researchers similarly exploit spectrogram inconsistencies and vocoder-specific artifacts to distinguish synthetic speech from human speech~\cite{sun2023ai, ren2025improving}.
A recent work~\cite{cozzolino2024zero} draws direct inspiration from LLM-based text detectors like DetectGPT~\cite{mitchell2023detectgpt}, which evaluate the log-likelihood or "surprise" of a token sequence under the model’s learned distribution. Similarly, their zero-shot image detector (ZED) treats an image as "statistically surprising" if it deviates from the distribution modeled by a lossless encoder trained solely on real images. This analogy demonstrates how distributional surprise can serve as a modality-agnostic principle for synthetic content detection, manifesting as log-perplexity in text, and as coding cost gaps in images.
Across modalities, detection systems exploit these distributional anomalies by identifying statistical irregularities or generative artifacts. 

\smallskip
\noindent
$\clubsuit$ \textbf{On the Use of Model Fingerprints and Manifold Traces}. 
Recent advances across modalities reveal that generative models often exhibit statistically or geometrically structured biases in their outputs, which can be exploited through perturbation or reconstruction-based analysis. In text detection, a growing body of work has used \textbf{log-probability curvature}, the second-order change in log-likelihood under small input perturbations, as a robust indicator of LLM-generated text. For example, DetectGPT~\cite{mitchell2023detectgpt} observes that AI-generated text tends to lie on flatter regions of the model's learned probability manifold and proposes a curvature-based detection criterion. Later works improve efficiency using Bayesian surrogate models~\cite{deng2023efficient}, conditional sampling~\cite{bao2023fast}, and sentence-level token probability aggregation~\cite{wang2023seqxgpt}. 
Analogously, \textbf{diffusion fingerprinting} techniques in image detection leverage the structured nature of generative image manifolds. Methods such as DIRE~\cite{wang2023dire} and SeDID~\cite{ma2023exposing} measure how well a pre-trained diffusion model can reconstruct an input image. The hypothesis is that generated images sampled directly from the model's distribution exhibit lower reconstruction error than real images. This mirrors the principle of log-probability curvature: real data tends to fall off the generative manifold, leading to higher deviation under model-guided reconstruction. Brokman et al.~\cite{brokman2025manifold} formalize this intuition by defining curvature and gradient-based geometric criteria on the score function manifold of a diffusion model. These quantities, computed over perturbed versions of the image, serve as zero-shot indicators of generation.
The shared insight across modalities is that the local geometry of the model’s generative manifold, whether measured via reconstruction error or curvature of probability estimates, encodes telltale signs of synthetic content. These techniques, rooted in manifold analysis, suggest promising future directions for cross-modal detection frameworks that generalize by probing the generative "fit" of a given sample rather than relying on domain-specific artifacts alone.

\smallskip
\noindent
$\clubsuit$ \textbf{Detect Watermark and Embedded Signals}. A widely adopted detection oriented strategy across modalities is to detect imperceptible signals, known as watermarks, that are embedded into AI-generated content at generation time. In text, this typically involves subtly biasing token selection toward a secret subset of vocabulary based on a hidden key, producing statistical patterns detectable through lightweight hypothesis tests without compromising fluency~\cite{kirchenbauer2023watermark, kuditipudi2023robust, fernandez2023three}. In images, watermarking advances embed signatures directly into the latent representations of diffusion models, allowing detection even after transformations like cropping or compression~\cite{fernandez2023stable, zhang2024attack}.
Detecting watermarks provides a modality-agnostic and efficient means of verifying AI-generated content. Beyond binary classification, watermark-based methods also support model attribution, enabling the identification of the specific generative model responsible.

\smallskip
\noindent
$\clubsuit$ \textbf{Disentangled Representation Learning for Generalization}. One technique transferring across modalities is disentangled representation learning aimed at isolating AI-generated artifacts from the underlying content. In image deepfake detection, researchers found that naive classifiers often overfit onto content features rather than the forgery artifacts, hurting generalization to new data~\cite{zhang2020face, liang2022exploring}. To address this, Liang et al.~\cite{liang2022exploring} proposed a framework that disentangles image features into content and artifacts, training detectors solely on the latter. A similar approach has since been applied to LAIM-generated audio. Ren et al.~\cite{ren2025improving} propose a disentanglement framework for synthetic speech, using multitask and contrastive learning to isolate vocoder-agnostic artifacts. This improves generalization across voice vocoders, with only vocoder-agnostic artifact features used for final classification. Extending this idea to text, Bhattacharjee et al.~\cite{bhattacharjee2023conda} propose ConDA, a contrastive domain adaptation framework that separates generator specific patterns from task relevant features. By training on labeled and unlabeled data from different models, it learns domain-invariant representations that generalize to unseen text generators.

\smallskip
\noindent
$\clubsuit$ \textbf{Semantic Context and Reasoning}. The use of high-level reasoning was traditionally the realm of text (logic and consistency in narrative). Now, image and video detection are incorporating semantic reasoning (with the help of MLLMs)~\cite{huang2024ffaa, liu2024forgerygpt, xu2024fakeshield, huang2024sida, song2024learning}, effectively importing NLP-style consistency checks into visual analysis. Conversely, text detection is starting to use “world knowledge” (via models) to ask if a stated fact is real or a hallucination (which could hint at AI generation)~\cite{zhang2024knowhalu}. This blending of contextual, semantic analysis means the line between modalities is blurring: detecting a fake might involve simultaneously reading text, analyzing image pixels, and using background knowledge from a knowledge base.

\subsection{Challenges}

\subsubsection{Common Challenges Across Modalities}
\noindent
$\clubsuit$ \textbf{Generalization to New Generators}. A persistent challenge is the limited generalizability of detectors to content from novel large AI models~\cite{wu2024detectrl, yan2024df40, li2024sonar}. Generative models evolve rapidly, but many detectors overfit to artifacts of the training data and struggle to detect content generated by new generators.

\smallskip
\noindent
$\clubsuit$ \textbf{Robustness Against Adversarial Attacks and Post-Processing}. Detection methods, regardless of the modality they target, currently struggle with adversarial manipulations, paraphrasing, and common post-processing operations, such as compression, resizing, which often obscure or erase the subtle artifacts that detectors rely on. Detected text can be paraphrased or lightly edited to evade classifiers~\cite{wahle2022large, becker2023paraphrase}, and audio deepfakes can be passed through filters to mask telltale traces~\cite{li2024sonar}. Minor perturbations, a slight rotation or noise addition to an image can fool many detectors~\cite{ju2023glff, cocchi2023unveiling}. 

\smallskip
\noindent
$\clubsuit$ \textbf{Open-Domain and Social Media Complexity}. The detection of LAIM-generated content across modalities on open-domain platforms, such as social networks, presents significant challenges due to the informal, noisy, and heterogeneous nature of social media content~\cite{liu2024mmfakebench, huang2024sida, yan2024sanity}. For text, posts are brief, laden with slang, emojis, or hashtags, diverging from the formal prose that detectors are usually trained on. This domain mismatch means a model effective on well-written text may fail on a jargon-filled tweet. Likewise, user-generated images and videos often contain filters, edits, or montages; one post might be entirely LAIM-generated while another only has an LAIM-edited segment. Consequently, the complexity and variability of open-domain social media content pose obstacles for detectors trained on idealized, tightly controlled laboratory data~\cite{walker2024merging}, necessitating more robust and adaptable detection frameworks.

\smallskip
\noindent
$\clubsuit$ \textbf{Lack of Interpretability}. Across all modalities, most detection systems operate as black boxes, providing binary real/fake outputs without human understandable explanations. This lack of transparency makes it difficult for users, moderators, and policymakers to trust or validate detection outcomes, especially in high stakes contexts like content moderation, journalism, or forensic analysis. While some recent efforts (\eg, IPAD~\cite{chen2025ipad} for text and SIDA~\cite{huang2024sida} for image) have begun incorporating explainable outputs, such as generating inverse prompts to reveal likely AI intent or highlighting tampered regions in images, these methods still fall short of accurately reflecting the detector’s underlying decision making process or providing comprehensive, human interpretable justifications for their judgments.

\subsubsection{Modality-Specific Challenges}
$\clubsuit$ \textbf{Text-Specific Challenges}.
AI-generated text detection faces unique challenges in terms of generalization, robustness, and interpretability. A key issue in generalization is the limited cross-lingual and domain adaptability of current detectors. Many detectors are trained primarily on English datasets, making them less effective on non-English texts~\cite{macko2023multitude}. Similarly, detectors trained on domain-specific datasets, such as academic articles, may struggle when applied to other domains like social media posts, news articles~\cite{li2023deepfake}.
From a robustness standpoint, paraphrasing attacks remain a major vulnerability~\cite{dugan-etal-2024-raid}. Light edits, such as synonym replacement, sentence restructuring, or rewording via another LLM, can obscure the statistical patterns or watermark signatures that detectors rely on, allowing LAIM-generated content to bypass detection even when its meaning remains unchanged.

Attribution poses additional difficulties. Current attribution models often suffer from domain dependence, showing sharp performance drops on unseen styles or topics~\cite{venkatraman2023gpt, uchendu2020authorship}. Furthermore, fine-tuning and sampling variability introduce ambiguity: the same underlying model can produce stylistically diverse outputs based on fine-tuning or sampling parameters, making attribution even harder~\cite{munir2021through} The growing number of LLM variants, both open and proprietary, also creates scalability challenges\cite{uchendu2021turingbench, uchendu2024topformer}, and representation overlap across models reduces discriminability\cite{ai-etal-2022-whodunit}.

Moreover, interpretability is notably lacking in most text detection systems. Many models offer binary predictions without articulating the reasoning behind their decisions, which limits their transparency and trustworthiness. Although recent approaches like IPAD~\cite{chen2025ipad} attempt to improve interpretability by inferring a plausible inverse prompt that explains the text's generative origin, such methods are still in early stages and not widely adopted.

\underline{Social Media.} Models trained on formal English text, such as news articles, often struggle with informal, multilingual, or domain-shifted content like social media posts or low-resource languages, revealing poor performance across diverse linguistic and contextual domains~\cite{li2023deepfake, wu2024detectrl}. The informal, brief, and multilingual nature of social media posts, often laden with slang, emojis, hashtags, or internet jargon, exacerbates this domain shift, rendering detectors trained on formal or long form data less effective on platforms like X or TikTok~\cite{macko2024multisocial}. Recent benchmarks like MultiSocial~\cite{macko2024multisocial} demonstrate that fine-tuning detectors on social media data can improve performance, but this success is often platform or language-specific, requiring retraining to maintain effectiveness across different platforms or languages.

\smallskip
\noindent
$\clubsuit$ \textbf{Image-Specific Challenges}. 
Detecting LAIM-generated images remains challenging due to the increasing visual fidelity of modern generative models and the fragility of detectors under real-world conditions. In addition to the common challenges discussed above, the image modality faces specific obstacles related to robustness against post-processing and partial manipulations, localization and attribution, limited interpretability of detection outputs, and growing concerns about demographic fairness. 

Robustness is a critical issue, 
benchmarks demonstrate that even mild JPEG compression or resolution reduction significantly degrades detection accuracy~\cite{lin2024ai, yan2024df40}. Social media platforms exacerbate this challenge by routinely applying such transformations to uploaded images, necessitating highly robust detectors capable of handling diverse, processed content.

Beyond binary detection, localization itself presents additional challenges. Pinpointing partially manipulated regions is harder than global detection, as edits are often small, well-blended, and vary by forgery type~\cite{guo2023hierarchical}. Fine-grained ground truth is also labor intensive to obtain~\cite{tantaru2023weakly}. Moreover, artifacts tend to cluster near object boundaries or semantic inconsistencies~\cite{zhang2023perceptual}, motivating multi-view or object-level modeling~\cite{Li_2024_CVPR}.
Attribution, which identifies the generative model behind a fake image, faces several challenges. It is highly sensitive to domain shifts, minor post-processing like compression or resizing can degrade performance~\cite{sha2023fake, keita2025fidavl}. Attribution models also struggle to generalize to unseen or updated generators, as the fingerprints they rely on often change~\cite{guarnera2024level}. Unlike binary detection, attribution requires multi-class classification, increasing model complexity and the risk of confusion between similar generators.

Interpretability also remains limited. Most vision-based detectors output binary labels or heatmaps that highlight suspicious regions but do not provide clear, human-readable explanations. While recent frameworks like SIDA have introduced explanation-aware detection by combining tampered region localization with textual justifications, such methods are not yet standard and often struggle to align model reasoning with human perceptual cues. Lastly, fairness in image detection is an emerging concern~\cite{lin2024ai}, as detectors may exhibit bias across demographic groups. For example, forensic artifacts can manifest differently depending on skin tone, facial features, or lighting conditions, potentially leading to disparate false positive rates across racial or gender groups. 

\underline{Social Media.} Common social media operations, such as compression, resizing, or filtering, can erase subtle artifacts or noise patterns that detectors rely on. As a result, detectors trained on clean, high-quality datasets often experience severe performance degradation when applied to social media content~\cite{huang2024sida, yan2024sanity}. Another challenge is that LAIM-generated images shared on social media are often open-domain, covering a wide range of subjects beyond specific categories like faces or objects. Detectors trained on domain-specific datasets, such as face-focused datasets~\cite{lin2024ai}, may fail to generalize to diverse content types in the open-world setting.

\smallskip
\noindent
$\clubsuit$ \textbf{Video-Specific Challenges}. In addition to inheriting the challenges faced by image-based detection, video introduces added complexity due to its temporal dimension. Convincing LAIM-generated deepfake videos often exhibit smooth motion, consistent lighting, and accurate lip-sync across frames, leaving only subtle temporal artifacts, such as unnatural blinking, jitter, or slight misalignments between facial expressions and speech, for detectors to identify. 
Additionally, the development of robust and generalizable video detection methods is hindered by the lack of large-scale, diverse datasets featuring LAIM-generated videos. Furthermore, the absence of standardized video deepfake benchmarks makes it difficult to compare methods fairly or assess their generalization to real-world deployment scenarios. Lastly, video is inherently multimodal, typically accompanied by audio that may itself be synthesized. This requires detection systems to go beyond visual cues and incorporate cross-modal consistency analysis, such as verifying alignment between speech and lip movement or between speaker identity and voice.

\underline{Social Media.} Social media platforms further complicate detection. Same as images, compression, resolution reduction, filtering, and watermarking, common on platforms like TikTok and X, can obscure the subtle forensic artifacts that detectors rely on. Moreover, the fast-paced, high-volume nature of social media demands near instantaneous analysis to flag LAIM-generated videos before they spread widely, yet processing high-resolution, temporally complex, and multimodal content in real-time remains computationally demanding. Current detection systems, often trained on curated datasets, struggle to achieve the speed and accuracy needed for real-time deployment.

\smallskip
\noindent
$\clubsuit$ \textbf{Audio-Specific Challenges}. Detecting LAIM-generated audio, such as synthetic speech or voice clones, is particularly difficult due to the subtlety of generative artifacts and the variability of real-world acoustic conditions. Forensic cues, such as unnatural prosody, spectrogram anomalies, or phase irregularities, are often very subtle and difficult to detect consistently. The broad range of recording qualities, from clean studio recordings to noisy mobile captures, exacerbates the generalization problem, as models trained on curated datasets often fail in uncontrolled settings. This challenge is compounded by the lack of large-scale, diverse LAIM-generated audio datasets and standardized benchmarks, which hinders both model development and fair evaluation across methods.

\underline{Social Media.} Social media platforms add additional challenges for LAIM-generated audio detection. Audio shared online is often compressed into low-bitrate formats, mixed with background noise, or embedded within videos, further masking forensic traces. Compression, platform specific filtering, and varying playback conditions can distort or erase subtle synthetic artifacts, making detection much harder. As with video, the fast dissemination speed on social media requires detectors to operate in real time, balancing accuracy with computational efficiency under noisy, degraded conditions.

\smallskip
\noindent
$\clubsuit$ \textbf{Multimodal-Specific Challenges}.LAIM-generated multimodal content, where text, images, audio, and video are combined, presents distinct detection difficulties beyond those seen in unimodal scenarios. One challenge lies in identifying cross-modal inconsistencies, such as mismatches between audio and lip movements or between image content and accompanying text. Single modality detectors often fail to capture such discrepancies. Further complicating detection is the presence of partially manipulated media, where only one modality (\eg, audio in a real video) is synthetic, requiring systems to localize which component is fake.
Moreover, there is a critical lack of large-scale LAIM-generated multimodal datasets and standardized benchmarks that cover richly entangled combinations of modalities. Recent evaluations, such as those in the LOKI benchmark~\cite{ye2024loki}, show that no universal detection model currently performs reliably across modalities: large multimodal models, while promising in reasoning and explainability, still exhibit modality imbalance and perform poorly on complex tasks involving audio or video. These challenges highlight the nascency of multimodal detection research and the urgent need for scalable, explainable, and cross-modality aware detection systems.

\underline{Social Media.} On social media platforms, multimodal content is especially prevalent, often subject to compression, stylization, or casual editing. These platform specific transformations obscure cross-modal cues and make it easier to conceal manipulations across one or more modalities. For example, AI-generated deception can subtly modify a single modality while preserving overall narrative coherence~\cite{cai2024av}. Furthermore, the need to analyze multiple data streams (\eg, video frames, audio tracks, and text captions) imposes substantial computational overhead, which is non-trivial at the scale of social media.

\subsubsection{Challenges of Detection Tools }
A primary challenge lies in the lack of interpretability and user trust. Many current detection tools operate as black boxes, offering binary outputs ( or confidence scores without any contextual explanation or rationale. This lack of interpretability can undermine user trust, particularly for non-technical users such as journalists, educators, or fact-checkers, who require transparency to make informed decisions.

A second limitation is the lack of transparency. As shown in Table~\ref{tab:my-table}, the majority of high performance tools are closed-source commercial platforms. These systems typically do not disclose details about their model architectures, training datasets, or potential biases, which hinders scientific reproducibility and independent evaluation.

Lastly, current tools often fall short in supporting real-time detection. Most existing tools are only accessible through web applications, which may not align well with user needs for fast and real-time detection. For example, users may need to analyze audio or video content on social media platforms in real time, but most current tools often require uploading files manually through dedicated portals.

\subsection{Future Directions}

\subsubsection{Common Future Directions Across Modalities}
\noindent
$\clubsuit$ \textbf{Enhance Generalizability, Robustness, and Interpretability}. 
Future detection systems must be more adaptable to evolving generative models, robust against real-world distortions, and transparent in their decision-making. Training on a diverse and large-scale dataset, building robust detectors with attacks, incorporating explainable AI (XAI) techniques~\cite{arrieta2020explainable}, considering strategies like online learning~\cite{epstein2023online} or transfer learning with new data to keep the model updated would contribute to building such a reliable detector.

\smallskip
\noindent
$\clubsuit$ \textbf{Social Media Adaptation}. 
As LAIM-generated content proliferates on platforms like X, TikTok, and Reddit, future detection systems must adapt to the unique constraints of social media environments. These platforms introduce compression, format conversions, and short form or informal content that degrade detection performance~\cite{macko2024multisocial, vahdati2024beyond}. To enhance robustness, future work should incorporate platform specific data and leverage domain adaptation or robustness oriented training strategies~\cite{xu2023exposing, seraj2025multi}. Additionally, integrating rapid provenance verification mechanisms, such as watermark detection~\cite{fernandez2023stable, Zhang_2024_CVPR} or cryptographic signatures like C2PA~\cite{c2pa2025}, can enable scalable real-time screening prior to deep forensic analysis. Several studies also highlight the need for social-media aware benchmarks to reflect realistic deployment conditions and avoid overfitting to clean or curated datasets~\cite{sun2024we, gameiro2024llm, papadopoulos2024verite}. Across modalities, social media adapted detection remains a crucial direction for practical, deployable solutions.

\subsubsection{Modality-Specific Future Directions}
\noindent
$\clubsuit$ \textbf{Text–Future Directions}.
Advances in LAIM text detection will center on making models generalizable across different languages. One approach is to train multilingual detectors so that detection is not limited to English texts.
Researchers are also investigating fine-grained stylometry and semantic consistency checks, for instance, does the text follow unnatural repetition or omit context in a way that humans typically would not? Addressing the text domain’s challenges requires embracing its diversity: models must become fluent in the spectrum of human and AI writing styles to flag machine-generated content reliably in any context.

\underline{Social Media.} Future detectors should also learn from social media data, mastering the idioms, shorthand, and irregular grammar that characterize those platforms. Incorporating training corpora from social media and building models that recognize human-like variability versus AI consistency will be critical for improving robustness. Given the casual and often fragmented nature of social media language, detectors must balance sensitivity to AI artifacts with tolerance for authentic human irregularities.

\smallskip
\noindent
$\clubsuit$ \textbf{Image – Future Directions}. 
As LAIM-generated images become increasingly photorealistic, future detection systems should integrate diverse forensic cues into a unified framework. Despite recent advances, current diffusion models still exhibit limitations in generating high frequency details~\cite{ricker2022towards, wolter2022wavelet}, rich textures~\cite{zhong2023rich}, repetitive patterns~\cite{borji2023qualitative}, and physically consistent lighting~\cite{farid2022lighting}. Effective detection strategies may leverage a combination of these shortcomings by incorporating multiple complementary signals. For example, frequency domain analysis can detect hidden spectral anomalies or periodic artifacts subtle generative “fingerprints” that persist even in natural looking images. Physics-based cues, such as inconsistencies in lighting, shadows, and reflections, can expose violations of real-world constraints that are difficult for generative models to replicate. Physiological cues, such as unnatural facial geometry or distorted proportions, further contribute to both robustness and interpretability. These heterogeneous cues can be effectively utilized through techniques such as ensemble learning\cite{sagi2018ensemble}, multitask learning, or by employing foundation models\cite{bommasani2021opportunities} trained across diverse visual forensic tasks.

\underline{Social Media.}
Rapid provenance verification, checking for cryptographic certificates of origin or embedded authenticity signals, will become increasingly important for detecting social media processed images. Initiatives like C2PA~\cite{c2pa2025} aim to attach secure provenance metadata to images at creation time, enabling detectors to validate content before resorting to computationally intensive forensic analysis. Similarly, watermarking strategies~\cite{fernandez2023stable, Zhang_2024_CVPR} are being developed to embed resilient signals that survive typical social media transformations. Future detection pipelines may thus adopt a two-tier approach: a fast check for provenance indicators (metadata or watermarks), followed by deeper forensic investigation if no authenticity signals are found. This strategy could enable scalable monitoring and early flagging of LAIM-generated images in the high-volume, fast-paced environment of social media.

\smallskip
\noindent
$\clubsuit$ \textbf{Video – Future Directions}. Future research should explore finer-grained modeling of temporal dynamics across frames. This could include the design of lightweight yet expressive spatiotemporal architectures that integrate intra and inter-frame attention mechanisms to capture inconsistencies in motion continuity. Such modeling is particularly important given that even advanced generative models like Sora often exhibit temporal artifacts such as jittery motion, abrupt transitions, and object flicker~\cite{chang2024matters, sun2024sora}-failures that cannot be reliably detected by frame-level analysis alone.
Real-time detection remains a bottleneck due to the high computational cost of analyzing full resolution video streams with temporal context. To meet this demand, future work should prioritize model compression techniques (\eg, quantization, pruning, knowledge distillation) and develop efficient early-exit mechanisms or key-frame selection strategies to reduce latency. Architectures optimized for edge inference can facilitate on-device deployment on mobile or streaming platforms.
Finally, the absence of large-scale, diverse, and standardized datasets of LAIM-generated videos hinders both generalization and benchmarking. Future initiatives should aim to create dynamic, open-source datasets that encompass a range of generative models, scene types, and manipulations. Alongside this, establishing unified evaluation benchmarks and protocols would enable consistent assessment of detection performance across varying video conditions, promoting methodological comparability and real-world readiness.

\underline{Social Media.} Compression and resolution degradation applied by social platforms can distort forensic traces. One promising direction is the development of preprocessing pipelines that learn to restore or enhance compression-corrupted artifacts relevant to forgery detection~\cite{ke2023df, shan2025mscscc}. Additionally, adversarial training and domain adaptation strategies~\cite{seraj2025multi, lv2024domainforensics} could improve generalization across post-processed and platform specific video formats. Integrating rapid provenance verification, such as detecting watermark signatures or metadata if available, before deep forensic analysis could further help scale detection systems.

\smallskip
\noindent
$\clubsuit$ \textbf{Audio – Future Directions}. 
One of the biggest challenges in detecting LAIM-generated audio is the lack of large-scale, diverse datasets and standardized benchmarks, which significantly hinders the development and fair evaluation of audio detection systems. To address this, future research should expand the scope of training data to include a wide variety of synthetic audio samples spanning multiple languages, speaker identities, generation techniques, and real-world distortions such as background noise or channel degradation. Exposure to such diversity will improve model generalization and robustness in uncontrolled environments.
Beyond dataset expansion, future efforts should also focus on more robust forensic features. Rather than relying solely on spectral or vocoder artifacts, which are often erased by compression or playback, detectors may increasingly target features that are harder for AI to replicate, such as breathing patterns~\cite{layton2024every}, micro-pauses~\cite{kulangareth2024investigation}, and subtle pitch or timing irregularities~\cite{warren2025pitch} that emerge from human cognitive and physiological processes.

\underline{Social Media.}  
Audio deepfake detection on social media faces unique challenges, as audio is often shared alongside video and subjected to platform specific compression and noise. A promising direction is the use of speaker verification and authentication techniques that leverage speaker embeddings to capture individual-specific vocal traits~\cite{tran2024spoofed, pianese2022deepfake}. By incorporating speaker verification frameworks, detectors can perform personalized authenticity checks, comparing a user's voice against reference samples to detect mismatches indicative of manipulation. On social media, identity-aware methods that jointly analyze audio and visual cues, such as fusing speaker recognition with facial analysis, can further enhance detection robustness, particularly when cross-modal inconsistencies between voice and appearance are present.

\smallskip
\noindent
$\clubsuit$ \textbf{Multimodal–Future Directions.} Future multimodal detection systems should jointly analyze text, image, audio, and video to uncover cross-modal inconsistencies and detect partial modality manipulations. This calls for unified or modular architectures capable of both localizing manipulated modalities and reasoning across them. Developing such systems requires large-scale, LAIM-generated multimodal datasets that reflect realistic, partially edited content. Recent benchmarks such as LOKI~\cite{ye2024loki} and MMFakeBench~\cite{liu2024mmfakebench} reveal that current large multimodal models still struggle, particularly on audio and video inputs, underscoring the need for modality-balanced training and fine-grained supervision. 
Advances in MLLMs (multimodal large language models), such as LLaMA 4~\cite{meta2025llama4}, GPT-4o~\cite{openai2024gpt4o}, and Gemini 2.5 Flash~\cite{deepmind2025gemini25flash}, combined with domain-specific forensic knowledge, offer promising foundations for building robust multimodal detection systems.

\underline{Social Media.}  To enable reliable verification in complex, real-world scenarios such as social media, future multimodal detectors must be scalable and resilient to noisy, user-generated content, including informal, stylized, or culturally nuanced formats like memes and TikTok videos. Achieving this may require specialized pre-processing pipelines and platform specific training strategies. Rapid provenance verification and modality-specific robustness enhancements will also be important for scalable deployment.

\subsubsection{Future Directions for Detection Tools} 
Our analysis of popular online detection tools discussed in Section~\ref{sec:tools}, reveals a common shortcoming: a lack of interpretability. 
Developing interpretable detection tools can therefore gain more social attention and user adoption. To enhance interpretability, future systems could integrate real source verification, physical and physiological cues, and multimodal large language model reasoning to provide contextual explanations for why content is identified as synthetic. Moreover, human-in-the-loop detection platforms (\eg, involving expert review of challenging cases) that combine AI analysis with expert review can ensure more accountable deployment. To improve transparency and reproducibility, open-sourcing model architectures, datasets, and evaluation protocols, where ethically and legally feasible, should be prioritized. In addition, to better support real-time detection needs, future tools should be designed with lightweight, integrable formats such as browser extensions, enabling seamless deployment across social media, messaging apps, and content platforms.
This is not only a technical upgrade but also makes these tools more accessible to broader user groups, particularly underrepresented populations such as non-technical individuals and elderly adults.

\section{Conclusion}\label{sec:Conclusion}
This paper provides the first systematic and comprehensive survey covering existing research on detecting multimedia from text, image, audio, and video to multimodal content generated by large AI models. We introduce a novel taxonomy for detection methods within each modality, categorizing them under two primary frameworks: \textit{pure detection} (focusing on improving detection accuracy) and \textit{beyond detection} (integrating attributes like generalizability, robustness, and interoperability to detectors). We further analyze them with modality-specific insights and recommendations. Additionally, we have 
outlined the sources contributing to detection, such as generation mechanisms of LAIMs, public datasets, online tools, evaluation metrics. We also emphasize the importance of detecting social media processed multimedia throughout the survey, systematically reviewing social media specific datasets, detection methods, and unique challenges across modalities. Finally, we pinpoint current challenges in this field and propose potential directions for future research in a modality-specific manner. We believe that this survey serves as the initial contribution to addressing a notable academic gap in this field, aligning with global AI security initiatives, thereby upholding the authenticity and integrity of digital information.

\section*{Acknowledgments}
This work is supported by the U.S. National Science Foundation (NSF) under grant IIS-2434967, and the National Artificial Intelligence Research Resource (NAIRR) Pilot withand TACC Lonestar6, and Purdue University Seed Funding for High-Impact Review Papers.
The views, opinions and/or findings expressed are those of the author and should not be interpreted as representing the official views or policies of NSF, and NAIRR Pilot, and Purdue University.



 

\ifCLASSOPTIONcaptionsoff
  \newpage
\fi

\bibliographystyle{IEEEtran}
\bibliography{main}
\vspace{-1.0cm}

\begin{IEEEbiography}
[{\includegraphics[width=1.0in,height=1.20in]{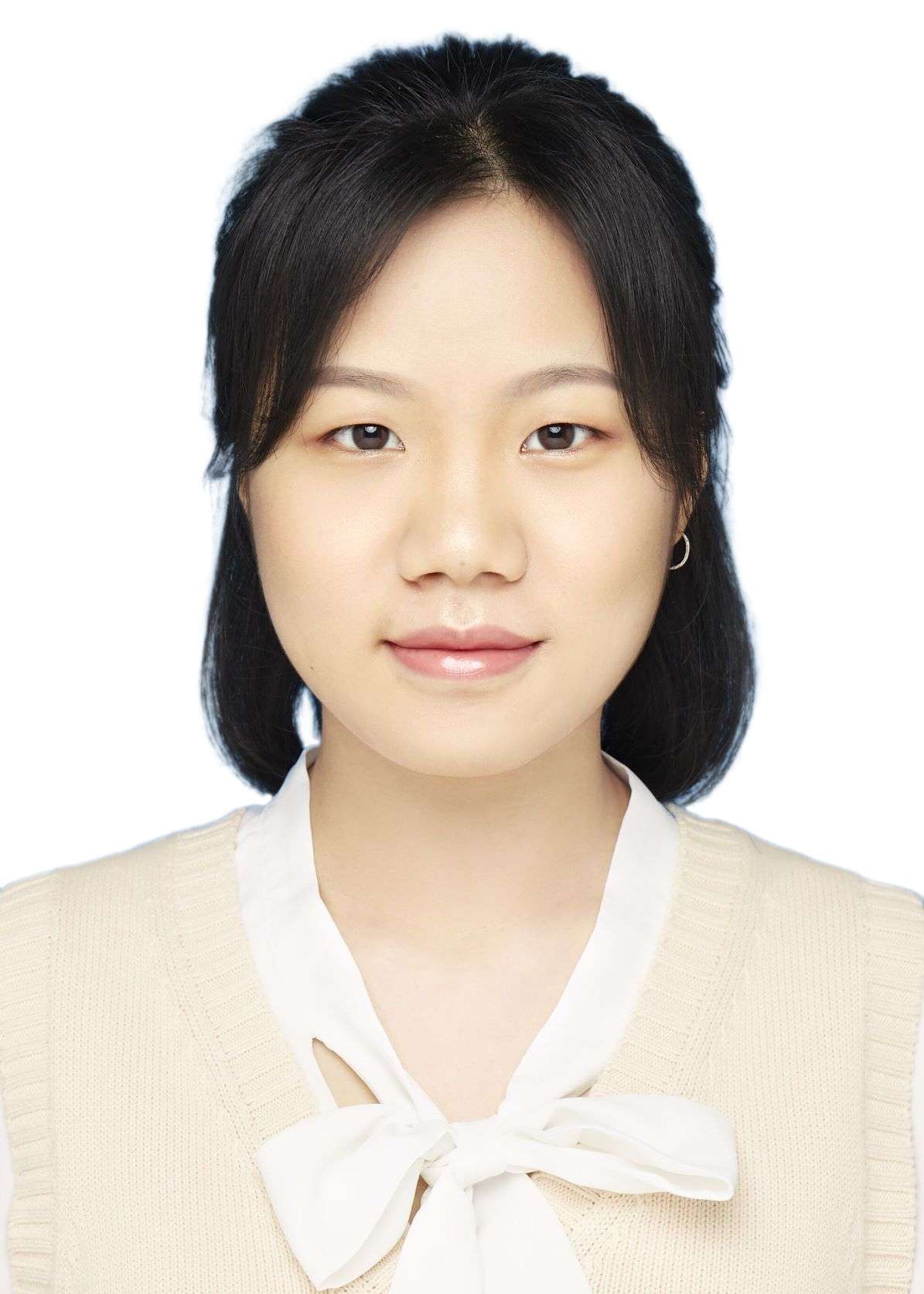}}]
{Li Lin} received the B.S. degree in communication engineering from Chongqing University, China, in 2020. She is a Ph.D. student in the Department of Computer Science, Purdue University. Her research interests include computer vision, digital media forensics, and deep learning.
  \vspace{-1.5cm}
\end{IEEEbiography}

\begin{IEEEbiography}
[{\includegraphics[width=1.0in,height=1.20in]{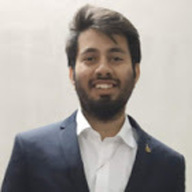}}]
{Neeraj Gupta} received the B.S. degree in Technology from the Indian Institute of Technology Kanpur, in 2019. He is a master student in the Department of Computer Science, Purdue University in Indianapolis. His research interests include machine learning, computer vision, and digital media forensics.
  \vspace{-1.5cm}
\end{IEEEbiography}

\begin{IEEEbiography}
[{\includegraphics[width=1.0in,height=1.20in]{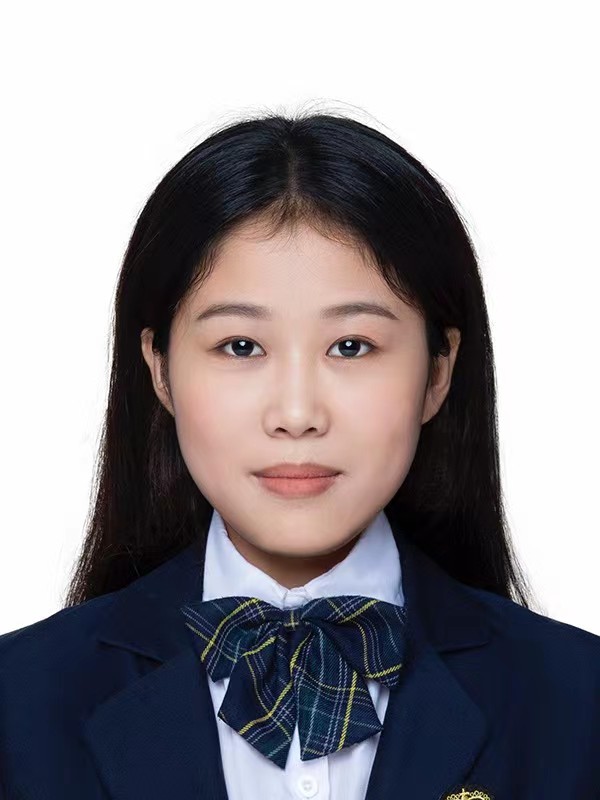}}]
{Yue Zhang} received the B.S. degree in network engineering from Anhui Jianzhu University, Hefei, China, in 2023. She is a master student in the School of Software, Nanchang University, Nanchang, China. Her research interests mainly include digital forensics, machine learning, and digital image processing.
  \vspace{-1.5cm}
\end{IEEEbiography}

\begin{IEEEbiography}
[{\includegraphics[width=1.0in,height=1.20in]{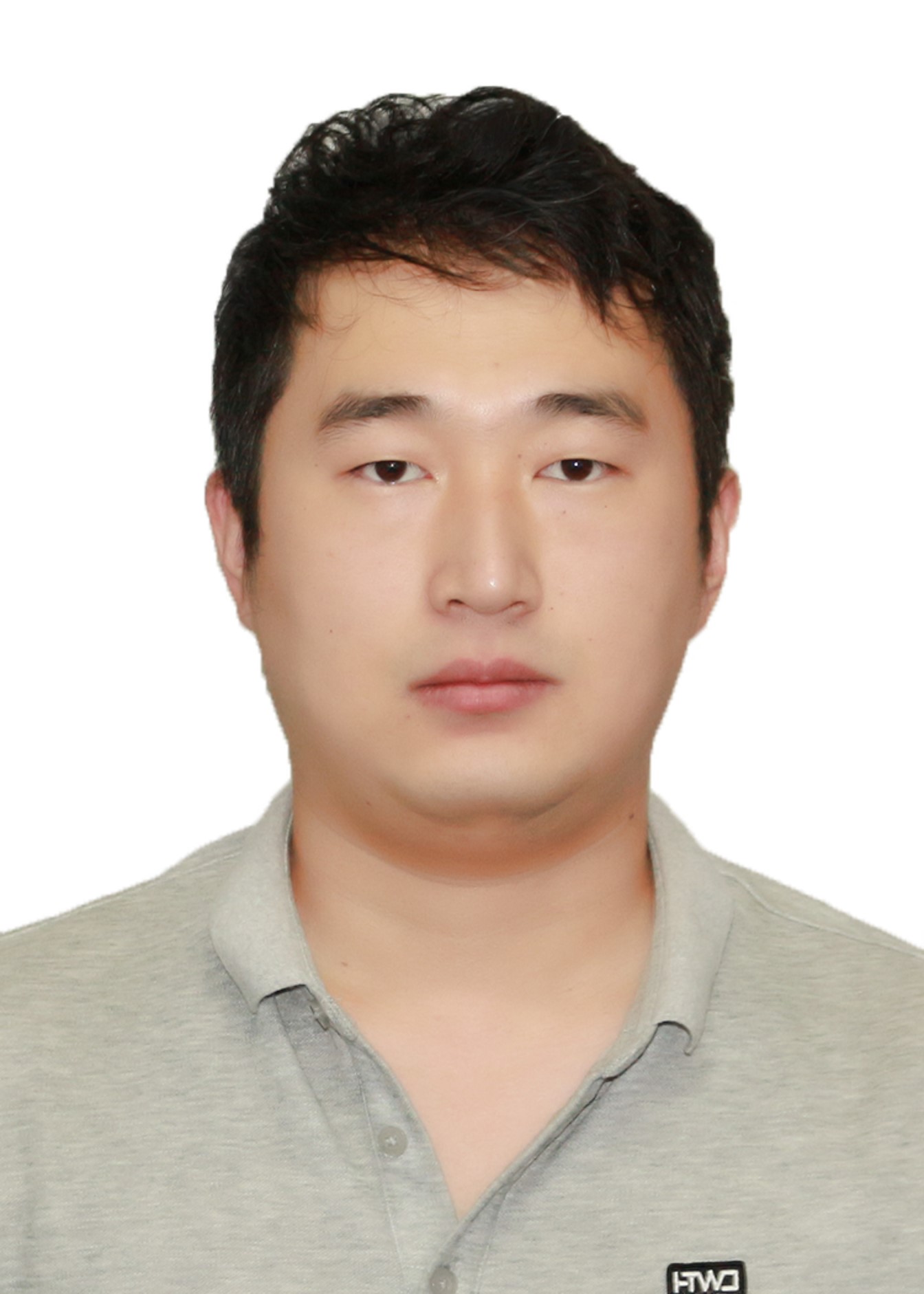}}]
{Hainan Ren} received the B.S. degree in automation engineering from Hebei University of Technology in 2012, and the M.S. degree in control engineering from Hebei University of Technology in 2015. He is a senior engineer in Algorithm Research, Aibee Inc. His research interests include face recognition, person re-identification, multi-modal learning, and generative models.
  \vspace{-1.0cm}
\end{IEEEbiography}

\begin{IEEEbiography} 
[{\includegraphics[width=1.0in,height=1.20in]{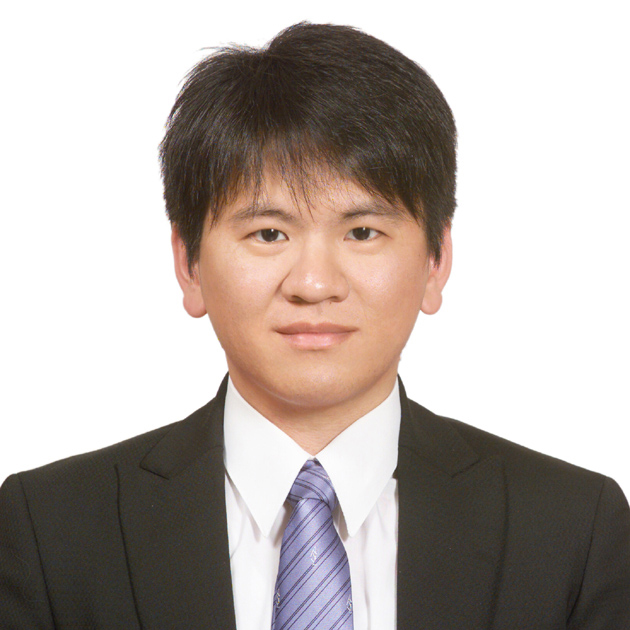}}]
{Dr. Chun-Hao Liu} received the B.S. degree in electronics engineering from National Chiao Tung University in 2007, M.S. degree in electronics engineering from National Taiwan University in 2009, and PhD degree in electrical engineering from University of California, Los Angles in 2015. He is currently with Amazon Prime Video as a senior applied scientist. His research interests are computer vision, deep learning, and signal processing.
  \vspace{-1.5cm}
\end{IEEEbiography}

\begin{IEEEbiography}
[{\includegraphics[width=1.0in,height=1.20in]{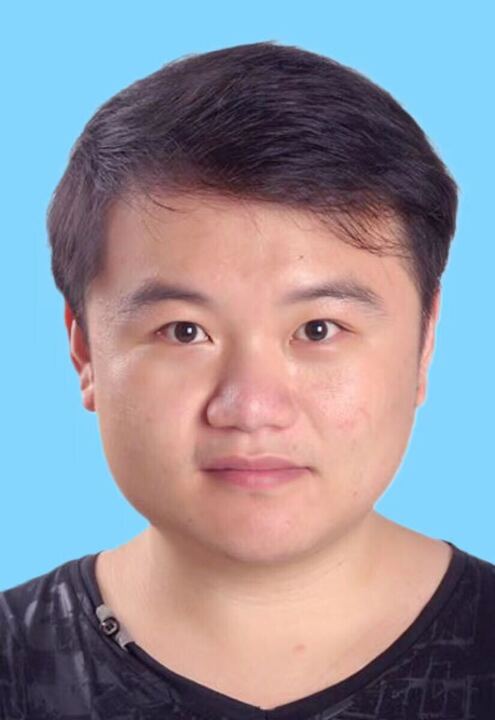}}]
{Dr. Feng Ding} received the B.S. degree from Huazhong University of Science and Technology, China, in 2007, and the M.S. and Ph.D. degrees in electrical and computer engineering from New Jersey Institute of Technology, Newark, NJ, USA, 2011 and 2017, respectively. He was a Postdoctoral Researcher with the University at Albany, SUNY, USA, from 2019 to 2020. He is currently a professor at the School of Software, Nanchang University, China. His current research interests mainly include digital forensics, machine learning, and digital image processing.
  \vspace{-1.0cm}
\end{IEEEbiography}

\begin{IEEEbiography}
[{\includegraphics[width=1.0in,height=1.20in]{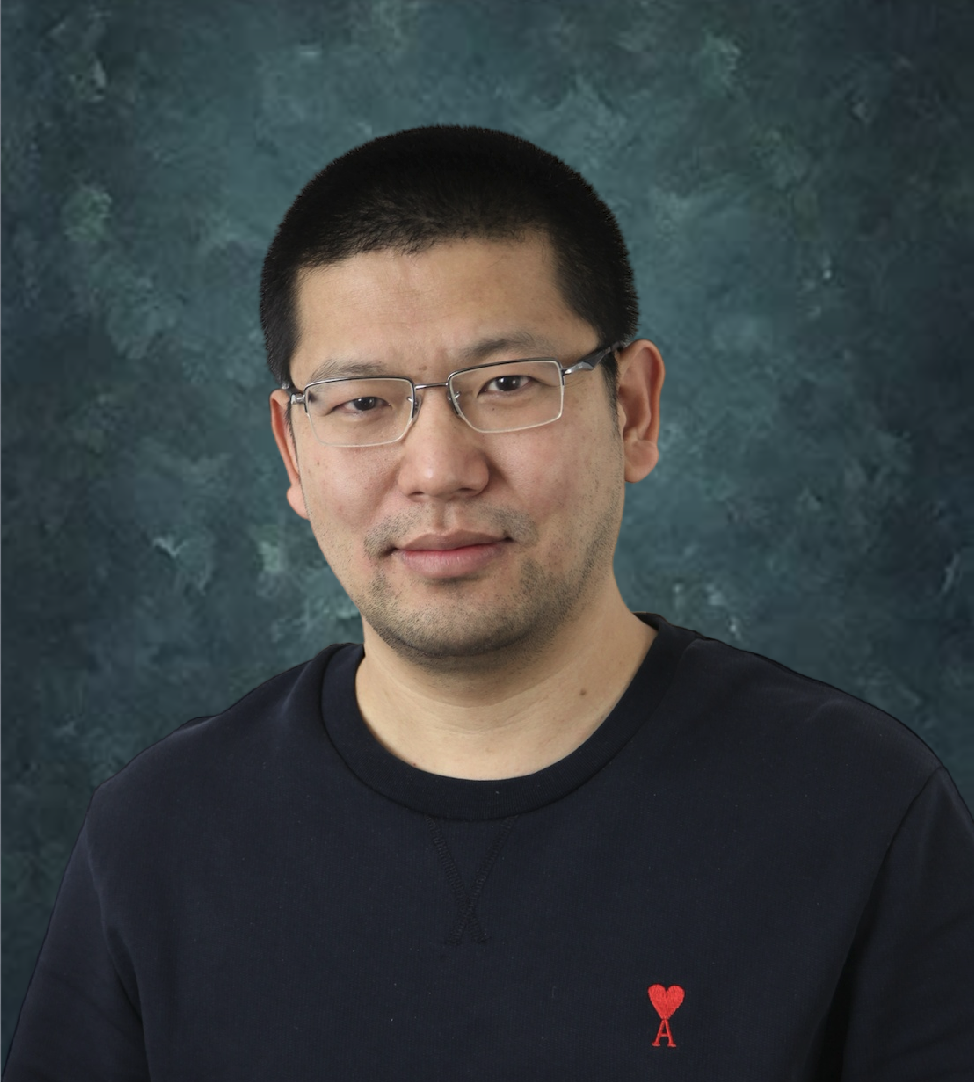}}]
{Dr. Xin Wang} received the PhD
degree in computer science from the University at
Albany, State University of New York (SUNY), in
2015. He is an assistant professor with the Department
of Epidemiology and Biostatistics, School of Public
Health, University at Albany, SUNY. His research
interests include artificial intelligence, reinforcement
learning, deep learning, and their applications. He is a Senior Member of IEEE.
  \vspace{-1.0cm}
\end{IEEEbiography}

\begin{IEEEbiography}
[{\includegraphics[width=1.0in,height=1.20in]{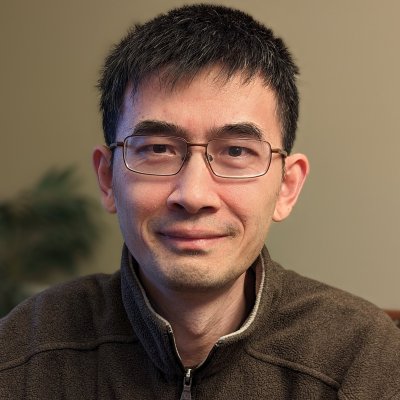}}]
{Dr. Xin Li} received the B.S. degree with highest honors in electronic engineering and information science from University of Science and Technology of China, Hefei, in 1996, and the Ph.D.degree in electrical engineering from Princeton University, Princeton, NJ, in 2000. 
He was a Member of Technical Staff with Sharp Laboratories of America, Camas, WA from Aug. 2000 to Dec. 2002. He was a faculty member in Lane Department of Computer Science and Electrical Engineering, West Virginia University from Jan.2003 to Aug. 2023. 
He is with the Department of Computer Science, University at Albany, Albany, USA. His research interests include image and video processing, computer vision and computational neuroscience. He is a Fellow of IEEE. 
  \vspace{-1.0cm}
\end{IEEEbiography}

\begin{IEEEbiography}
[{\includegraphics[width=1.0in,height=1.20in]{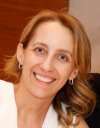}}]
{Dr. Luisa Verdoliva} a professor with the Department of Electrical Engineering and Information Technology, University Federico II, Naples, Italy. She is Editor-in-Chief for IEEE Transactions on Information Forensics and Security (2025–2027) and senior area editor for IEEE Signal Processing Letters. She is the recipient of a Google Faculty Research Award for Machine Perception (2018) and a TUM-IAS Hans Fischer Senior Fellowship (2020–2024). She was chair of the IFS TC (2021–2022). Her scientific interests are in the field of image and video processing, with main contributions in the area of multimedia forensics. She is a Fellow of IEEE.
  \vspace{-1.0cm}
\end{IEEEbiography}

\begin{IEEEbiography}
[{\includegraphics[width=1.0in,height=1.20in]{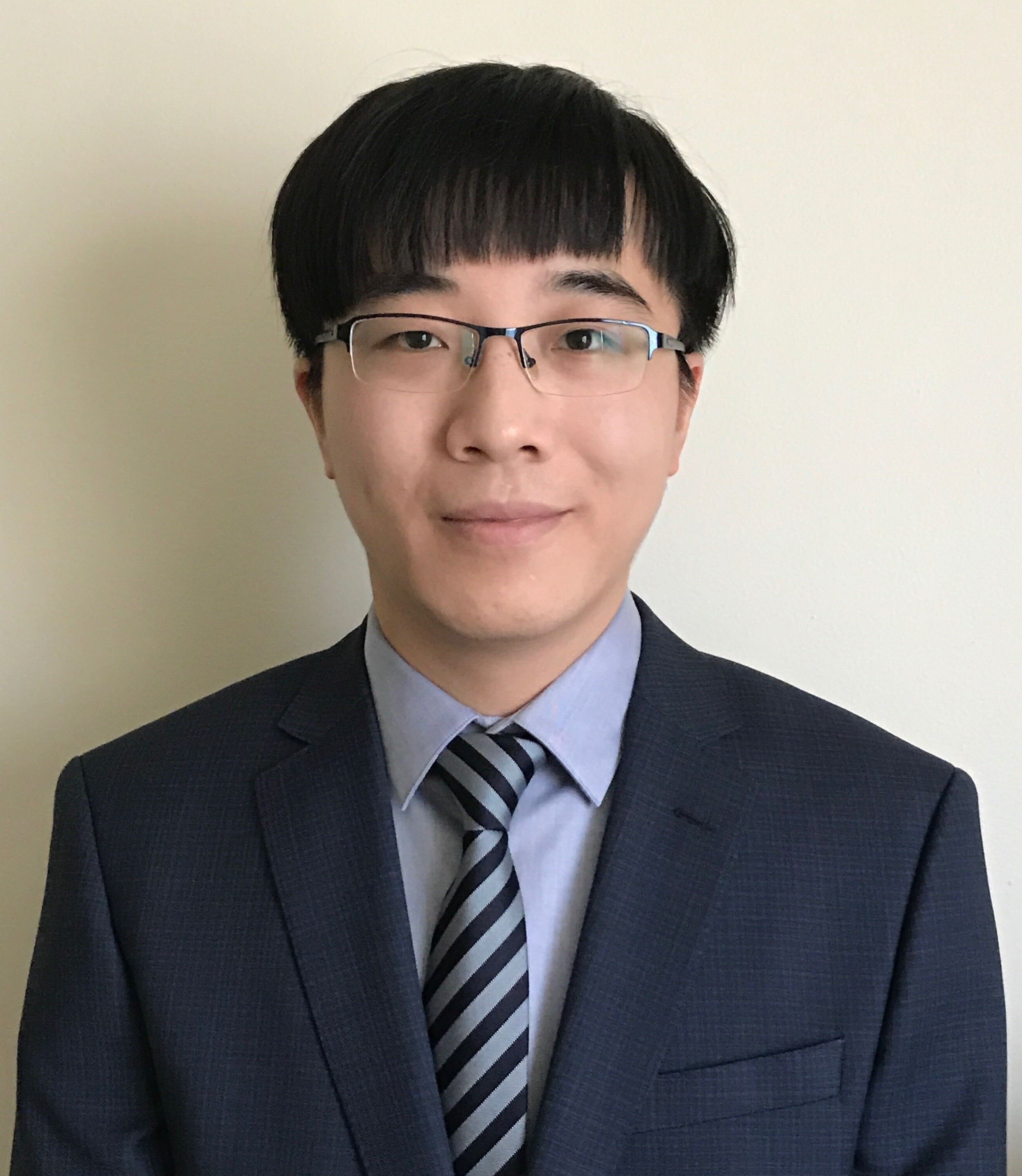}}]
{Dr. Shu Hu} received the MEng degree in software engineering from the University of Science and Technology of China, in 2016, the MA degree in mathematics
from University at Albany, SUNY, in 2020, and the PhD
degree in computer science and engineering from
University at Buffalo, SUNY, in 2022. He is an assistant professor in the
Department of Computer and Information  
Technology, Purdue University. He was a postdoc at Carnegie
Mellon University. His research interests include machine learning, multimedia
forensics, and computer vision. He is a member of IEEE.  
  \vspace{-1.0cm}
\end{IEEEbiography}





\end{document}